\documentclass[11 pt]{article}
\usepackage[T1]{fontenc}
\usepackage[left=1 in,right=1 in, top=0.75 in, bottom=0.85 in]{geometry}
\usepackage[english]{babel}
\usepackage{fourier} 
\usepackage{amstext}
\usepackage{amsmath}
\usepackage{adjustbox}
\usepackage{calc}
\usepackage{caption}
\usepackage{setspace}
\usepackage[dvipsnames]{xcolor}
\usepackage[
  colorlinks=true,linkcolor=blue, urlcolor=red]{hyperref}
  
\AtBeginDocument{\hypersetup{citecolor=Blue,linkcolor=BrickRed}}
\usepackage{color, colortbl}
\usepackage{booktabs}
\usepackage{threeparttable}
\usepackage{multirow,array}
\usepackage[dvipsnames]{xcolor}

\usepackage{tikz}
\usepackage{pgfplots}
\pgfplotsset{compat=newest,
samples=500,
every axis/.append style={
                    axis line style={thick},
                    label style={font=\scriptsize},
                    tick label style={font=\footnotesize},
                    xlabel style={at={(ticklabel* cs:1)},anchor=north west},
                    ylabel style={at={(ticklabel* cs:1)},anchor=south west}
                 }
}

\tikzset{
    myarrow/.style = {-{Triangle[length = 1.5mm, width = 1.5mm]}}
}

\pgfplotsset{yticklabel style={text width=3em,align=right}}

\usetikzlibrary{decorations.pathreplacing}

\usetikzlibrary{calc,patterns,positioning}
\usepgfplotslibrary{fillbetween}
   \usetikzlibrary{arrows,snakes,backgrounds,automata,positioning,arrows.meta}
\usepackage[font=footnotesize]{caption}
\usetikzlibrary{arrows.meta}
\usepackage[outline]{contour} 
\tikzset{>=latex} 
\colorlet{myred}{BrickRed!70!purple}
\colorlet{mygreen}{green!50!olive}
\colorlet{myblue}{blue!70!gray}
\colorlet{mypurple}{purple!70!black}

\usepackage{adjustbox}
\usetikzlibrary{decorations.pathmorphing}
\tikzset{discont/.style={decoration={zigzag,segment length=12pt, amplitude=3pt},decorate}}
\def\discontarrow(#1)(#2)(#3)(#4);{
  \draw[discont] (#2) -- (#3);
  \draw[thick][->] (#1) -- (#2) (#3) -- (#4);
}
\renewcommand{\thesubsection}{\thesection.\Alph{subsection}}

\usepackage[redeflists]{IEEEtrantools}
\newtheorem{theorem}{Theorem}[section]
\newtheorem{lemma}[theorem]{Lemma}

\newtheorem{assm}[theorem]{Assumption}
\newtheorem{defn}[theorem]{Definition}

\newtheorem{prop}[theorem]{Proposition}

\usepackage{graphicx} 

\usepackage{graphicx} 
\usepackage{pgf,tikz}
\usetikzlibrary{babel} 
\usepackage{circuitikz}
\usepackage{siunitx}
\usepackage{hyperref} 
\usetikzlibrary{arrows,calc}
\usepackage{relsize}

\usepackage{tcolorbox}
\usepackage[notocbib]{apacite}
\usepackage{adjustbox}
\usepackage{float}
\usepackage{enumitem}
\usepackage{multibib}

\title{Back to the Surplus: An Unorthodox Neoclassical Model of Growth, Distribution and Unemployment with Technical Change }

\author{Juan Jacobo\thanks{I thank Anwar Shaikh and Duncan Foley for insightful comments that helped improve the paper. } \\ Economics Department, Externado University of Colombia. }

\date{\today}

\begin{document}
\maketitle

\begin{abstract}
The article examines  how institutions,  automation, unemployment and income distribution interact in the context of a neoclassical growth model where profits are interpreted as a surplus over costs of production.   Adjusting the  model to the experience of the US economy, I show that joint variations in labor institutions and technology are required to provide reasonable explanations for the behavior of income shares, capital returns, unemployment, and the  ``big ratios'' in macroeconomics. The model offers new perspectives on recent trends by showing that they can be analyzed by the interrelation between the profit-making capacity of capitalist economies and the political environment determining labor institutions. 

\bigskip
\smallskip
\noindent \emph{Keywords.}  Aggregate surplus, equilibrium unemployment, automation, equilibrium growth, labor institutions. 


\end{abstract}


 
\clearpage
\pagenumbering{arabic}

\section{Introduction}

Over the past 50 years, the US economy  experienced a large decline in  the labor share, a considerable rise   in  capital returns, and a surge in income and wage  inequality. Isolating for short-run fluctuations,  this  all occurred almost simultaneously with a fall in the rate of unemployment, an increase in the intensive use of capital, and  a low and stable  rate of inflation. It also coincided with institutional changes which affected the balance of power between workers and firms reflected,  among other things, in a reduction in the incidence of unions, a fall in the real value of minimum wages,  and  smaller  benefits per unemployed relative to average labor productivity. 
 
 These empirical regularities have attracted a great deal of attention but, given the complexity of the phenomena,  have been studied for the most part in isolation.  A case in point is the  study of the falling  labor share, which has been explained by four main hypotheses: technological change, represented in the declining relative price of capital \cite{karabarbounis2014global,piketty2014capital}; automation and offshoring, normally described by a task-based formalism  \cite{acemoglu2018race,grossman2008trading}; enhancing market power by large firms, usually measured by rising markups  \cite{autor2020fall, de2020rise}; and the  eroding bargaining  power of workers \cite{DiNardo1996,stansbury2020declining,farber2021unions}.

While each one of these hypotheses has been successful in explaining part of story behind the declining labor share, they are inconsistent or silent about other key empirical regularities.  Narratives based on technological change, for instance,  depend on the existence of gross substitution between labor and capital, which is at odds with the findings of numerous studies; see \citeA{chirinko2008sigma} for a survey review. The automation hypothesis, though it provides a general equilibrium rationale for the decline in the labor share, cannot account for the rising return of capital.\footnote{\citeA{hubmer2021not} extend the task-based framework to include positive markups. In doing so, however, they maintain similar problems as those encountered by the market power hypothesis.} The market power hypothesis, as noted by \citeA{stansbury2020declining}, is hard to reconcile  with  the falling rates of unemployment and inflation given that an increasing monopoly or monopsony power is likely associated with less hiring and a higher pass-through of profits to prices. Lastly, the hypothesis of the  eroding  bargaining power of  workers, though it offers a unified explanation for the rise in profitability, the decreasing wage share, the falling rate of unemployment, and the low and stable  inflation, cannot account for  the increase in the intensive use of capital and provides no endogenous theory for the determination of the rate of return of capital.

The main contribution of this paper is to offer a framework for an endogenous  theory of aggregate profits  in a  competitive general equilibrium environment with which to understand the aforementioned empirical regularities. This approach builds upon Sraffa's \citeyear{sraffa1960production}  critique of the mainstream interpretation of "costs of production,"  which  is known to be logically inconsistent with the presence of equilibrium prices with uniform rates of return \cite{garegnani1990quantity,eatwell2019cost}.\footnote{It is commonly thought that the criticisms towards the logical consistency of the marginalist approach  are confined to the theories of distribution  using aggregate production functions (see, for example, \citeNP{hahn1982neo}), when in  fact  the critiques that followed Sraffa's \citeyear{sraffa1960production} contribution show that neoclassical theories which attempt to explain equilibrium prices and income distribution based on supply and demand equations  alone are generally indeterminate in competitive environments with a uniform rate of return \cite{eatwell1990walras, garegnani1990quantity}. This conclusion also covers the  Arrow-Debreu  model, which only provides a coherent solution to the price equations by abandoning the effort of determining a long-run equilibrium with a single rate of profit.}  By rejecting the neoclassical interpretation of costs, the paper restores the view of aggregate profits as a surplus and with it the need of referring to political and institutional factors as  key determinants of income distribution. In essence, though the model accepts  the premise of competitive-rational behavior of neoclassical economics, it rejects the view that commonly accepted concepts like  ``aggregate capital,''  ``marginal productivity,'' and ``costs of production''  have a coherent economic interpretation  independently of how  the aggregate surplus is divided in society.\footnote{In this paper,  instead of attempting to construct multi-sectoral microfoundations to aggregate production functions like, for example, \citeA{baqaee2019jeea}, I reject from the start the principle that income distribution can be  determined by the neoclassical theory of value based on the equilibrium of demand and supply equations. This does not state that neoclassical models using a single capital good are necessarily inconsistent, but given that the conclusions  derived from a single good economy cannot be extended to disaggregated systems, it seems appropriate  to start  from a foundation which recognizes the logical inconsistencies that arise in general when economic analysis is  conducted in isolation of the institutions of society.}

 \textbf{Framework.}  The  argument of the paper is divided in three  parts.  First, I present an environment with a chain of intermediate and final good producers who hire labor services from workers and buy new machines from capital good firms for the production of the final good (net aggregate output). Here the final good can be used for consumption and as a means of production of new capital goods,  which helps  highlight the principle that capital is generally a produced commodity and that the costs of production of the final good cannot be defined independently of the rate of return  of capital. This leads to an indeterminacy problem which is analogous to that described by \citeA{sraffa1960production} in the sense that the price equations of the system hold a degree of freedom that cannot be determined by the technology of production.

The second part of the model presents a solution to the indeterminacy problem by introducing a ``closure'' to the price equations based on the  dynamic interaction between unemployment, technical change (described by the mechanization and creation of tasks) and income distribution.   This is represented by merging the task-based formalism of \citeA{zeira1998workers} and \citeA{acemoglu2018race},  the equilibrium unemployment literature, and the capital adjustment cost theory of investment.  The task-based formalism provides a microfoundation to the cost structure of the final good and it helps examine how the mechanization and creation of tasks is interrelated with the dynamics of unemployment and  aggregate profits.  The equilibrium unemployment formalism presents a rationale explaining how the labor market interacts with income distribution and establishes clear economic principles based   on bargaining processes  for the determination of the rate of return of capital. Lastly, the theory of capital cost adjustments  is used to highlight the fact that capital is generally owned by firms and that profit maximization problems can be best understood by explicitly acknowledging that production that takes place in time \cite{kydland1982time,lucca2007resuscitating}.

 Third,  in order to show how labor market institutions  affect the rate of unemployment and the rate of return, I  follow  \citeA{smith1776}  and \citeA{becker1997endogenous}  and characterize the relative bargaining power of labor   in terms of heterogenous discount factors for capitalists and workers.\footnote{ In discussing the determination of  average wages as a result of the negotiation  between capitalists and workers, \citeA[p. 84]{smith1776} summarized the importance of time preferences noting that ``in all such disputes the masters can hold out much longer. A landlord, a farmer, a master manufacturer, or merchant, though they did not employ a single workman, could generally live a year or two upon the stocks which they have already acquired. Many workman  could not subsist a week, few could subsist a month, and scarce any a year without employment. In the long-run the workman may be as necessary to his master as his master is to him; but the necessity is not so immediate.'' } This is formalized using the principle that bargaining processes can take the form of   alternating offers models \cite{binmore1986nash}, which has the desirable property of portraying labor power  in terms of   endogenous discount factors  determined by current institutional and political settings.  The model  then shows that the relative welfare condition of social classes is central for the determination of bargaining strengths and these are, in turn,   key determinants  of  aggregate profits.

\textbf{Contributions.}  Building on this  unorthodox neoclassical synthesis, the model  characterizes the restrictions under which the economy reaches a balanced growth path with automation and creation of new tasks \cite{acemoglu2018race}, equilibrium unemployment, positive rates of return, and falling investment-good prices with less-than-unitary elasticity of substitution \cite{grossman2017balanced}.   A key  feature of this characterization of  balanced growth paths is that the the economy is represented as a circular flow of values sustained by its capacity of generating aggregate profits. In this respect, it is shown that specific forms of labor institutions and technology are required to  guarantee sustainable growth, i.e., it cannot be taken for granted that the economy will always and forever generate sufficiently high profits to reproduce itself in time. 

The steady-state general equilibrium   offers  a rich but tractable framework that illustrates  how automation and  varying institutional support to workers shape the asymptotic  aggregate outcomes of the economy. Under suitable institutional settings, automation reduces the stationary values of employment, wages and the wage share, while it raises the steady-state values of the capital-output ratio and the share of investment expenditures to aggregate output.  Respectively, a rising institutional support to workers---expressed by higher unemployment benefits  or other related factors which increase the outside options to employment---result in a long-run decline in employment and profitability, but it increases stationary wages, the labor share, the capital-output ratio and the share of investment expenditures to aggregate output.

The theoretical framework also establishes a three-way interaction between labor institutions, technology and income distribution. It highlights, for example, how varying support to workers  may indirectly affect income distribution by  impacting the  assignment of tasks between labor and capital. This provides an endogenous theory showing that if, for instance, labor is ``overpriced" because of institutional related factors, firms can respond by investing in new technologies that can effectively replace workers \cite{acemoglu2011skills}.    Additionally, it provides  explicit  bounds defining  the extent to which labor institutions can support workers before they become a threat to the reproduction of capital, suggesting there may exist a dichotomy between the social and economic sphere of citizenship---
recognized by elements of the welfare state---and the profit-making capacity of capitalist societies. Lastly, it highlights the key role  that institutions  can have in protecting workers from the impact of unregulated market forces by relating the potential  negative effects of  automation to increasing rates of unemployment, lower real wages, and greater income inequality. 

From an empirical perspective, the model shows that a combination of institutional and technological factors  is  required to provide reasonable explanations for the  behavior  of income distribution, profitability,  unemployment  and the big ratios in macroeconomics in the postwar US economy.  The technology hypothesis, expressed by a reduction in the measure of tasks perform by labor,  can account for a large bulk of the decrease in the labor share following the 2000s, but  is inconsistent with the surge in profitability and the reduction in the rate of unemployment in the wake of the 1980s. Conversely, the  hypothesis based on labor institutions  provides a plausible story for the fall in the profit share following the policies of the Great Society in the 1960s by relating the rise in the welfare state with an increase in  workers' outside options to employment, and for the reduction in the rate of unemployment and the rise in profitability since the early 1980s by relating these economic outcomes with the conservative retrenchment that followed \cite{pierson1994dismantling}. The labor institutions hypothesis, however,   falls short in explaining the variations of the capital-output ratio, suggesting it is only part of the story describing the main changes in the US economy. 

  To evaluate the plausibility of each hypothesis, I compare the inferred changes in technology and labor institutions derived from the model and show that they are consistent with the history of  welfare and technical change   described by \citeA{pierson1994dismantling},  \citeA{noble1997welfare}, \citeA{frey2019technology},  \citeA{Dechezlepretre2019} and \citeA{mann2021benign}. This not only shows that the model can provide a basis for interpreting some of the key empirical regularities of the US economy,  but  also  that a thorough understanding of macroeconomic trends can strongly benefit from a careful examination of social policies associated with the rules of liberal democracy defining the accord between capital and labor. 

\textbf{Interpretation of the Contributions.} The logic behind this article is based on the principle that capitalist  economies are ``open" systems which cannot be detached from the specific institutional context of society. Formally, this is captured by interpreting profits as surplus over costs of production, since it implies that: (i) firms cannot take profits as given when initiating production; (ii) profits cannot be determined by the technology of firms; and (iii) costs of production cannot be determined independently of the class distribution of income. 

Altogether,  (i)-(iii) create  an \emph{analogy} of Sraffa's \citeyear{sraffa1960production} critique of the neoclassical theory of distribution. It is from this perspective  that the model is perceived as unorthodox, even though  it maintains the use of production functions and rational agents as a tool for counterfactual analysis. In this interpretation, however,  the aggregate production function resulting from the task-based formalism is merely expressing an accounting identity in an economy with  time-varying wage-shares  and capital-output ratios; \footnote{The proof of this result follows from the works of \citeA{shaikh1974laws} and \citeA[p. 84-86]{felipe2013aggregate}, and it is shown in detail in Appendix \ref{appendix:ces}.}  meaning that the inherent social element characterizing the creation of profits is not hidden under the shadow of the  production function, but is rather treated explicitly as a social outcome resulting from bargaining processes between capitalists and workers. Ultimately, this  structure entails that no single contribution of the model can be derived independently of the historic-specific power relations defining the accord between capital and labor---all of which may depend on the rate of unemployment, technical change, minimum wages, union density, austerity policies, etc.

\textbf{Related  Literature.}  This paper contributes   to different areas of the literature. First, the theoretical framework builds from the task models of \citeA{zeira1998workers}, \citeA{acemoglu2011skills}, \citeA{acemoglu2018race} and \citeA{nakamura2018}. Relative to this literature, this paper proposes a bridge to reconcile the equilibrium unemployment literature with the economic decision of task automation. As part of this contribution, the model treats the return of capital as an endogenous variable determined by bargaining processes between capitalists and workers,  similar to  \citeA{shimer2005cyclical}, \citeA{
hall2008limited},\citeA{pissarides2009}, \citeA{petrosky2018endogenous}, and others. The combination of these two lines of research provides a basis for understanding how unemployment, automation, and income distribution are jointly determined in a general equilibrium setting.  

Second, this work extends on the literature attempting to explain the trends of key macroeconomic variables over the past 50 years \cite{karabarbounis2014global,piketty2014capital,FARHI_GOURIO2018,
autor2020fall,barkai2020declining,de2020rise,stansbury2020declining,eggertsson2021kaldor}. Most closely related to this paper is \citeA{stansbury2020declining}, who also identify the changes in the bargaining power of labor as a leading cause of the changes in the labor share of income, unemployment and profitability. The main difference with the current literature is that by incorporating automation technologies, the analysis can identify the extent to which the changing trends in macroeconomic variables are caused by technology or institutional related factors.\footnote{In this paper,  I omit referring to the market power hypothesis for two reasons: (a) it is hard to distinguish rising markups with a falling bargaining power of labor \cite[p. 6]{stansbury2020declining}; and (b) a rising markup  is  the subject matter one is interested in understanding,  not the assumption that one should be imposing to justify the changes in the macroeconomy.}

Third, the empirical narrative of this work builds on the literature highlighting the central role of institutions on market related outcomes. 
Particularly, extending on the works of  \citeA{dinardo2000unions}, \citeA{piketty2013capital,piketty2020capital} \citeA{ahlquist2017labor}, \citeA{farber2021unions}, among others,  the model emphasizes the potential impact that changes in the support  to workers can have on income distribution, profitability and employment. Additionally, the paper  shows that the impact of automation  on the economy can be best understood in the context of specific institutional settings which either enforce or attenuate the displacing effects of technology on labor \cite{lemieux2008changing, levy2011inequality}.

\textbf{Outline.}    Section \ref{sec:model} presents the theoretical  structure of the model. In Section \ref{sec:mod_dyn}, I show that the task-based formalism can be linked to the traditional search and matching model, and that  simple bargaining models can be used to determine the rate of return of capital and the class distribution of income in a general equilibrium environment. Section \ref{sec:steady_state} presents the conditions for steady-state growth and the  analysis on  comparative statics.  Next, in Sections \ref{sec:emp_analysis} and \ref{sec:hist_analysis}, I present the main empirical results of the model and the historical investigations associated with institutional and technical change in the postwar US economy.  Finally, I offer some concluding remarks in Section \ref{sec:conclusions}.

\textbf{Notation.}  The partial derivative of any function $\tilde{g}_{t+h}(x_{1t},...,x_{nt})$ with respect to any $x_{it}$ $(i=1,...,n)$ is denoted as $\tilde{g}_{x_{it},t+h}$. If a function $g(x)$  depends on a single variable, $g'$ denotes its  derivative with respect to $x$.

\section{Model setup}\label{sec:model}

The model  follows a task-based framework along the lines of  \citeA{zeira1998workers}  and \citeA{acemoglu2018race}.  There is a chain of intermediate and final good firms producing a final good, which can be consumed by households or used as an input for the production of capital goods. Time is discrete and is indexed by $ t \in \mathbb{N}$.

\subsection{Final and Intermediate Goods Production}

I consider a closed economy with  a single final good  produced using the technology

\begin{equation}
Y_{t}=B \Big(\int_{M_{t}-1}^{M_{t}} \;y_{t}(j)^{\frac{\sigma-1}{\sigma}} \; dj \Big)^{\frac{\sigma}{\sigma-1}}, 
\end{equation} 

where $\sigma>0$ is the elasticity of substitution  and  $y_{t}(j)$ is an  intermediate output produced with a task $j \in [M_{t}-1,M_{t}]$. Similar to \citeA{acemoglu2018race},  the measure of tasks used in production is always equal to 1, meaning that newly-created tasks represent higher productivity versions of the existing ones. 

Intermediate outputs are   produced  with labor or capital using a linear production function $  y_{t}(j) = \Gamma^{x} x_{t}(j)$, where $x_{t}(j)=k_{t}(j)$ and $\Gamma^{x}_{t}=\Gamma^{k}_{t}$ if $j \leq J_{t}$, and $x_{t}(j)=l_{t}(j)h_{t}$ and $\Gamma^{x}_{t}=\Gamma^{N}_{t}$ otherwise.  Here $J_{t}$ represent the available number of mechanized tasks, $l_{t}(j)$ is the employed labor,  $h_{t}$  represent the  number of hours per worker of any type,   $\Gamma^{K}_{t}(j)$ and $\Gamma^{N}_{t}(j)$ are  capital and labor-augmenting technologies, and $k_{t}(j)$ are the units of capital needed in the production of task $j$.  

The unit cost of  each task is represented by a linear system

\begin{equation}
  p_{t}(j) = \left\{ \,
    \begin{IEEEeqnarraybox}[][c]{l?s}
      \IEEEstrut
     \frac{\delta P^{k}_{t}}{\Gamma^{K}_{t}(j)} & \text{if}  $j \leq J_{t}$ \\
   \frac{w_{t}}{\Gamma^{N}_{t}(j)}& \text{if}  $j > J_{t}$.
      \IEEEstrut
    \end{IEEEeqnarraybox}
\right.
  \label{eq:leontief_prices}
\end{equation}

The rate of depreciation is   $\delta \in (0,1)$, $P^{k}_{t}$ is the  price of capital units and $w_{t}$ is the wage rate. 

Throughout, I will use the following assumption on the technical coefficients.

\begin{assm}\label{ass1} (i)  $\Gamma^{N}_{t}(j) =e^{\alpha j}$ and  $\Gamma^{K}_{t}(j)=\Gamma^{K}_{t}$ for all $ j \in [M_{t}-1,M_{t}]$.  (ii)  $e^{\alpha M_{t}} > \Gamma^{K}_{t} w_{t}/(\delta P^{k}_{t})$. 
\end{assm}

Assumption \ref{ass1}(i) says that labor has a comparative advantage in higher-indexed tasks, meaning there is a threshold $\tilde{J}_{t}$ such that $e^{\alpha \tilde{J}_{t}}/\Gamma^{K}_{t} = w_{t}/\delta P^{k}_{t}$. At $\tilde{J}_{t}$, intermediate good producers are indifferent between producing with capital or labor. In particular, for all $j \leq \tilde{J}_{t}$, tasks will be produced with capital since $\delta P^{k}_{t}/ \Gamma^{K}_{t} < w_{t} e^{\alpha j}$. However, if $j > \tilde{J}_{t}$, intermediate good producers are bounded by the existing technology and will  only be able to mechanize tasks up to $J_{t}$. The unique threshold is consequently given by 

\begin{equation}\label{eq:threshold_tech}
J^{*}_{t} =\text{min}\big\{J_{t}, \tilde{J}_{t} \big\},
\end{equation}

such that all tasks in $[M_{t}-1, J^{*}_{t}]$ are produced with capital and the remaining are produced with labor. Assumption \ref{ass1}(ii), in turn, implies that an increase in the number of tasks will increase aggregate output \cite{acemoglu2018race}.

As usual, the demand  function for task $j$ is given by $y_{t}(j)=B^{\sigma-1} Y_{t} \Big(p_{t}(j)/P^{c}_{t}\Big)^{-\sigma}$, such that the aggregate demand for capital and labor can be expressed as

\begin{subequations}
\begin{align}
\begin{split}\label{eq:ag_cap_demand}
&\int_{M_{t}-1}^{J^{*}_{t}} k_{t}(j) \; dj \equiv K_{t} = B^{\sigma-1} Y_{t} \Big(\delta P^{k}_{t}/P^{c}_{t}\Big)^{-\sigma} (1-m^*_{t}) {\Gamma^{K}_{t}}^{\sigma-1} 
\end{split}\\
\begin{split}\label{eq:ag_lab_demand}
&\int_{J^{*}_{t}}^{M_{t}} l_{t}(j) \; dj \equiv L_{t} =B^{\sigma-1} \Big(Y_{t}/h_{t}\Big)  \Big(w_{t}/P^{c}_{t}\Big) ^{-\sigma}  e^{\alpha(\sigma-1)J^{*}_{t}} \Big(\frac{e^{\alpha(\sigma-1)m^*_{t}}-1}{\alpha(\sigma-1)}\Big).
\end{split}
\end{align}
\end{subequations}

Where $m_{t} \equiv [M_{t}-J_{t}] \in [0,1]$ is the measure of tasks produced by labor,   $m^*_{t}$ is the equilibrium technology measure, and $P^{c}_{t}$ is the price index of costs of production  satisfying

\begin{equation}\label{eq:price_cost_index}
P^{c}_{t}=B^{-1} \Bigg[(1-m^*_{t}) \Bigg(\frac{\delta P^{k}_{t}}{\Gamma^{K}_{t}}\Bigg)^{1-\sigma}  + \Bigg(\frac{w_{t}}{e^{\alpha J^{*}_{t}}}\Bigg)^{1-\sigma} \Bigg(\frac{e^{\alpha(\sigma-1)m^*_{t}}-1}{\alpha(\sigma-1)}\Bigg)  \Bigg]^{\frac{1}{1-\sigma}}.
\end{equation}

Replacing the solution of $w_{t}$ and $\delta P^{k}_{t} $ from \eqref{eq:ag_cap_demand}-\eqref{eq:ag_lab_demand}  in \eqref{eq:price_cost_index},   the aggregate production function takes a simple CES form described by

\begin{equation}\label{eq:agg_prod_fun}
Y_{t}=\tilde{B}_{t} \Big[ \omega^{k}_{t}  \Big(\Gamma^{K} _{t} K_{t}\Big) ^{\frac{\sigma-1}{\sigma}} + (1-\omega^{k}_{t}) \Big(e^{\alpha J^{*}_{t}} N_{t}\Big)^{\frac{\sigma-1}{\sigma}} \Big]^{\frac{\sigma}{\sigma-1}}
\end{equation}

where $\tilde{B}_{t} \equiv  B \bar{\omega}_{t}^{\sigma/(\sigma-1)}$  and  $ \bar{\omega}_{t} \equiv  (1-m^{*}_{t})^{1/\sigma}+ \big( (e^{\alpha(\sigma-1)m^{*}_{t}}-1)/(\alpha(\sigma-1))\big)^{1/\sigma}$.  Here $N_{t} \equiv L_{t} h_{t}$ represent total hours of work, and  $\omega^{k}_{t} \equiv (1-m^{*}_{t})^{1/\sigma}/\bar{\omega}_{t}$ is the capital distribution parameter.

\subsection{Capital Good Producers}

Similar to \citeA{lucca2007resuscitating}, I assume that  the capital stock increases with the maturity of a large number of symmetric and complementary investment projects. Each investment project of type $i$ at time $t$ is denoted as $\mathcal{I}_{t}(i)$ and it reaches maturity if $ i \in \mathcal{H}_{t} \subseteq [0,1]$. Firms choose the desired scale of investment when initiating each project and cannot modify it until the period of maturity, in which case it can start a new project with a new scale of investment the following period.

The time of maturity of investment projects  is described by a Poisson-process with arrival rate $\pi_{I} \in (0,1)$ and the production of investment goods  is described by a technology

\begin{equation}\label{eq:tech_inv}
I_{t} = \Big(\int_{i \in \mathcal{H}_{t}} \mathcal{I}_{t}(i)^{\frac{\upsilon-1}{\upsilon}} \; di \Big)^\frac{\upsilon}{\upsilon-1}, \quad \;\; \upsilon \in  (1,\infty)
\end{equation}

 The final good is used in the production of  investment projects using a linear technology in which one unit of $Y_{t}$ is transformed into $\Psi_{t}$ units of $\mathcal{I}_{t}(i)$ for all $i \in [0,1]$.  Denoting  $P_{t}$ as the selling price of the final good and working in a competitive economy with a uniform  rate of return, the price of  investment projects satisfies $P^{I}_{t}/P_{t} =\Psi_{t}^{-1}$,  where $\Psi_{t}=\Psi_{t-1}e^{z^{\psi}}$ is a non-stationary   investment-specific technology   with growth rate $z^{\Psi}$.

 Assuming that investment firms minimize the expenditure on investment projects $\mathcal{X}_{t} \equiv P^{I}_{t} \int_{0}^{1} $ $\mathcal{I}_{t}(i)\; di$ subject to \eqref{eq:tech_inv} (see  Appendix \ref{appendix:AppendixA}):

\begin{equation}\label{eq:cap_exp_inv}
\mathcal{X}_{t} = P^{I}_{t}  \Bigg\{  \Omega(I_{t},I_{t-1})+  (1-\pi_{I}) \Big( \mathcal{X}_{t-1}/P^{I}_{t-1}\Big)\Bigg\}
\end{equation}

with 

\begin{equation*}
\Omega(I_{t},I_{t-1}) \equiv \pi_{I}^{\frac{1+\upsilon}{1-\upsilon}} \Bigg( I_{t}^{\frac{\upsilon-1}{\upsilon}} -(1-\pi_{I}) I_{t-1}^{\frac{\upsilon-1}{\upsilon}}\Bigg)^{\frac{\upsilon}{\upsilon -1}},\;\; \Omega_{I_{t},t} \geq 0,\;\;  \Omega_{I_{t-1},t} \leq 0.
\end{equation*}

 In the limit when $\pi_{I} \rightarrow 1$, the investment expenditure is $\mathcal{X}_{t} = P^{I}_{t} I_{t}$, which can be interpreted as the limiting case where the time-to-build period approaches zero.  On the contrary, if $\pi \rightarrow 0$, then $\mathcal{X}_{t}  \rightarrow \infty$ since $\upsilon >1$. Intuitively, this represents a scenario with an infinite time-to-build period. 
 
 Given the value of investment expenditures, capital  producers sell new capital goods to the chain of intermediate and final good firms at a price $P^{k}_{t}$  to maximize the  discounted value

\begin{equation}\label{eq:cap_prod_problem}
\Lambda^{c}(X_{0}) =  \sum_{t=0}^{\infty} \Big(\prod_{i=0}^{t} \tilde{\beta}^{c}_{i} \Big)  \Big( P^{k}_{t}  I_{t} - P^{I}_{t}  X_{t} \Big) \equiv  \sum_{t=0}^{\infty} \Big(\prod_{i=0}^{t} \tilde{\beta}^{c}_{i} \Big) \Pi^{c}_{t}
\end{equation}

subject to \eqref{eq:cap_exp_inv} and $X_{t} \equiv \mathcal{X}_{t}/ P^{I}_{t}$. Here  $\tilde{\beta}^{c}_{t}$ is the discount factor derived  from a utility maximization problem of capitalist households.  As usual,  $P^{k}_{t} = P^{I}_{t}$ when $\pi_{I} \rightarrow 1$ for all $t \geq 0$, which resembles the case of no investment adjustment costs. 

\subsection{Price System} 

Similar to \citeA[p. 8]{sraffa1960production}, the model presents an interrelation between capital and final good producers  showing that costs of production  cannot be defined independently of the price of capital  and consequently of the price of the final good itself. This link  is established using the notion of own-rates of return, described here as

\begin{equation}\label{eq:sale_price}
\mu_{t} = P_{t}/P^{c}_{t} -1. 
\end{equation}

In this case, even if we normalize the system by fixing $P^{c}_{t}=1$, we still have 
 four unknowns $\{P^{k}_{t}, P^{I}_{t}, P_{t}, \mu_{t}\}$ and  three independent equations since $P_{t}$ and $\mu_{t}$ both depend on \eqref{eq:sale_price}.\footnote{The price of investment projects follows from $P^{I}_{t}=P_{t}\Psi_{t}^{-1}$, the price of new capital goods is obtained from the first order conditions of \eqref{eq:cap_prod_problem}, but $\mu_{t}$ and $P_{t}$ cannot both be determined from \eqref{eq:sale_price}.} This indeterminacy issue is well captured by  the equations of the marginal productivity of  capital and labor, which satisfy

\begin{equation}\label{eq:marg_prods}
\begin{split}
\frac{\partial Y_{t}}{\partial K_{t}}  &= \tilde{B}^{\frac{\sigma-1}{\sigma}}_{t} \Big(\frac{Y_{t}}{K_{t}}\Big)^{1/\sigma} \big(\Gamma^{K}_{t}\big)^{\frac{\sigma-1}{\sigma}} \omega^{k}_{t} =\frac{\delta P^{k}_{t}}{P^{c}_{t}}=\frac{\delta P^{k}_{t}}{P_{t}} (1+\mu_{t}) \\
\frac{\partial Y_{t}}{\partial N_{t}}  &=\tilde{B}^{\frac{\sigma-1}{\sigma}}_{t} \Big(\frac{Y_{t}}{N_{t}}\Big)^{1/\sigma} \Big(e^{\alpha J^{*}_{t}}\Big)^{\frac{\sigma-1}{\sigma}} (1-\omega^{k}_{t})=  \frac{w_{t}}{P^{c}_{t}}=\frac{w_{t}}{P_{t}}(1+\mu_{t}).
\end{split}
\end{equation}

Equation \eqref{eq:marg_prods}  helps highlight the principle that, in general,   costs  and marginal productivity equations cannot be measured independently of, and prior to,  the determination of rate of return of capital  \cite[p. 9]{sraffa1960production}.  In this respect, though I use marginal productivity theory as a by-product of the CES aggregator,  this cannot be considered  the  determinant of profitability and  the distribution of income without  being trapped in a circular argument.

Ultimately, the indeterminacy of the system brings back to the surface the interpretation  of the rate of return as a surplus, and with it the importance of referring to political and institutional factors  in order to determine how the   surplus is divided in society. In the following section I illustrate this principle  by describing the rate of return of capital as an endogenous outcome of the dynamic interaction between unemployment and  income distribution using an extended version of the  search-and-matching model. 

\section{Model dynamics}\label{sec:mod_dyn}

I consider a general equilibrium  model with unemployment  with three  variants. First,  profits are a surplus over  costs of production.  Second, employment and capital dynamics are integrated with  a technological unemployment component resulting from the automation of tasks. Third, the relative bargaining power of workers is an endogenous outcome determined by  the  discount factors of workers and capitalists. 

\subsection{Search, Matching and State Dynamics} In each period a measure $1$ of economically active agents are either employed or unemployed workers, and can be hired by capitalists for the purpose of creating profits in exchange for wages.  Employed workers are represented by $L_{t}$ and the remaining $U_{t}=1-L_{t}$ represent the unemployed.  Vacancies are filled via the \citeA{den2000job} matching function $G(U_{t}, V_{t}) $ $= (U_{t} V_{t})/(U^{\iota}_{t}+V^{\iota}_{t})^{1/\iota}$, with  $\iota >0$. Define  $\theta_{t} \equiv V_{t}/U_{t}$ as the vacancy-unemployment ratio (labor market tightness).  The job finding rate is $f(\theta_{t}) \equiv G(U_{t},V_{t})/U_{t} = (1+\theta^{-\iota}_{t})^{-1/\iota}$  and the vacancy filling rate  is $q(\theta_{t}) \equiv G(U_{t}, $ $V_{t})/V_{t}= (1+\theta^{\iota}_{t})^{-1/\iota}$, with $q'(\theta_{t})<0$. The real unit cost per vacancy is modeled as

\begin{equation}\label{eq:vacancy_costs}
\frac{\kappa_{t}}{P_{t}} = e^{\alpha J^{*}_{t}} \big[ \kappa_{0} +\kappa_{1} q(\theta_{t})\big],\;\; \text{with}\;\; \kappa_0, \kappa_1 >0.
\end{equation}

Like \citeA[p. 2215]{petrosky2018endogenous}, unit costs  per vacancy contain fixed costs ($\kappa_{1}$), which capture training and administrative costs of adding workers to the payroll, and proportional costs $(\kappa_{0})$, which increase in relation to the expected duration of vacancies, $1/q(\theta)$. 

Introducing changes  in the mechanization and  the creation of new tasks, and assuming that workers can only transit within the working population,  the evolution of employment can be described as\footnote{Though this is  an unrealistic assumption, it portrays the  difficulty that workers have in bettering their condition up to the point where they can become either self-employed or employers; see, for example, \citeA[p. 115]{lamont2000}.}

\begin{equation*}
L_{t+1}=(1-\lambda) L_{t} +q(\theta_{t}) V_{t} - \underbrace{\int_{J^{*}_{t}}^{J^{*}_{t+1}} l_{t}(j) dj}_{\text{displacement effect}} + \underbrace{\int_{M_{t}}^{M_{t+1}} l_{t}(j) dj}_{\text{reinstatement effect}}.
\end{equation*}

As usual,  $\lambda$ is an exogenous Poisson rate defining the probability that an employed worker becomes  unemployed per unit of time.  The displacement and reinstatement effects of mechanization and the creation of new tasks is represented in the last term on the right-hand side of the equation. As noted by \citeA{acemoglu2018race}, mechanizing existing tasks creates, on one hand,  a displacement effect by replacing labor for machines. On the other hand, the creation of new tasks generates a reinstatement effect by expanding the demand for labor,  consequently reducing unemployment. 

Correspondingly, the dynamics of the value of the capital stock is described as

\begin{equation*}
P^{k}_{t+1} K_{t+1}= \big[ (1-\delta) K_{t} +I_{t} + \int_{J^{*}_{t}}^{J^{*}_{t+1}} k_{t}(j) dj - \int_{M_{t}}^{M_{t+1}} k_{t}(j) dj \big] P^{k}_{t}.
\end{equation*}

Using  equations \eqref{eq:ag_cap_demand}-\eqref{eq:ag_lab_demand} and  \eqref{eq:marg_prods},    technological unemployment is defined as the sum of the displacement and reinstatement effects, such that

 \begin{equation}
  U^{A}_{t} \equiv \int_{J^{*}_{t}}^{J^{*}_{t+1}} l_{t}(j) dj - \int_{M_{t}}^{M_{t+1}} l_{t}(j) dj=L_{t} \;\Bigg(1-  \frac{e^{\alpha(\sigma-1)(J^{*}_{t+1}-J^{*}_{t})}\Big[e^{\alpha(\sigma-1)m^{*}_{t+1}}-1\Big]}{e^{\alpha(\sigma-1)m^{*}_{t}}-1}\Bigg).
  \label{eq:tech_unemploy}
\end{equation}

Similarly,  the addition of capital resulting from the mechanization and creation of tasks is given by

 \begin{equation}
  A_{t} \equiv \int_{J^{*}_{t}}^{J^{*}_{t+1}} k_{t}(j) dj-\int_{M_{t}}^{M_{t+1}} k_{t}(j) dj=-K_{t} \; \Big(\frac{m^{*}_{t+1}-m^{*}_{t}}{1-m^{*}_{t}}\Big).
  \label{eq:tech_unemploy2}
\end{equation}

One of the  insights of \eqref{eq:tech_unemploy} and \eqref{eq:tech_unemploy2} is that technological unemployment can be interpreted as the net result of the mechanization and creation of additional tasks. Altogether, the dynamics of aggregate employment and aggregate capital can  be expressed concisely as

\begin{equation}\label{eq:ag_lab_dyn}
L_{t+1}=(1-\hat{\lambda}_{t}) L_{t} +q(\theta_{t}) V_{t},
\end{equation}

\begin{equation}\label{eq:ag_cap_dyn}
K_{t+1}=\Big[(1-\hat{\delta}_{t}) K_{t} +I_{t}\Big] P^{k}_{t}/P^{k}_{t+1}
\end{equation}

The addition of technological unemployment can be interpreted as introducing an endogenous separation rate $\hat{\lambda}_{t}=\lambda + U^{A}_{L_{t},t}$,  which in equilibrium is related  to the costs of labor relative to capital. On the capital side, $\hat{\delta}_{t}=\delta-A_{K_{t},t}$ can be interpreted as the effective depreciation rate, since it balances the value of fixed capital lost from wear and tear, and the value acquired from the automation and creation of tasks. 

\textbf{Workers.} I assume a continuum of ex ante identical workers of measure 1 who  are   perfectly insured against variations in labor income.   All employed workers receive $w_{t}N_{t}$ and  unemployed workers  receive  $U_{t} b_{t}$, where $b_{t}$ is the nominal value of public benefits that unemployed forgo upon employment \cite{chodorow2016cyclicality}.   

The worker chooses  the consumption of the employed and unemployed,  $C^{we}_{t}$ and $C^{wu}_{t}$,  to maximize the expected sum of discounted utility flows

\begin{equation}\label{eq:obj_workers} \Phi(L_{0},U_{0})= \sum_{t=0}^{\infty} \Big(\prod_{i=0}^{t} \beta^{w}_{i}\Big)\; \Big[  L_{t} \;\mathcal{U}^{we}(C^{we}_{t},h_{t}) +  U_{t}\; \mathcal{U}^{we}(C^{wu}_{t},0)\Big],
\end{equation}

subject to \eqref{eq:ag_lab_dyn},  $U_{t+1}=1- L_{t+1}$ and the budget constraint

\begin{equation}\label{eq:bud_c_work}
P_{t} C^{w}_{t} =P_{t} C^{we}_{t}\; L_{t} + P_{t} C^{wu}_{t} \;U_{t} \leq w_{t} N_{t} + U_{t} b_{t}.
\end{equation}

The flow utility of the employed and unemployed are represented by  $\mathcal{U}^{we}(C^{we}_{t},h_{t})$  and $\mathcal{U}^{wu}(C^{wu}_{t},0)$, respectively.  Expressing the present-discounted value of  income streams for an employed and unemployed worker in consumption units  by dividing by the marginal utility of consumption, the first-order conditions can be expressed as:

\begin{subequations}\label{eq:foc_worker}
\begin{align}
\begin{split}
 &\Phi_{L_{t},t}= \frac{w_{t}h_{t}}{P_{t}}  + \frac{\mathcal{U}^{we}(C^{we}_{t},h_{t})}{\phi^{w}_{0t}} -C^{we}_{t} + \tilde{\beta}^{w}_{t+1} \big[ \big(1-\hat{\lambda}_{t}\big) \Phi_{L_{t},t+1}   +  \hat{\lambda}_{t}  \Phi_{U_{t},t+1}\big]
\end{split}\\
\begin{split}
& \Phi_{U_{t},t}=\frac{b_{t}}{P_{t}} + \frac{\mathcal{U}^{wu}(C^{wu}_{t},0)}{\phi^{w}_{0t}} -C^{wu}_{t} + \tilde{\beta}^{w}_{t+1} \big[ f(\theta_{t}) \Phi_{L_{t},t+1}+(1-f(\theta_{t}))  \Phi_{U_{t},t+1} \big]
 \end{split}
\end{align}
\end{subequations}

where $ \tilde{\beta}^{w}_{t+1} \equiv \beta^{w}_{t+1} (\mathcal{U}^{w}_{C^{w}_{t+1},t+1}/\mathcal{U}^{w}_{C^{w}_{t},t})$ and $\phi^{w}_{0,t}=  \mathcal{U}^{wi}_{C^{wi}_{t}} /P_{t}$ for $i=\{e,u\}$.   These optimality conditions are standard in all respects but on the treatment of the discount factor, which is  interpreted as a time-varying  variable $\beta^{w}_{t}(\Gamma_{t})$ expected to satisfy  $\beta^{w}_{\Gamma_{it}} \geq 0$ if $\Gamma_{it}$ is an institutional,  political or economic variable favoring  the relative welfare condition of workers, and $\beta^{w}_{\Gamma_{it}} <0 $ if the contrary is the case. For example, an increase in labor unions, a higher real value of minimum wages, or higher unemployment benefits are  variables in $\Gamma_{t}$ that probably have a positive effect on  $\beta^{w}_{t}$.  In contrast,  variables related to globalization, labor outsourcing, automation, lower top-income tax rates, or the increasing hiring of managers with business education may have a negative  effect on the discount factor by improving the relative welfare condition of capital relative to labor.

\textbf{Capitalists.} A representative capitalist chooses the amount of  vacancies $V_{t}$ and consumption $C^{c}_{t}$ that maximize the discounted utility flows:

 \begin{equation*}
\Lambda(L_{0})= \sum_{t=0}^{\infty} {\beta^{c}}^{t} \;  \mathcal{U}^{c}_{t}(C^{c}_{t})
\end{equation*}

subject to  \eqref{eq:ag_lab_dyn} and  the financial constraint\footnote{It is worth noting that all financial constraints are derived from the flows of value of Marx's circuit of capital as formalized by \citeA{foley1986understanding}. The connection between Marx's accounting structure and the general equilibrium model in this article is presented in  Appendix \ref{appendix:accounting_app}. }

\begin{equation}\label{eq:cap_finac_constraint}
P_{t}  C^{c}_{t}  \leq P_{t} Y_{t} - w_{t} N_{t} - P^{K}_{t} I_{t} -\kappa_{t} V_{t} - T_{t}
\end{equation}

As usual,  $T_{t}=U_{t}b_{t}$ are lump-sum taxes used to finance unemployment benefits.  Expressing  the first-order conditions in consumption units by dividing by the marginal utility of consumption:

\begin{equation}\label{eq:labor_demand_foc}
 \Lambda_{L_{t},t}  = h_{t} \Big( Y_{N_{t},t} - w_{t}/P_{t} \Big)   + \tilde{\beta}^{c}_{t+1}   (1-\hat{\lambda}_{t})  \Lambda_{L_{t+1},t+1},
 \end{equation}

where $\tilde{\beta}^{c}_{t+1}=\beta^{c} (\mathcal{U}^{c}_{C^{c},t+1}/\mathcal{U}^{c}_{C^{c},t})$ and  $\Lambda_{L_{t+1},t+1}= (\kappa_{t}/P_{t})/\tilde{\beta}^{c}_{t+1} q(\theta_{t})$ is the ``zero-profit" condition.  Equation \eqref{eq:labor_demand_foc} introduces a time-varying separation rate and shows that technological unemployment lowers the marginal value of an employed worker since it increases the probability of unemployment per unit of time.

\textbf{Functional Forms.} Similar to \citeA{chodorow2016cyclicality}, the preferences of workers and capitalists are described by:

 \begin{equation}\label{eq:util_sep}
 \mathcal{U}^{j}(C^{j}_{t},h^{j}_{t}) = \mathrm{log}\big(C^{j}_{t}\big) - \frac{\epsilon_{0} \; \epsilon_{1}}{1+\epsilon_{1}} {h^{j}_{t}}^{1+(1/\epsilon_{1})}. 
 \end{equation}

Where $j=\{we,wu,c\}$,   $we$ represents employed workers, $wu$  the unemployed, and $c$ the capitalists. By assumption $h^{j}_{t}=0$ for $j=\{wu,c\}$ since the labor of employed productive workers is  the only one directly relevant for the creation of profits.

\subsection{Wage Bargaining, the Rate of Return and the Class Distribution of Income}

  Following the approach of the Classical economists, the rate of return  is  interpreted as a social outcome resulting from wage-bargaining processes  between capitalists and workers. This can be formalized using a variety of game-theoretic models of bargaining; see, e.g.,  \citeA{hall2008limited}, \citeA{gertler2009unemployment}, and \citeA{christiano2016unemployment}. Here, however, I use the Nash bargaining solution  for its simplicity and interpret the solution of the model  as the limiting case of an alternating offers model with heterogenous discount factors. Assuming no side payments per round of negotiation,  $\{w_{t}/P_{t}, h_{t}\}$ is  the solution of:

\begin{equation}\label{eq:nash_problem}
\underset{w_{t}/P_{t},\; h_{t}}{ \mathrm{arg\; max}} \; \Big(\Phi_{L_{t},t}- \mathcal{R}_{w,t}\Big)^{\eta_{w,t}} \Big(\Lambda_{L_{t},t}\Big)^{1-\eta_{w,t}},
\end{equation}

with 

\begin{equation*}
\eta_{w,t} = \frac{\mathrm{log}\; (1-\lambda)+\mathrm{log}\;\tilde{\beta}^{c}_{t+1} }{2  \mathrm{log} (1-\lambda)+\mathrm{log}\; \tilde{\beta}^{c}_{t+1} + \mathrm{log}\;\tilde{\beta}^{w}_{t+1} },\; \mathcal{R}_{w,t}=\frac{\lambda\; \Phi_{U_{t},t}}{1-\big[\tilde{\beta}^{w}_{t+1}\big]^{\Delta}(1-\lambda)}
\end{equation*}

and a bargaining time delay  $\Delta \rightarrow 0$.\footnote{Similar to \citeA{hall2008limited}, here I assume that the probability that the job opportunity will end in a given round during bargaining is equal to the Poisson separation rate.}  One of the  advantages of presenting the  Nash bargaining solution as a limiting case of the alternating offers model is that it provides a clear interpretation of the relative bargaining power of workers  in terms of the discount factors.  If  workers  have less resources or fewer outside options than capitalists, it can be expected that $\tilde{\beta}^{w}_{t+1} < \tilde{\beta}^{c}_{t+1}$, i.e., the relative bargaining power of workers, $\eta_{w,t},$ will generally be less than 0.5.\footnote{This formalizes Smith's \citeyear[p. 84]{smith1776}   observation that  capitalists are generally  in a better bargaining position than workers. \citeA[p. 124]{lichtenstein2002} makes a similar observation by noting that regardless on how much wages or benefits may rise, capital will systematically have an upper hand over labor.  }

\textbf{Rate of return.}  Considering the alternating offer bargaining model as a benchmark for an interpretation of the Nash solution, the rate of return of capital is solved using   \eqref{eq:sale_price},  \eqref{eq:marg_prods}, \eqref{eq:util_sep},  and  the first order conditions of  \eqref{eq:nash_problem} with respect to $w_{t}/P_{t}$, such that

\begin{equation}\label{eq:mark_up_nash}
\mu_{t} =  \frac{(1-\eta_{w,t})\big(h_{t} Y_{N_{t}} - Z_{t}  \big) - \eta_{w,t}  \frac{\kappa_{t}}{P_{t}} \Bigg(\Big(1-\frac{\tilde{\beta}^{w}_{t+1}}{\tilde{\beta}^{c}_{t+1}}\Big)  \frac{(1-\hat{\lambda}_{t} )}{q(\theta_{t})} + \frac{\tilde{\beta}^{w}_{t+1}}{\tilde{\beta}^{c}_{t+1}} \theta_{t} \Bigg) }{(1-\eta_{w,t}) Z_{t} + \eta_{w,t} \Bigg[ h_{t} Y_{N_{t},t}  + \frac{\kappa_{t}}{P_{t}} \Bigg(\Big(1-\frac{\tilde{\beta}^{w}_{t+1}}{\tilde{\beta}^{c}_{t+1}}\Big)  \frac{(1-\hat{\lambda}_{t})}{q(\theta_{t})} + \frac{\tilde{\beta}^{w}_{t+1}}{\tilde{\beta}^{c}_{t+1}} \theta_{t} \Bigg) \Bigg] }
\end{equation}

and

\begin{equation*}
Z_{t} = \frac{b_{t}}{P_{t}} + \frac{\epsilon_{1}}{1+\epsilon_{1}}h_{t} Y_{N_{t}}
\end{equation*}

As usual,  $Z_{t}$ is the opportunity cost of employment and it represents  the sum of real unemployment benefits  and  the utility differential from nonworking time of unemployed and employed workers.

The first term in \eqref{eq:mark_up_nash}  depicts the share that capitalists can extract in the production process if they were to pay workers the opportunity cost of employment. The second term shows that the rate of return decreases with a tighter labor market  in relation to the the average hiring costs of each unemployed and with a rise in the  power of workers.  From this  term it is also  clear that a decline in $\tilde{\beta}^{w}$ relative to $\tilde{\beta}^{c}$  generally lowers  the negative impact of employment on profitability, since it will tend to reduce the relative bargaining power of workers and  will consequently reduce the variations of real wages  to changes in $\theta_{t}$. An interpretation of this result is that a tighter labor market is required when the relative bargaining power of workers is low in order to support real wage growth.\footnote{This problem was recently noted by  \citeA{stansbury2020declining} as an explanation of the falling NAIRU.}

\textbf{Corridor of Economic and Political Stability.} A simple inspection of equation \eqref{eq:mark_up_nash}  shows that $\mu_{t}$ is negatively affected by a rise in the relative bargaining power of workers and by an increase in the opportunity costs of employment.  This  juxtaposition between the economic sphere of citizenship---represented by welfare related factors raising the outside options to employment---and the profitability of capital is  used in the following definition. 

\begin{defn} The corridor of economic and political stability is defined by values of $\eta_{w,t} \in (0, \eta^{U}_{w,t})$, such that

\begin{equation}
 \left\{ \,
    \begin{IEEEeqnarraybox}[][c]{l?s}
      \IEEEstrut
     \mu^{\text{max}}_{t} = \frac{h_{t} Y_{N_{t}} - Z_{t}}{\tilde{Z}_{t}}  & \text{if}  \; $\eta_{w,t}=0$ \\
 \mu^{\text{min}}_{t} \approx   \frac{(1-\Omega^{c}_{t})(g_{t} + \tau_{t} + \zeta_{t})}{\delta } & \text{if} \; $\eta_{w,t} =\eta^{U}_{w,t}$
      \IEEEstrut
    \end{IEEEeqnarraybox}
\right.
\label{eq:corridor}
\end{equation}

where  $\Omega^{c}_{t}\equiv w_{t} N_{t}/(P^{c}_{t}Y_{t})$ is the labor share on costs of production, $ \tau_{t} \equiv T_{t}/(P^{k_{t}} K_{t})$  represents the  value of taxes over the capital stock, $\zeta_{t} \equiv \kappa_{t} V_{t}/(P^{k_{t}} K_{t})$ is the ratio of vacancy costs to capital, and 

 \begin{equation}\label{eq:upper_power}
 \eta^{U}_{w,t} = \frac{h_{t} Y_{N_{t}} - Z_{t}(1+\mu^{\text{min}}_{t})}{(1+\mu^{\text{min}}_{t}) \Bigg\{ h_{t} Y_{N_{t}}-Z_{t}+   \frac{\kappa_{t}}{P_{t}} \Bigg(\Big(1-\frac{\tilde{\beta}^{w}_{t+1}}{\tilde{\beta}^{c}_{t+1}}\Big)  \frac{\big(1-\hat{\lambda}_{t} \big)}{q(\theta_{t})} + \frac{\tilde{\beta}^{w}_{t+1}}{\tilde{\beta}^{c}_{t+1}} \theta_{t} \Bigg) \Bigg\}} <1. 
 \end{equation}

\end{defn}

The lower bound $\mu^{\text{min}}_{t}$ is set to satisfy the condition that $C^{c}_{t} \geq 0$ for all $t \geq 0$  in equation \eqref{eq:cap_finac_constraint}.  Intuitively,  if $\mu_{t} > \mu^{\text{min}}_{t}$, capitalists can use a share of  net aggregate profits  for consumption and new additions to productive capital. However,  if $\mu_{t} = \mu^{\text{min}}_{t}$, capitalist consumption becomes zero because all retained profits must be used to finance capital outlays, taxes, and vacancy costs.  In the opposite extreme, $ \mu^{\text{max}}_{t} $  offers a clear view of the principle that a reduction in the opportunity cost of employment tends to raise the profitability of capital.  Equation \eqref{eq:corridor}  also shows that in the limit when $\eta_{w,t} = 0$, the rate of return does not depend on the conditions of the labor market, meaning that a tighter labor market  will generally have a lower  negative pressure on $\mu_{t}$ as $\eta_{w,t} \rightarrow 0$. 
 
 The corridor of economic and political stability has major implications from a political economy perspective. It shows, on one hand, that policies raising the support to workers and promoting full employment  conditions may reduce the return of capital to the point that $\mu_{t} \leq  \mu^{\text{min}}_{t}$---making the  economy unsustainable.  On the other hand, policies that severely harm the  bargaining power of workers may lead to politically fragile societies, which may manifest in different forms such as democracies favoring  populist movements (\citeNP{frey2017political};  \citeNP[p. 130]{frey2019technology}),  social instability and political unrest \cite{dal2011workers,caprettini2020rage}, or acts of desperation reflected in a rise of alcoholism, suicide and drug addiction (\citeNP[p. 204]{hobsbawm1996}; \citeNP{case2021deaths}).\footnote{The rise of populist movements and social unrest can be interpreted as specific political choices where workers exercise their ``voice" to make a change in society. The rise of alcoholism, suicide and drug addiction, in turn, can be seen as an ``exit" to the  precarious conditions exerted by society \cite{hirschman1970exit}. }

\textbf{Class Distribution of Income.}    Using \eqref{eq:marg_prods} and \eqref{eq:mark_up_nash}, the labor share on  gross and net aggregate income can be expressed as\footnote{It is straightforward to show that if $J^*_{t} = \tilde{J}_{t}$, then  $\Omega^{w}_{t} \equiv \frac{w_{t}N_{t}}{P_{t}Y_{t}} = \frac{\Omega^{C}_{t} }{1+\mu_{t}} = \frac{1}{1+\mu_{t}} \times \Big[1 + \Big(\frac{\omega^{k}_{t}}{(1-\omega^{k}_{t}) }\Big)^{\sigma}  \; \Big]^{-1}$.}

\begin{equation}\label{eq:labor_share_inc}
\Omega^{w}_{t} \equiv \frac{w_{t}N_{t}}{P_{t}Y_{t}} = \frac{\Omega^{C}_{t} }{1+\mu_{t}} = \frac{1}{1+\mu_{t}} \times \Bigg[1 + \frac{\omega^{k}_{t}}{(1-\omega^{k}_{t}) } \Bigg(\frac{e^{\alpha J^{*}_{t}} N_{t}}{\Gamma^{K}_{t} K_{t}}\Bigg)^{\frac{1-\sigma}{\sigma}} \; \Bigg]^{-1}
\end{equation}

 Here the class distribution of income is a function of automation and the rate of return. All factors contributing to an increase in the rate of return or to an increase  in the mechanization of tasks will, holding everything else equal, decrease the labor share on  aggregate   income. This result generalizes  the commonly posited explanations of the declining share of wages based on technological change and rising monopoly power (see, e.g.,  \citeNP{acemoglu2018race,autor2020fall,de2020rise})  by explicitly considering the role institutional and political factors in the distribution of income. In this respect,  the model is   capable of reconciling the evidence of an eroding bargaining power of workers found by \citeA{ahlquist2017labor} and \citeA{stansbury2020declining}, among others,  as an additional explanation of the declining  labor share.

\subsection{ Aggregation and Equilibrium}\label{sec:equil_agg}

Aggregate consumption is defined as a weighted average of the corresponding variables for capitalists and workers:

\begin{equation}\label{eq:ag_consumption}
C_{t} = U_{t}\; C^{we}_{t} + L_{t} C^{we}_{t} +  C^{c}_{t} + \Pi^{c}_{t}/P_{t}
\end{equation}

Summing over the financial restrictions of workers and capitalists, together with the profits of capital good producers, the aggregate resource constraint satisfies $
P_{t} Y_{t}= P_{t} C_{t} + \mathcal{X}_{t} + \kappa_{t} V_{t}$ since it is assumed that unemployment insurance benefits are entirely financed by lump-sum taxes on capitalists.

 \textbf{Equilibrium: Definition and Characterization.} The equilibrium properties of the economy are described by the following definition.

\begin{defn}\label{def:recursive_equilibrium} A recursive equilibrium is a solution for (a) a list of functions $\{\Phi_{L_{t},t}, \Phi_{U_{t},t}, \Lambda_{L_{t},t}, \Lambda^{c}_{X_{t},t}\} $; (b) prices $\{w_{t},  P_{t}, P^{c}_{t}, P^{I}_{t}, P^{k}_{t}\}$; (c)  rates $\{\mu_{t}, r_{t}, s_{t}\}$;  and (d) allocations $\{I_{t}, K_{t}, X_{t},  $ $\theta_{t},  V_{t}, L_{t},$ $ U_{t}, h_{t}, U^{A}_{t+1}, A_{t+1}, Y_{t}, $ $ C^{c}_{t},  C^{w}_{t}, C_{t}, $ $J^{*}_{t}\}$, such that for an initial set of values $\{P^{k}_{0}, K_{0}, L_{0},$ $ X_{0},  J_{0},M_{0}\}$:  (i) The value functions satisfy the Bellman equations;  (ii) $w_{t}$  satisfy \eqref{eq:marg_prods}; (iii) $P^{c}_{t}$ is normalized to 1; (iv) $P^{I}_{t}$ is equal to $\Psi_{t}^{-1}P_{t}$;  (v) $P^{k}_{t}$ is solved from \eqref{eq:cap_prod_problem}; (vi) $P_{t}$ satisfy \eqref{eq:sale_price}; (vii) the rate of return $\mu_{t}$ is derived from \eqref{eq:mark_up_nash};  (viii) $r_{t}=\mu_{t} Y_{t}/(P^{k}_{t}K_{t})$ ;  (ix) the capitalist savings rate $s_{t}$ is determined from \eqref{eq:cambridge_eq};  (x) $I_{t}$ is obtained from \eqref{eq:ag_cap_dyn}; (xi) $K_{t}$  satisfy \eqref{eq:marg_prods}; (xii) $X_{t}$ follows from \eqref{eq:cap_exp_inv}; (xiii) $\theta_{t}$ satisfies \eqref{eq:labor_demand_foc}; (xiv) $V_{t}$  is obtained from the matching function; (xv) $L_{t}$ satisfies \eqref{eq:ag_lab_dyn};  (xvi) $U_{t}=1-L_{t}$; (xvii) $h_{t}$ is obtained from \eqref{eq:nash_problem}; (xviii) technological unemployment $U^{A}_{t}$ satisfy \eqref{eq:tech_unemploy}; (xix) $A_{t}$  is given by \eqref{eq:tech_unemploy2}; (xx) $Y_{t}$ satisfy \eqref{eq:agg_prod_fun}; (xxi) $C^{c}_{t}$ satisfies \eqref{eq:cap_finac_constraint}; (xxii) $C^{w}_{t}$ follows from \eqref{eq:bud_c_work}; (xxiii) $C_{t}$ is given by \eqref{eq:ag_consumption}; and (xxiv) $J^{*}_{t}$  follows from \eqref{eq:threshold_tech}. 
\end{defn}

As usual, the model admits no closed-form solution. However, in order to build our intuition about the properties of the system, in the next section I provide a complete characterization of the steady-state equilibrium and  some key results of comparative statics.

       \section{Steady-State  Growth Analysis and Comparative Statics}\label{sec:steady_state}
       
  The analysis in this section extends on Uzawa's \citeyear{uzawa1961neutral}  seminal paper and the more recent works of \citeA{grossman2017balanced} and \citeA{acemoglu2018race},  showing that the economy can  reach an equilibrium growth path while allowing for falling investment-good prices and less-than-unitary elasticity of substitution between capital and labor. 
       
       \subsection{Steady-State Growth}\label{subsec:steady_state}
       
       To clarify terms and set the basis for the analysis, it is convenient to start with the following definition. 
       
       \begin{defn} A  balance growth path for the economy is a path along which $\{Y_{t}, C^{we}_{t},$ $  C^{wu}_{t}, C^{c}_{t}, \Pi^{c}_{t},$ $ w_{t}, \kappa_{t}\}$ grow at  constant rates,  $\{\mu_{t}, \theta_{t}, L_{t}, U_{t}, h_{t}, V_{t},  m^*_{t}, \frac{P^{k}_{t}K_{t}}{P_{t} Y_{t}},  \frac{P^{k}_{t}I_{t}}{P_{t} Y_{t}}, s_{t},  U^{A}_{t}, A_{t}\}$ are constant (possibly zero), and the capitalists savings rate $s_{t} \in (0,1)$   for all $t \geq 0$.  
       \end{defn}
       
       This definition of  balance-growth paths is  intended to describe  an economy capable of producing sufficiently large profits so that capitalists can finance their  consumption, expenses, and leave a positive remnant for the continuous expansion of capital. 
       
 To ensure balance growth, I  impose some additional structure on the model using  the following assumption. 
       
       \begin{assm}\label{ass:steady_state_det}  (i) The capital-augmenting technology satisfies $\Gamma^{k}_{t}=\gamma_{k}\Psi^{-1}$ for all $t \geq 0$.(ii) The labor-augmenting technology satisfies $e^{\alpha(J_{t}-J_{t-1})}=e^{g}$, where $g$ is the long-run rate of growth. (iii) The creation of tasks satisfies $M_{t}=m^{*}_{t} + J^{*}_{t}$, where $m^{*}_{t}$ is the equilibrium measure of automation. 
       \end{assm}

       Assumption \ref{ass:steady_state_det} (i) is meant to satisfy the condition that the value of capital measured in units of the final output is constant in equilibrium. Combining  Assumptions \ref{ass:steady_state_det} (i)-(ii) presents a purely labor-augmenting technological change, which is necessary for balance growth paths \cite{uzawa1961neutral}. Lastly, Assumption \ref{ass:steady_state_det} (iii) imposes the condition that the creation of tasks evolves in time at the same rate that the equilibrium mechanization of tasks. 
       
 Before proceeding   to the key results of the section, it is useful to introduce a modified version of Lemma A2 of \citeA{acemoglu2018race} to characterize the effects of automation as a function of the rate of return of capital.

\begin{lemma}\label{lemma:lemma_2A_AR} Suppose that Assumption \ref{ass:steady_state_det} (i) holds.  Setting $\gamma_{k} = B^{-1} \delta \hat{P}^{k}_{t}(0)$, with $\hat{P}^{k}_{t} = P^{k}_{t}/\Psi_{t}$, there exists an increase function $\bar{m}(\mu_{t}): [\mu^{\text{min}}_{t},  $ $\mu^{\text{max}}_{t}] \rightarrow [m^{\text{min}}_{t}, m^{\text{max}}_{t}]$, with $0 < m^{\text{min}}_{t} < m^{\text{max}}_{t} <1$, such that for all $m_{t} > \bar{m}(\mu_{t})$, $w_{t} e^{-\alpha J^{*}_{t}} > \delta \hat{P}^{k}_{t}/\gamma_{k} > w_{t} e^{-\alpha M_{t}}$, meaning that automated tasks are immediately produced with capital. In turn, if $w_{t} e^{-\alpha \tilde{J}_{t}} = \delta \hat{P}^{k}_{t}/\gamma_{k} $, small changes in automation do not affect $m^*_{t}$. 
\end{lemma} 

\begin{figure}
\begin{center}
\begin{tikzpicture}
\begin{axis}[
   axis x line=middle,
                    axis y line=middle,
 width=11.5cm, height=7.25cm,
ymin=0.1, ymax = 0.9,
xmin=0, xmax=1,
xtick={0.25,0.85},
xticklabels={$\mu^{\text{min}}$, $\mu^{\text{max}}$},
ytick={0.8*0.25^0.75,0.8*0.85^0.75,0.82},
yticklabels={$\bar{m}(\mu^{\text{min}})$,$\bar{m}(\mu^{\text{max}})$,1 },
extra x ticks={0.98}, extra x tick labels={$\mu$},
 extra y ticks={0.88}, extra y tick labels={$m$},
]

\addplot[domain=0:0.85,
thick,
gray,
dashed,
]
{0.8*x^0.75};

\draw[name path=B,
 dotted,
 thick,
]
(0,0.8*0.85^0.75) -- (0.9,0.8*0.85^0.75);

\addplot[name path=A, domain=0.25:0.85,
thick,
]
{0.8*x^0.75}
node[black,above, pos=0.1,xshift=1.25em,yshift=1.25em,font=\footnotesize]{$\bar{m}(\mu)$}
;

\draw[
 dashed,
 red,
->
]
(0,0.82) node[myred,above, xshift=8em,font=\footnotesize]{$m=1$}--(1,0.82);

\draw[
thick,
->
]
(0.85,0.8*0.85^0.75)--(1,0.8*0.85^0.75);

\addplot [lightgray!50] fill between [of= A and B, soft clip={domain=0.25:0.85}];

\draw[
 dotted,
  thick,
]
(0.85,0.8*0.85^0.75) -- (0.85,0);

\draw[
 dotted,
  thick,
]
(0,0.8*0.25^0.75) -- (1,0.8*0.25^0.75);

\draw[
 dotted,
  thick,
]
(0.25,0.8*0.85^0.75) -- (0.25,0);

\node [right,font=\tiny,align=left] at (0.28,0.8*0.74^0.75) {Region 1:  $w_{t} e^{-\alpha J^{*}_{t}} > \frac{\delta \hat{P}^{k}_{t}}{\gamma_{k}}>w_{t} e^{-\alpha M_{t}} $\\
$m^*=m$};

\node [right,font=\tiny,align=left] at (0.475,0.8*0.35^0.75) {Region 2: $\frac{\delta \hat{P}^{k}_{t}}{\gamma_{k}}>w_{t} e^{-\alpha J^{*}_{t}}$\\
 $m^*_{t}=\bar{m}(\mu_{t})$};

\end{axis}
\end{tikzpicture}
\end{center}
\caption{Automation regions. \emph{Notes---} The capital-augmenting technology is expressed using Assumption \ref{ass:steady_state_det} (i). \label{fig:automation_reg}}
\end{figure}
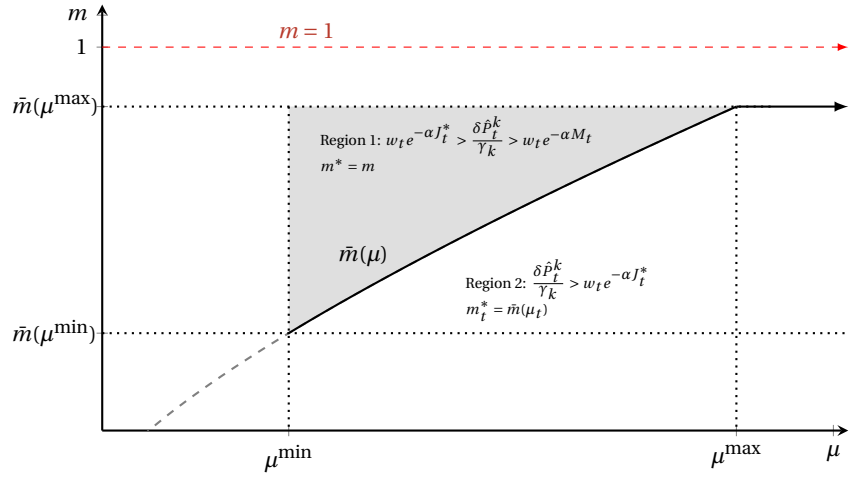

The initial statement setting $\gamma_{k} = B^{-1} \delta \hat{P}^{k}_{t}(0)$ is  used to guarantee that  $\bar{m}(\mu)\geq 0$ for all $\mu\geq 0$.\footnote{Here $ \hat{P}^{k}_{t}(0)$ represents the price of capital conditional on $\mu=0$. Generally, $\bar{m}(\bar{\mu})=0$ for some $\bar{\mu}<\mu^{\text{min}}$, so it makes little difference to set $\bar{\mu}=0$ as in Lemma \ref{lemma:lemma_2A_AR}. } Lemma \ref{lemma:lemma_2A_AR} has the intuitive appeal of linking the effects of automation to the rate of return of capital, and consequently on the institutional variables which may affect $\mu_{t}$. This is well portrayed in Figure \ref{fig:automation_reg}, where it can be deduced that  policy measures leading to a reduction in the rate of return of capital may have the unintended consequence of making automation a viable option for the reduction of unit labor costs. Ultimately,  Lemma \ref{lemma:lemma_2A_AR}  shows  that the effects of automation on the economy  always depend  on the specific institutional arrangements of society and cannot be properly understood independently of the rate of return of capital.

The minimum and maximum values of $m$  are defined by the the corridor of economic and political stability  in equation \eqref{eq:corridor}. These bounds rule out the possibility of an equilibrium where $m=1$, which is reasonable given that it is meaningless to refer to capitalist societies without capital. Correspondingly, the viable values of $\mu$  also rule out an equilibrium where $m=0$, since without the institution of wage-labor there is no basis for determining aggregate profits in capitalist societies.

\begin{theorem}\label{theorem:equil_harrod} Suppose that Assumption \ref{ass:steady_state_det}  holds. Given  an initial value of capital assets $K_{0}P^{k}_{0}$, and a bargaining power $\eta_{w,t} \in (0, \eta^{U}_{w,t})$, the economy admits a  balance growth path with:

\begin{enumerate}[label={(\Alph*)}]

\item  An equilibrium growth rate equal to:

\begin{equation}\label{eq:cambridge_eq}
g_{t} =s_{t} (r_{t}-\tau_{t}-\zeta_{t}) \rightarrow g
\end{equation}

where   $r_{t}\equiv \Pi_{t}/(P^{k}_{t}K_{t})$ is the aggregate rate of profit  and $P^{k}_{t}K_{t}$ is the money value of capital assets.

\item A steady-state equilibrium satisfying:

\begin{subequations}
\begin{alignat}{5}
&(\text{Capital marginal productivity}):  Y_{K_{t}} \Psi_{t} \rightarrow \frac{\delta (1+\mu)  \Omega_{I}\big( 1 -  (1-\pi_{I}) e^{\bar{z}(\upsilon-1)/\upsilon}\big)}{1-(1-\pi_{I}) e^{-\bar{z}}} \label{subeq:steady_marg_prod_cap}\\
& (\text{Rate of profit}): r_{t} \rightarrow \frac{\delta \mu}{1-\Omega^{c}}= \frac{\delta \mu}{(1-m^*)\big(\hat{Y}_{K}/(B\gamma_{k})\big)^{1-\sigma}}\label{subeq:steady_profit_rate}\\
& (\text{Investment-output ratio}):  \frac{\mathcal{X}_{t}}{P_{t} Y_{t}}   \rightarrow  \frac{(\delta +g)\pi_{I}^{1+\upsilon} \Omega^{\upsilon}_{I} }{\big(1-(1-\pi_{I})e^{\bar{z}}\big) } \Bigg(\frac{\hat{Y}_{K}}{B \gamma_{k}}\Bigg)^{-\sigma} \Bigg(\frac{1-m^*}{B \gamma_{k}}\Bigg) \label{subeq:steady_inv_output}
\end{alignat}
\end{subequations}

Where $m^{*} = m \in \big(\bar{m}(\mu^{\text{min}}),\bar{m}(\mu^{\text{max}})\big)$, $\bar{z} =g+ z^{\psi}$,  $\hat{Y}_{K} = \text{lim}_{ t \rightarrow \infty}  \; Y_{K_{t}} \Psi_{t} $, and $\Omega_{I} = \text{lim}_{ t \rightarrow \infty}  \; \Omega_{I_{t}}$. 

\end{enumerate}
\end{theorem}

Equation \eqref{eq:cambridge_eq} characterizes the equilibrium rate of growth under the assumption that all savings are made by capitalists, which is a first order approximation of reality intended to identify the sources of income by the role that individuals play in the production process of commodities. The decomposition of equation \eqref{eq:cambridge_eq} shows that changes in the equilibrium rate of growth must act through changes in the equilibrium  rate of return and the share of retained profits recommitted in the form of capital outlays \cite{foley1986understanding}. That is, the sources of  growth are found in the expansion of the value of capital outlays in the process of production and by how much this value is recommitted as productive capital.

A distinctive feature of Theorem \ref{theorem:equil_harrod} (A), which is not always explicit in balance growth-path analyses, is that the existence of an equilibrium rate of growth depends on specific institutional settings allowing the reproduction of sufficiently large profits. Particularly, it is necessary  that $\eta_{w,t} \in (0, \eta^{U}_{w,t})$ to obtain equilibrium aggregate profits that surpass the value of capitalist expenses. The key matter in this respect is that the relative bargaining power of workers  is not determined by technology or preferences,  but rather by institutional and political factors.  The steady-state equations in \eqref{subeq:steady_marg_prod_cap}-\eqref{subeq:steady_inv_output}, in turn,  provide valuable information for understanding the  results on comparative statics  in subsection \ref{subsec:comp_stat} and the empirical findings in subsection \ref{sub:empirical_results}.

\textbf{Labor Market Equilibrium.} The ``closure" for determining the steady-state equations in Theorem \ref{theorem:equil_harrod}  is  obtained using the equilibrium  of the labor market. In this setting, the Nash solution in \eqref{eq:mark_up_nash} replaces the  usual labor supply equation since it draws a negative relation between profitability and the vacancy-unemployment ratio. Correspondingly, the first order conditions of capitalists in \eqref{eq:labor_demand_foc} can be used as the labor demand equation since it presents a positive relation between labor market tightness and the rate of return of capital. Expressing both equations in steady-state form, it follows that:

\begin{equation}\label{eq:steady_mu}
\begin{split}
&(\text{Labor demand}):\mu^{D} = \Bigg( \frac{\beta^{c} \hat{h} \hat{Y}_{N} q(\theta)}{(1-\beta^{c}(1-\hat{\lambda}))\hat{\kappa}}-1\Bigg)^{-1} \\
&(\text{Labor supply}): \mu^{S} = \frac{(h  \hat{Y}_{N} - \hat{Z})(1-\eta_{w}) - \eta_{w} \hat{\kappa}\Big[\Big(1-\frac{\beta^{w}}{\beta^{c}}\Big) \frac{(1-\hat{\lambda})}{q(\theta)} + \frac{\beta^{w}}{\beta^{c}}\theta\Big] }{(1-\eta_{w})\hat{Z} +\eta_{w}\Big\{h  \hat{Y}_{N} + \hat{\kappa}\Big[\Big(1-\frac{\beta^{w}}{\beta^{c}}\Big) \frac{(1-\hat{\lambda} )}{q(\theta)} + \frac{\beta^{w}}{\beta^{c}}\theta\Big] \Big\}   }
\end{split}
\end{equation}

Where $\hat{Z}  =\text{lim}_{ t \rightarrow \infty} Z_{t}e^{-\alpha J^{*}_{t}}$,  $\hat{Y}_{N}  = \text{lim}_{ t \rightarrow \infty} Y_{N_{t}} e^{-\alpha J^{*}_{t}}$,  $\hat{\kappa}   = \text{lim}_{ t \rightarrow \infty}  (\kappa_{t}/P_{t}) e^{-\alpha J^{*}_{t}}$ and $U^{A}_{L} =  \text{lim}_{ t \rightarrow \infty}  U^{A}_{L_{t},t} =  1-e^{(\sigma-1)g}$.   Under fairly general conditions the intersection of $\mu^{D}$ and $\mu^{S}$ defines a unique equilibrium of $\mu$ and $\theta$ which can then be used to identify all other variables in the economy.

       \subsection{Comparative Statics}\label{subsec:comp_stat}

       We  now  study the long-run implications of  permanent changes in technology and labor institutions. For this purpose I will consider the effects of a decline in $m$, which represents a situation where automation runs ahead of the creation of new tasks;  a permanent reduction in $\hat{b} = (b_{t}/P_{t}) e^{-\alpha J^{*}_{t}}$, which corresponds to lower public benefits that unemployed forgo upon employment relative to labor productivity;  and a decline in $\beta^{w}$, representing permanent  institutional changes worsening the relative welfare conditions of workers.

The next proposition characterizes the long-run impacts of   technological and institutional changes on employment, profitability, income distribution and big ratios in the economy.

       \begin{prop}\label{prop:comp_stat} Suppose that the  assumptions of Theorem \ref{theorem:equil_harrod} hold. For a value of  $m^* \in (m^{\text{min}},m^{\text{max}})$ satisfying the conditions of a balance growth path:
       
       \begin{enumerate}[label=(\roman*)]
       \item  (Automation)
       \begin{itemize}
       \item For $m<\bar{m}(\mu)$, small changes in $m$ do not affect the equilibrium of the economy. 
       \item For $m>\bar{m}(\mu)$, a decrease in $m$ lowers the asymptotic stationary values of $w_{t}$, $\theta_{t}$, $L_{t}$ and $\Omega^{c}_{t}$. Respectively,  a permanent reduction in $m$ raises the asymptotic stationary values of $\mathcal{X}_{t}/(P_{t} Y_{t})$, and $K_{t}/Y_{t}$. The effects on the rate of return of capital $\mu_{t}$ depend on the model parameters. 
       \end{itemize}

           \item  (Unemployment benefits and relative welfare of workers)  A reduction in $\hat{b}$ or  $\beta^{w}$ raises the asymptotic values of $\mu_{t}$, $\theta_{t}$,  $L_{t}$, and  $Y_{K_{t}}$.  Correspondingly,  lower values of  $\hat{b}$ or  $\beta^{w}$ reduce  the asymptotic stationary values of  $w_{t}$, $K_{t}/Y_{t}$, $\mathcal{X}_{t}/(P_{t}Y_{t})$ and $\Omega^{c}_{t}$ if $\sigma \in (0, 1)$. 
       \end{enumerate}
      
       \end{prop}

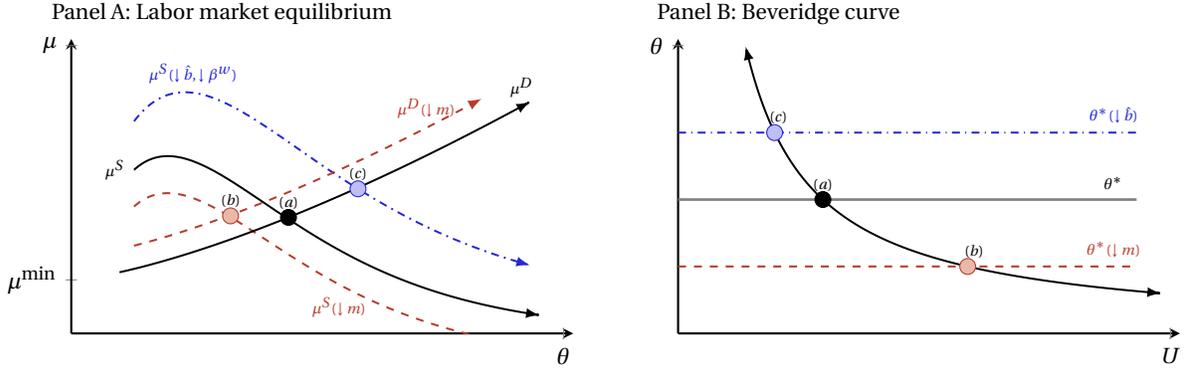
\begin{figure}
\begin{center}
\begin{tikzpicture}
     \begin{axis}[
        axis x line=middle,
                    axis y line=middle,
         name=plot1,
          title={\footnotesize Panel A: Labor market equilibrium}, 
              title style={at={(axis description cs:0.3,0.965)}, anchor=south} , 
 width=0.5*\textwidth, height=5.5cm,
ymin=0, ymax = 1.1,
xmin=0, xmax=5.2,
xtick={0},
extra x ticks={5.1}, extra x tick labels={$\theta$},
ytick={0.2},
yticklabels={$\mu^{\text{min}}$ },
 extra y ticks={1.075}, extra y tick labels={$\mu$},
scatter/classes={
    a={mark=o,draw=black, mark size = 3pt},
    b={mark=*, mark size = 3pt,draw=myred, fill = myred!30},
        c={mark=*, mark size = 3pt,draw=black, fill = black},
               d={mark=*, mark size = 3pt,draw=myblue, fill = myblue!30},
                   e={mark=*, mark size = 3pt,draw=myblue, fill = myblue}
    }
]

\addplot[domain=0.65:4.85,
->,
  line width=0.75pt,
black
]
{1.8*x*exp(-x)}
node [pos=0,left,font=\tiny]{$\mu^S$};

\addplot[domain=0.65:4.75,
->,
  line width=0.75pt,
color=myblue,
dashdotted,
]
{0.1+1.85*x*exp(-x*0.85)}
node [pos=0.15,above,font=\tiny]{$\mu^S( \downarrow \hat{b}, \downarrow \beta^{w})$};

\addplot[domain=0.65:4.65,
->,
  line width=0.75pt,
color=myred,
dashed,
]
{-0.12+1.75*x*exp(-x)}
node [pos=0.55,xshift=-0.25em,below,font=\tiny]{$\mu^S(\downarrow m)$};

\addplot[domain=0.5:4.75,
->,
  line width=0.75pt,
]
{0.2+0.075*x^(1.4)}
node [pos=0.98,above,font=\tiny]{$\mu^D$};

\addplot[domain=0.65:4.25,
->,
  line width=0.75pt,
dashed,
myred
]
{0.285+0.078*x^(1.4)}
node [pos=0.95,left,font=\tiny]{$\mu^D(\downarrow m)$};


\addplot[scatter,only marks, scatter src=explicit symbolic]
  coordinates {
    (1.65, 0.439 )     [b]
     (2.25,0.4334)      [c]
          (2.97,0.54)      [d]
  };

\node [above,font=\tiny,align=left] at (1.65,0.439) {$(b)$};
\node [above,font=\tiny,align=left] at (2.25,0.4334) {$(a)$};
\node [above,font=\tiny,align=left] at (2.97,0.54) {$(c)$};

\end{axis}

 \begin{axis}[%
    axis x line=middle,
                    axis y line=middle,
    name=plot2,
     title={\footnotesize Panel B: Beveridge curve           },
                title style={at={(axis description cs:0.2,0.965)}, anchor=south} , 
    at=(plot1.right of south east), anchor=left of south west,
 width=0.5*\textwidth, height=5.5cm,
ymin=0, ymax = 1.1,
xmin=0, xmax=5.2,
xtick={0},
extra x ticks={5.1}, extra x tick labels={$U$},
ytick={0},
yticklabels={},
 extra y ticks={1.075}, extra y tick labels={$\theta$},
scatter/classes={
    a={mark=o,draw=black, mark size = 3pt},
    b={mark=*, mark size = 3pt,draw=myred, fill = myred!30},
        c={mark=*, mark size = 3pt,draw=black, fill = black},
               d={mark=o, mark size = 3pt,draw=myblue, fill = white},
                   e={mark=*, mark size = 3pt,draw=myblue, fill = myblue!30}
    }]
    
    \addplot[domain=0.7:5,
<->,
  line width=0.75pt,
black
]
{0.75/x};

\draw[
 dashdotted,
  thick,
  line width=0.75pt,
  myblue
]
(0.0,0.75) -- (4.75,0.75)
node [pos=0.95,above,font=\tiny]{$\theta^*(\downarrow \hat{b})$};

\draw[
  thick,
  gray,
  line width=1 pt,
]
(0.0,0.5) -- (4.75,0.5)
node [black,pos=0.95,above,font=\tiny]{$\theta^*$};

\draw[
 dashed,
  thick,
  line width=0.75 pt,
myred
]
(0.0,0.25) -- (4.75,0.25)
node [pos=0.95,above,font=\tiny]{$\theta^*(\downarrow m)$};

\addplot[scatter,only marks, scatter src=explicit symbolic]
  coordinates {
    (3, 0.25 )     [b]
     (1.5,0.5)      [c]
               (1,0.75)      [e]
  };

\node [above,xshift=0.25em,font=\tiny,align=left] at (3,0.25) {$(b)$};
\node [above,font=\tiny,align=left] at (1.5,0.5) {$(a)$};
\node [above, xshift=0.15em,font=\tiny,align=left] at (1,0.75) {$(c)$};

    \end{axis}
    
\end{tikzpicture}
            \caption{Steady-state equilibrium labor market. \emph{Notes---}  The labor supply equation is based on the Nash solution and changes on automation assume $m^*=m>\bar{m}(\mu)$.\label{fig:labor_market_comp_stat}}
                \end{center}
            \end{figure}

      \textbf{Effects of automation.}  Starting with the effects of automation in Figure \ref{fig:labor_market_comp_stat}, it is clear that if $m>\bar{m}(\mu)$,  a permanent reduction in $m$ creates a negative effect on employment from two fronts: it lowers the labor supply equation since, by Lemma \ref{lemma:lemma_2A_AR}, wages decrease relative to the long-run expansion of the economy; and, it raises the  demand for labor because for any given vacancy-to-unemployment ratio, firms will be able to extract a greater surplus over wages. Graphically, the steady-state travels from $(a)$ to $(b)$, with the resulting equilibrium of $\mu$ depending on the model parameters, but ultimately leading  to an increasing rate of unemployment  in an amount which depends on the form of the Beveridge curve.

   The equilibrium in the capital market is also affected by changes in automation. Drawing on the results of Theorem \ref{theorem:equil_harrod}, Figure \ref{fig:cap_market_comp_stat} shows how the demand for capital changes in relation to variations in $m$. Particularly, for a given value of $\mu$, a higher automation rate increases the asymptotic stationary capital-output ratio since new tasks are produced using capital when $m^*=m$. The increase in capital intensity ($\hat{K}/\hat{Y}$) leaves the equilibrium marginal productivity of capital unaltered, so the ultimate effects of automation on the capital market are a reduction in the equilibrium rate of profit, an increase in the capital cost share and a rise in the investment expenditure to output ratio; see the transition from $(a)$ to $(d)$ in Panels A and B of Figure  \ref{fig:cap_market_comp_stat}. 
   
          The contrasting behavior between the marginal productivity of capital and the rate of profit is a matter of significant importance and deserves special attention. The difference between these two variables is well represented in Panel B of Figure \ref{fig:cap_market_comp_stat}, where it can be observed  that the marginal productivity of capital generally differs from the rate of profit. Additionally, Figure \ref{fig:cap_market_comp_stat} shows that whereas an increase in the automation  of tasks reduces the rate of profit because it increases the value of capital outlays relative to the cost of final output, the marginal productivity of capital stays the same because the asymptotic stationary relative price of capital does not depend on $m$. In this respect, the argument not only shows that the rate of profit is generally  not a proxy for the marginal productivity of capital, but also that their behavior may differ depending on the factors causing their changes in time.

            \begin{figure}
\begin{center}
\begin{tikzpicture}
     \begin{axis}[
        axis x line=middle,
                    axis y line=middle,
         name=plot1,
          title={\footnotesize Panel A: Capital market equilibrium},
              title style={at={(axis description cs:0.35,0.975)}, anchor=south} , 
 width=0.48*\textwidth, height=5.5cm,
ymin=0, ymax = 2.1,
xmin=0, xmax=5.2,
xtick={0},
extra x ticks={5.1}, extra x tick labels={$\hat{K}/\hat{Y}$},
ytick={1},
yticklabels={$\frac{\delta \hat{P}^{k}}{P^{c}}$ },
 extra y ticks={2.05}, extra y tick labels={$\hat{Y}_{K}$},
scatter/classes={
    a={mark=*,draw=gray, fill=gray!30, mark size = 3pt},
    b={mark=*, mark size = 3pt,draw=myred, fill = myred!30},
        c={mark=*, mark size = 3pt,draw=black, fill = black},
               d={mark=*, mark size = 3pt,draw=myblue, fill = myblue!30},
                   e={mark=*, mark size = 3pt,draw=myblue, fill = myblue}
    }
]

\addplot[domain=0.5:4.755,
 line width=0.75 pt,
black
]
{1/x}
node [pos=0.95,below,font=\tiny]{$\hat{Y}_{K}$};

\addplot[domain=0.9:4.75,
 line width=0.75 pt,
gray
]
{1.85/x}
node [pos=0.05,right,font=\tiny]{$\hat{Y}_{K}(\downarrow m)$};

\draw[
  thick,
    line width=0.75 pt,
]
(0.0,1) -- (4.75,1);

\draw[
  thick,
  myblue,
  dashed,
    line width=0.75 pt,
]
(0.0,1.45) -- (4.75,1.45)
node [pos=0.95,above,font=\tiny]{$\frac{\delta \hat{P}^{k}}{P^{c}}(\uparrow \mu)$};


\addplot[scatter,only marks, scatter src=explicit symbolic]
  coordinates {
    (0.97,1 )     [c]
               (0.7,1.45)      [e]
                              (1.85,1)      [a]
  };

\node [above, xshift=0.35em, font=\tiny,align=left] at (0.97,1) {$(a)$};
\node [above, xshift=-0.75em,font=\tiny,align=left] at (0.7,1.45) {$(c)$};
\node [above,xshift=0.35em, font=\tiny,align=left] at (1.85,1) {$(d)$};

\end{axis}

 \begin{axis}[%
    axis x line=middle,
                    axis y line=middle,
    name=plot2,
     title={\footnotesize Panel B: Automation versus $r$ and $\hat{Y}_{K}$  },
           title style={at={(axis description cs:0.35,0.975)}, anchor=south} , 
    at=(plot1.right of south east), anchor=left of south west,
 width=0.51*\textwidth, height=5.5cm,
ymin=0.95, ymax = 5.5,
xmin=0.65, xmax=3.3,
xtick={1,2.5,3},
xticklabels={$\bar{m}(\mu^{\text{min}})$, $\bar{m}(\mu^{\text{max}})$,1},
extra x ticks={3.2}, extra x tick labels={$m$},
ytick={0},
yticklabels={},
 extra y ticks={5.4}, extra y tick labels={$\hat{Y}_{K,}\;\; r$},
scatter/classes={
    a={mark=*,draw=gray, fill=gray!30, mark size = 3pt},
    b={mark=*, mark size = 3pt,draw=myred, fill = myred!30},
        c={mark=*, mark size = 3pt,draw=black, fill = black},
               d={mark=o, mark size = 3pt,draw=myblue, fill = white},
                   e={mark=*, mark size = 3pt,draw=myblue, fill = myblue!30}
    }]

    \addplot[name path=A, domain=1.75:2.5,
thick,
black,
 line width=1.15 pt,
]
{x^1.5}
node [pos=0.5,left,font=\tiny]{$r(m)$};

\path[name path=axis] (axis cs:1.75,0.96) -- (axis cs:2.5,0.96);

\addplot [lightgray!50] fill between [of= A and axis, soft clip={domain=1.75:2.5}];

  \addplot[domain=0:1.75,
 line width=0.75 pt,
gray,
dashed
]
(0,1.75^1.5)--(1.75,1.75^1.5);

  \addplot[domain=0:1.75,
thick,
 line width=1.15 pt,
]
(1,1.75^1.5)--(1.75,1.75^1.5);

  \draw[domain=2.5:3,
   line width=0.75 pt,
gray,
dashed
]
(2.5,2.5^1.5)--(3,2.5^1.5);

\draw[
 dotted,
   line width=0.75 pt,
  black
]
(2.5,0) -- (2.5,2.5^1.5)
 node[pos=0.55,align=right, xshift=-1.85em,font=\tiny]{Region 1: \\
 $m^*=m$};

\draw[
 dotted,
 line width=0.75 pt,
  black
]
(1.75,0) -- (1.75,1.75^1.5);

\draw[
 dotted,
   line width=0.75 pt,
  black
]
(1,0) -- (1,1.75^1.5)
 node[pos=0.75,align=center, xshift=2em,font=\tiny]{Region 2: \\
 $m^*=\bar{m}(\mu)$};

\draw[
 dashed,
   line width=0.75 pt,
  gray
]
(3,0) -- (3,5.1);

\draw[
 dashed,
  thick,
  line width=0.75pt,
  myred
]
(0.0,4.25) -- (3,4.25)
node [pos=1,right,font=\tiny]{$\hat{Y}_{K}$};

\addplot[scatter,only marks, scatter src=explicit symbolic]
  coordinates {
    (1.85,4.25 )     [a]
     (1.85,1.85^1.5)      [a]
               (2.35,4.25)      [c]
                              (2.35,2.35^1.5)      [c]
  };

\node [above,xshift=0.25em,font=\tiny,align=left] at (1.85,4.25) {$(d)$};
\node [above, xshift=-0.35em, font=\tiny,align=left] at (1.85,1.85^1.5) {$(d)$};
\node [above, xshift=0.25em,font=\tiny,align=left] at (2.25,4.25) {$(a)$};
\node [above,xshift=-0.35em, font=\tiny,align=left] at (2.35,2.35^1.5) {$(a)$};

    \end{axis}
    
\end{tikzpicture}
            \caption{Steady-state equilibrium in capital market. \emph{Notes---}  Changes on automation in Panel A assume $m^*=m>\bar{m}(\mu)$.\label{fig:cap_market_comp_stat}}
                \end{center}
            \end{figure}

   \textbf{Effects  of variations in unemployment benefits and the discount factor of workers.} Changes in institutions yield qualitatively similar results regardless on whether they are expressed by changes  in unemployment benefits or in the relative welfare condition of workers. For instance,  the expected result following a permanent decrease in unemployment benefits  is a rise in the profitability of capital and the vacancy-unemployment ratio; see Panel A in  Figure \ref{fig:labor_market_comp_stat}. This, in turn, lowers the unemployment rate in relation to the form of the Beveridge curve,  and depresses the stationary real value of wages. Additionally, the reduction of unemployment benefits lowers the labor share from two fronts. First, it decreases the wage share on costs if $\sigma \in (0,1)$. In this case the increase in employment does not surpass the fall in stationary real wages. Second, it increases the  price of the final good, which reduces the participation of labor on the value of sales.

The bottom line is that  changes in institutions can act as powerful methods that alter the balance of power between capital and labor.  In some cases, as shown by the reductions in unemployment benefits, governments can increase the profit-making capacity of the economy by worsening the relative condition of workers.\footnote{This is easier if workers  have a negative attitude towards  government assistance.  It has been well documented, e.g., that the media has played a key role shaping how a large number of white Americans see the welfare state by linking on one hand welfare with laziness, and laziness with being African American on the other \cite{lamont2000}. } The extent to which each one of the technology and institutional factors have altered the structure of the US economy is an empirical matter that I explore in the follow section.

  \section{Empirical Analysis}\label{sec:emp_analysis}
  
  This section presents an empirical exercise to measure the effects of technological and institutional changes on the steady-state equilibrium of  income shares, capital returns, the rate of unemployment, and the big ratios in macroeconomics previously explored  by \citeA{FARHI_GOURIO2018} and \citeA{eggertsson2021kaldor}.  The empirical analysis is divided in two parts. First, I employ a baseline calibration on some model parameters using US data and related literature.  Second, I show that the data calls for specific changes in institutions and technology  in order to match the time averages of some key variables in the postwar US economy.

  \subsection{Parameterization} All parameters are calibrated in monthly frequency. The growth parameters   $\delta$, $g$ and $z^{\Psi}$ are set to match a 10 percent annual depreciation rate, a 2 percent annual growth rate of labor productivity and a 2 percent decline in the relative price of investment. Following the empirical findings summarized in \citeA{chirinko2008sigma} and  \citeA{grossman2021elusive}, the elasticity of substitution is  equal to  0.6.  Similar to \citeA{altug1989time} and \citeA{lucca2007resuscitating}, I set $\pi_{I}=0.12$, implying  that the  average time required for completing investment projects is close to three quarters. The complementarity of investment projects is set to 5.84, which is not only 
 the elasticity of substitution parameter across products estimated by \citeA{christiano2005nominal}, but it also implies a steady-state marginal productivity of capital in line with the estimates of \citeA{caselli2007marginal} of about 0.14 (see Figure \ref{fig:counter_2} in Appendix \ref{appendix:data_c}).  The capital-augmenting parameter $\gamma_{k}$ is obtained according to Lemma \ref{lemma:lemma_2A_AR} and $\alpha$ is calibrated so that real wages are close to 1.

I follow \citeA{hall2009reconciling}  and set the  Frisch elasticity equal to 0.75. The labor supply parameter is set at 1.1  so that the equilibrium number of hours is about 1.   Similar to  \citeA{den2000job}, the matching function parameter is set at 1.27. The separation rate is equal to 0.034,  which is close to the values used by  \citeA{shimer2005cyclical} and \citeA{hagedorn2008cyclical} based on JOLTS data.  Lastly,  fixed hiring costs are matched to the value proposed by \citeA[p. 2215]{petrosky2018endogenous} and $\kappa_{0}$ is calibrated  to 8.25, which is about 4.5 times the average productivity of labor.\footnote{The calibration of $\kappa_{0}$ is significantly higher than the values normally used in the literature. However, as shown in Appendix \ref{appendix:cavancy}, this is necessary in order to satisfy the conditions for steady-state growth paths in Theorem \ref{theorem:equil_harrod}. Table \ref{table:implied_rate_of_return} shows that the rate of return of capital obtained from the usual calibrations in the literature are too low in relation to the data (e.g., Panel A Figure \ref{fig:counterfactual_test}) and do not even satisfy the minimum requirement for steady-state growth, which is that $\mu_{t} > \mu^{\text{min}}_{t}$.}

  \begin{table}\caption{Baseline calibration with SEP Preferences and Nash Solution}
  \centering
\resizebox{14 cm}{!}
 { \begin{tabular}{c c l l}
  \hline \hline
  Parameter & Value & Description & Source/target\\
    \hline 
  Growth  & & &\\
  $\delta $ & 0.0079 & Depreciation rate & 10$\%$ annual rate \\
    $ g$& 0.0167 & Labor productivity growth  & 2$\%$ annual rate \\
   $ z^{\Psi} $ & 0.0167 & Decline in  investment prices  & 2$\%$ annual rate \\
     Technology  & & &\\
  $\sigma $ &0.6 &  Elasticity of substitution & \citeNP{chirinko2008sigma}\\
       $\gamma_{k}$ & 0.10 & Capital-augmenting parameter & Lemma \ref{lemma:lemma_2A_AR}\\ 
                          $\alpha$ & 1.15 & Labor-augmenting parameter & $w/P \approx 1$\\ 
       $\upsilon $ & 5.84 & Complementarity of investments  &  \citeNP{christiano2005nominal} \\
       $\pi_{I}$ & 0.12 & Probability of investment projects & \citeNP{altug1989time}\\
          Preferences & & & \\      
           $\epsilon_{1}$ & 0.75 &  Frisch elasticity  & \citeNP{hall2009reconciling}\\
    $\epsilon_{0} $ & 1.1 &  Labor supply parameter & $ h \approx 1$ \\
               Search and matching & & & \\      
          $\iota$ & 1.27 & Matching function parameter & \citeNP{den2000job}\\
    $\lambda$ & 0.034 & Separation rate & \citeNP{shimer2005cyclical}\\
       $b^{\beta}_{1}$ & 0.5 & Steady-state reaction of UI & Figure \ref{fig:kalman_post} \\
              $\kappa_{1}$ & 0.5 & Fixed hiring costs & \citeNP{petrosky2018endogenous} \\
    \hline 
  \end{tabular}}
\label{table:calibration_steady_state1}
  \end{table}

  \textbf{Policy equation.} To specify the dynamics of  $b_{t}$ in the model,  I follow \citeA{hagedorn2015impact} and \citeA{chodorow2019macro} and set a policy equation mapping real unemployment benefits to the most recent value of the rate of unemployment. This is captured using an unobserved components model described by\footnote{The regression in \eqref{eq:tvp_ub} uses the same data as \citeA{chodorow2016cyclicality},  but unlike them the dependent variable is normalized with respect to the current value of labor productivity. }

  \begin{equation}\label{eq:tvp_ub}
  \begin{split}
  \frac{b_{t}/P_{t}}{(Y/N)_{t}} &= \beta^{b}_{0,t} +  \beta^{b}_{1,t}U_{t-1}  + \sum_{l=1}^{p} \beta^{b}_{l+1,t} \frac{b_{t-l}/P_{t-l}}{(Y/N)_{t-l}} + \epsilon^{b}_{t}, \;\;\;  \epsilon^{b}_{t} \sim N.I.I.D(0,\varphi^2_{b})\\
  \beta^{b}_{i,t} &=  \beta^{b}_{i,t-1} + \epsilon^{\beta^{b}_{i}}_{t},\;\;\;\;  \epsilon^{\beta^{b}_{i}}_{t}  \sim N.I.I.D(0,\varphi^2_{\beta^{b}_{i}})\;\; (\text{for}\; i=0,1,...,p+1)
  \end{split}
\end{equation}   

The time-varying elements of the equation portray changes in policy behavior related to real UI  extensions. These changes may occur as a consequence of recessions \cite{chodorow2019macro},  high periods of inflation \cite[p. 118]{pierson1994dismantling}, or active policies seeking to reduce the burden of welfare costs (\citeNP[p. 116]{pierson1994dismantling}; \citeNP[p.120]{noble1997welfare}). 

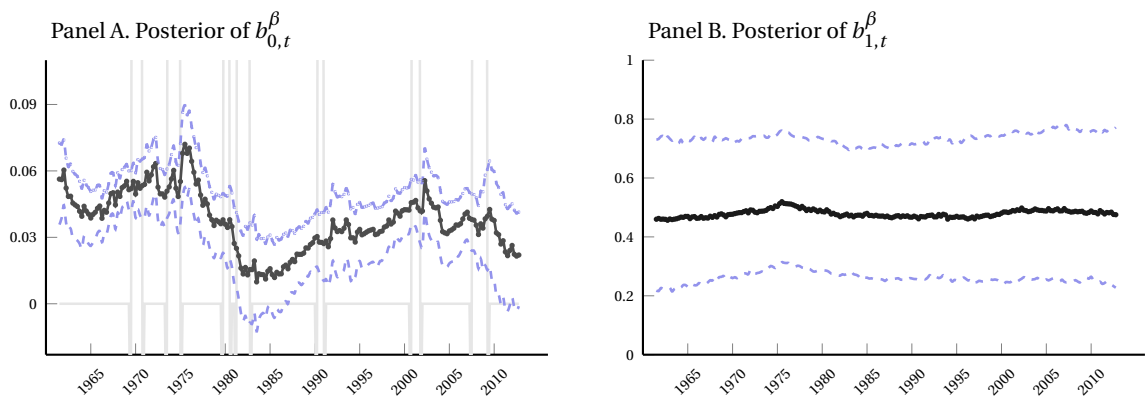
\begin{figure}
\begin{center}
\pgfplotstableread[col sep=comma,]{dt_kalman_UI.csv}\datatable
\pgfplotstableread[col sep=comma,]{data_UI.csv}\datatablee
  \begin{tikzpicture}

  \begin{axis}[
  name=plot1,
  width=0.5*\textwidth, height=5.5cm,
    y label style={at={(axis description cs:-0.065,.5)},rotate=90,anchor=south},
  xmin=1960,
  xmax=2016,
  ymin=-0.023,
  ymax=0.11,  
  axis x line*=bottom,
axis y line*=left,
  tick label style={font=\tiny,/pgf/number format/fixed,
                                 /pgf/number format/precision=3},
  title={\footnotesize Panel A.   Posterior  of $b^{\beta}_{0,t}$},
    title style={at={(axis description cs:0.25,0.95)}, anchor=south} , 
  xticklabel style={/pgf/number format/set thousands separator={}},
   every y tick scale label/.style={at={(yticklabel cs:1)},anchor=south west},
  xtick={1965,1970,1975,...,2010},
  x tick label style={rotate=45},
    ytick={-0.03,0,0.03,...,0.09}   ]

  \addplot [mark=*, mark size = 0pt,mark options={fill=myblue!30}, draw=gray!20,line width=0.95, smooth] table[x index = {0}, y index = {7}]{\datatable};

  \addplot [mark=*, mark size = 0.5pt,mark options={fill=myblue!30}, draw=black!70,line width=1, smooth] table[x index = {0}, y index = {4}]{\datatable};
  
  \addplot [mark=*, dashed,  mark size = 0.1pt,mark options={fill=myblue!30}, 
draw=myblue!50,line width=1, smooth] table[x index = {0}, y index = {5}]{\datatable};
  
  \addplot [mark=*, dashed, gray!30,, mark size = 0pt,mark options={fill=myblue!30}, 
draw=myblue!50,line width=1, smooth] table[x index = {0}, y index = {6}]{\datatable};

  \end{axis}

   \begin{axis}[%
  name=plot2,
  width=0.5*\textwidth, height=5.5 cm,
      at=(plot1.right of south east), anchor=left of south west,
    y label style={at={(axis description cs:-0.065,.5)},rotate=90,anchor=south},
  xmin=1960,
  xmax=2016,
  ymin=0,
  ymax=1,
    axis x line*=bottom,
axis y line*=left,
  tick label style={font=\tiny},
  legend pos=south west,
  legend style={draw=none,font=\footnotesize},
   legend cell align={left}, 
  title={\footnotesize Panel B. Posterior of $b^{\beta}_{1,t}$},
    title style={at={(axis description cs:0.25,0.95)}, anchor=south} , 
  xticklabel style={/pgf/number format/set thousands separator={}},
  xtick={1965,1970,1975,...,2010},
  x tick label style={rotate=45}  ]

  \addplot [mark=*,  mark size = 0.5 pt,mark options={fill=myblue!30}, draw=black!90,line width=1, smooth] table[x index = {0}, y index = {1}]{\datatable};
  
  \addplot [mark=*, dashed, mark size = 0pt,mark options={fill=myblue!30}, draw=myblue!50,line width=1, smooth] table[x index = {0}, y index = {2}]{\datatable};
  
  \addplot [mark=*, dashed, mark size = 0pt,mark options={fill=myblue!30}, draw=myblue!50,line width=1, smooth] table[x index = {0}, y index = {3}]{\datatable};
  
    \end{axis}

    \end{tikzpicture}
  \end{center}
  \caption{Time-series of labor opportunity costs. \emph{Notes---}  Panels A and B report the 68$\%$ probability intervals of the posterior densities  of $b^{\beta}_{0,t}$ and $b^{\beta}_{1,t}$.   The grey bars are the NBER  recession dates. The estimation details of equation  \eqref{eq:tvp_ub} are presented in Appendix \ref{appendix:kalman}.  \label{fig:kalman_post}}
  \end{figure}

The results of \eqref{eq:tvp_ub} are reported in  Figure  \ref{fig:kalman_post} using the reduced form parameters $b^{\beta}_{0,t} \equiv \beta^{b}_{0,t}/(1-\sum_{l=1}^{p} \beta^{b}_{l+1,t})$ and $b^{\beta}_{1,t} \equiv \beta^{b}_{1,t}/(1-\sum_{l=1}^{p} \beta^{b}_{l+1,t})$, which represent the relevant information for the steady-state  analysis below. Here, in particular, \eqref{eq:tvp_ub} is transformed to
 
 \begin{equation*}
 \hat{b}_{t} = b^{\beta}_{0,t}  + b^{\beta}_{1} U^*_{t},
 \end{equation*}
 
 where $ \hat{b}_{t}  = (b_{t}/P_{t}) e^{-\alpha J^{*}_{t}}$, $U^{*}_{t}$ is the equilibrium rate of unemployment,  $b^{\beta}_{1}=0.5$ is   a high density point of the posterior distribution of $b^{\beta}_{1,t}$, and  $b^{\beta}_{0}$ is fixed at a high density point of the posterior distribution of $b^{\beta}_{0,t}$  in relation to the relevant periods of investigation discussed in the following subsection. 
 
  \subsection{Quantitative Results}\label{sub:empirical_results}
  
  In the remaining part of the section I evaluate  the extent to which automation, measured by changes in $m_{t}$, and labor institutions, represented by variations in the discount factor of workers $\beta^{w}_{t}$, explain the changes in the steady-state equilibrium of income shares, capital returns, unemployment and the investment-output ratio in the postwar US economy.

 \textbf{Data.} The data used for the analysis  comes from the BEA-BLS integrated industry-level production account \cite{eldridge2020toward}, the Fixed Assets Accounts Tables, and  the Bureau of Labor Statistics (BLS).\footnote{The BEA-BLS integrated   data  is still under construction and is built with  some sketchy assumptions. However,  it is the only source covering the entire postwar period and including detailed data for 63 industries. } To keep the empirical exercise consistent with the structure of the theoretical model,  I exclude all ``imputed" outputs that are not actually marketed and realized as money revenue.  Essentially, I follow \citeA{basu2013dynamics}  and exclude Finance, insurance and real state (FIRE) sectors, Education and Health Services,  and Professional and Business Services.\footnote{The BEA, for example, calculates the value added of the banking sector from interest rate spreads between lending and deposits rates, which has no direct relation with the production of goods and services in the economy. Parts of the payments to professional and business services can also be regarded as costs of reproduction of society, rather than  direct contributions to net output. Ultimately, the inclusion of FIRE and other related sectors to GDP is the result of convention, not of clear and uncontroversial economic reasoning. } The corresponding data on value added, compensation, depreciation, investment, fixed assets, and gross operating surplus for the remaining sectors are based  on  BEA industry codes. Appendix \ref{appendix:data_c} reports additional data of the wage share segmented by sectors and type of labor; showing that the labor share only continues to rise after the 1980s in sectors with questionable imputations in value added. 
 
All the measures are constructed net of the nominal value of  intermediate inputs. For example, the measure of costs of production is the sum of nominal college labor input, nominal non-college labor input and the  Current-Cost Depreciation of Private Fixed Assets.  Correspondingly, aggregate output is measured as nominal gross output minus nominal intermediate inputs.  Further details on the data and model results are contained in Appendix \ref{appendix:data_c}.

\begin{figure}
\begin{center}
\pgfplotstableread[col sep=comma,]{dt_latex_shares.csv}\datatable
\pgfplotstableread[col sep=comma,]{dt_latex_un.csv}\datatablee
\pgfplotstableread[col sep=comma,]{dt_latex_wages.csv}\datatablew
\pgfplotstableread[col sep=comma,]{dt_results_nash_SEP.csv}\datatableN
\pgfplotstableread[col sep=comma,]{dt_savings.csv}\datatableS
  \begin{tikzpicture}
  \begin{axis}[
  name=plot1,
  width=0.5*\textwidth, height=5 cm,
    y label style={at={(axis description cs:-0.065,.5)},rotate=90,anchor=south},
  xmin=1945,
  xmax=2019,
  ymin=0.2,
  ymax=0.55,
      axis x line*=bottom,
axis y line*=left,
  tick label style={font=\footnotesize},
  title={\footnotesize Panel A.   Rate of return of capital},
    title style={at={(axis description cs:0.3,0.95)}, anchor=south} , 
  xticklabel style={/pgf/number format/set thousands separator={}},
  xtick={1950,1960,1970,...,2020},
  ytick={0.25,0.3,...,0.55}  ]

  \addplot [mark=o, mark size = 1.5pt,mark options={fill=lightgray!30}, draw=gray,line width= 1, smooth] table[x index = {0}, y index = {1},
  each nth point={2}]{\datatable};
  
 
 
 
         \addplot [Red!70,thick,smooth,mark=square*,mark size=2pt, dash dot, line width=1,
    mark options={fill=Red!70,draw opacity=0}] coordinates{
    (1949, 0.4650)
    (1962.5,      0.466076) 
  };

          \addplot [NavyBlue!70,thick,smooth,mark=triangle*,mark size=3pt, dashed , line width=1.5,
    mark options={fill=NavyBlue!70, draw opacity=0}] coordinates{
    (1949,0.4650)
    (1962.5,  0.39776 )
   };

          \addplot [Red!70,thick,smooth,mark=square*,mark size=2pt, dash dot, line width=1,
    mark options={fill=Red!70,draw opacity=0}] coordinates{
    (1962.5,  0.3770)
    (1980,  0.388146) 
    }; 
    
       \addplot [Red!70,thick,smooth,mark=square*,mark size=2pt, dash dot, line width=1,
    mark options={fill=Red!70,draw opacity=0}] coordinates{
    (1980,  0.2970)
    (1998,  0.2965) 
    }; 
  
     \addplot [Red!70,thick,smooth,mark=square*,mark size=2pt, dash dot, line width=1,
    mark options={fill=Red!70,draw opacity=0}] coordinates{
    (1998, 0.33536)
    (2009,  0.3502560) 
    };

          \addplot [NavyBlue!70,thick,smooth,mark=triangle*,mark size=3pt, dashed, line width=1.25,
    mark options={fill=NavyBlue!70,draw opacity=0},  forget plot] coordinates{
             (1962.5,    0.37704) 
              (1980,    0.30244)  
   };

         \addplot [NavyBlue!70,thick,smooth,mark=triangle*,mark size=3pt, dashed, line width=1.25,
    mark options={fill=NavyBlue!70,draw opacity=0},  forget plot] coordinates{
             (1980,    0.29707) 
              (1998,      0.338308)  
   };
   
       \addplot [NavyBlue!70,thick,smooth,mark=triangle*,mark size=3.5pt, dashed, line width=1.25,
    mark options={fill=NavyBlue!70,draw opacity=0},  forget plot] coordinates{
             (1998,     0.33536) 
              (2009,      0.439519)  
   };

  \addplot  [Green!70,smooth,mark=diamond*,mark size=5pt, dotted, line width=1.25, 
    mark options={fill=Green!70,draw opacity=0}]  table[x index = {0}, y index = {7}]{\datatableN};

  \end{axis}

   \begin{axis}[%
  name=plot2,
  width=0.5*\textwidth, height=5 cm,
      at=(plot1.right of south east), anchor=left of south west,
    y label style={at={(axis description cs:-0.065,.5)},rotate=90,anchor=south},
  xmin=1945,
  xmax=2019,
  ymin=0.49,
  ymax=0.69,
        axis x line*=bottom,
axis y line*=left,
  tick label style={font=\footnotesize},
  legend pos=south west,
  legend style={draw=none,font=\footnotesize},
   legend cell align={left}, 
  title={\footnotesize Panel B. Labor share},
    title style={at={(axis description cs:0.17,0.975)}, anchor=south} , 
  xticklabel style={/pgf/number format/set thousands separator={}},
  xtick={1950,1960,1970,...,2010},
  ytick={0.52,0.57,...,0.7} ]

  \addplot [mark=o, mark size =1.5pt,mark options={fill=gray!30}, draw=gray,line width=1, smooth,forget plot] table[x index = {0}, y index = {1},
 each nth point={2}]{\datatablew};

         \addplot [Red!70,thick,smooth,mark=square*,mark size=2pt, dash dot, line width=1,
    mark options={fill=Red!70,draw opacity=0}] coordinates{
    (1949, 0.61485)
    (1962.5, 0.607202) 
  };
       \addlegendentry{Automation }

          \addplot [NavyBlue!70,thick,smooth,mark=triangle*,mark size=3pt, dashed , line width=1.5,
    mark options={fill=NavyBlue!70, draw opacity=0}] coordinates{
    (1949,0.6148)
    (1962.5,0.653241  )
   };
       \addlegendentry{Labor institutions }

    \addplot  [Green!70,smooth,mark=diamond*,mark size=4pt, dotted,line width=1.35,
    mark options={fill=Green!70,draw opacity=0}]  table[x index = {0}, y index = {9}]{\datatableN};
       \addlegendentry{Automation and institutions }

          \addplot [Red!70,thick,smooth,mark=square*,mark size=2pt, dash dot, line width=1,
    mark options={fill=Red!70,draw opacity=0}] coordinates{
    (1962.5, 0.6433)
    (1980, 0.6205) 
    }; 
    
       \addplot [Red!70,thick,smooth,mark=square*,mark size=2pt, dash dot, line width=1,
    mark options={fill=Red!70,draw opacity=0}] coordinates{
    (1980, 0.662642)
    (1998, 0.649362) 
    }; 
  
     \addplot [Red!70,thick,smooth,mark=square*,mark size=2pt, dash dot, line width=1,
    mark options={fill=Red!70,draw opacity=0}] coordinates{
    (1998, 0.62833)
    (2009, 0.6015) 
    };

          \addplot [NavyBlue!70,thick,smooth,mark=triangle*,mark size=3pt, dashed, line width=1.25,
    mark options={fill=NavyBlue!70,draw opacity=0},  forget plot] coordinates{
             (1962.5,   0.6433) 
              (1980,   0.68895)  
   };

         \addplot [NavyBlue!70,thick,smooth,mark=triangle*,mark size=3pt, dashed, line width=1.25,
    mark options={fill=NavyBlue!70,draw opacity=0},  forget plot] coordinates{
             (1980,   0.6626) 
              (1998,     0.64016)  
   };
   
       \addplot [NavyBlue!70,thick,smooth,mark=triangle*,mark size=3.5pt, dashed, line width=1.25,
    mark options={fill=NavyBlue!70,draw opacity=0},  forget plot] coordinates{
             (1998,    0.6283) 
              (2009,      0.579458)  
   };
   
       \addplot  [Green!70,smooth,mark=diamond*,mark size=4pt, dotted,line width=1.35,
    mark options={fill=Green!70,draw opacity=0}]  table[x index = {0}, y index = {9}]{\datatableN};
  
    \end{axis}
    
    \begin{axis}[
  name=plot3,
  width=0.5*\textwidth, height=5 cm,
   at=(plot1.below south east), anchor=above north east, 
    y label style={at={(axis description cs:-0.065,.5)},rotate=90,anchor=south},
  xmin=1945,
  xmax=2019,
  ymin=0.025,
  ymax=0.17,
       axis x line*=bottom,
axis y line*=left,
  tick label style={font=\footnotesize,/pgf/number format/fixed},
  title={\footnotesize Panel C. Capital cost share:  $\delta P^{k}K/PY$ },
      title style={at={(axis description cs:0.3,0.95)}, anchor=south} , 
  xticklabel style={/pgf/number format/set thousands separator={}},
  xtick={1950,1960,1970,...,2020},
  ytick={0.05,0.075,...,0.15}  ]  

  \addplot [mark=o, mark size = 1.5pt,mark options={fill=lightgray!30}, draw=gray,line width= 1, smooth] table[x index = {0}, y index = {2},
  each nth point={2}]{\datatable};

 
        \addplot [Red!70,thick,smooth,mark=square*,mark size=2pt, dash dot, line width=1.25,
    mark options={fill=Red!70,draw opacity=0}] coordinates{
    (1949,0.06769)
     (1962.5, 0.0797) 
   };
            \addplot [Red!70,thick,smooth,mark=square*,mark size=2pt, dash dot, line width=1.25,
    mark options={fill=Red!70,draw opacity=0}] coordinates{
             (1962.5,  0.08282) 
             (1980, 0.104)
   };
         \addplot [Red!70,thick,smooth,mark=square*,mark size=2pt, dash dot, line width=1.25,
    mark options={fill=Red!70,draw opacity=0}] coordinates{
             (1980,  0.10832) 
             (1998, 0.12268)
   };
         \addplot [Red!70,thick,smooth,mark=square*,mark size=2pt, dash dot, line width=1.25,
    mark options={fill=Red!70,draw opacity=0}] coordinates{
             (1998,  0.12052) 
             (2009, 0.1390)
   };
  
        \addplot [NavyBlue!70,thick,smooth,mark=triangle*,mark size=2.5pt, dashed, line width=1.25,
    mark options={fill=NavyBlue!70,draw opacity=0}] coordinates{
    (1949,0.06769)
     (1962.5, 0.0696) 
   };
            \addplot [NavyBlue!70,thick,smooth,mark=triangle*,mark size=3pt, dashed, line width=1.25,
    mark options={fill=NavyBlue!70,draw opacity=0}] coordinates{
             (1962.5,  0.08282) 
             (1980, 0.08564)
   };
         \addplot [NavyBlue!70,thick,smooth,mark=triangle*,mark size=3pt, dash dot, line width=1.25,
    mark options={fill=NavyBlue!70,draw opacity=0}] coordinates{
             (1980,  0.10832) 
             (1998, 0.10630)
   };
         \addplot [NavyBlue!70,thick,smooth,mark=triangle*,mark size=3pt, dashed, line width=1.25,
    mark options={fill=NavyBlue!70,draw opacity=0}] coordinates{
             (1998,  0.12052) 
             (2009, 0.11520)
   };

  \addplot  [Green!70,smooth,mark=diamond*,mark size=4pt, dashed,line width=1.25,
    mark options={fill=Green!70,draw opacity=0}]  table[x index = {0}, y index = {10}]{\datatableN};

  \end{axis}

    \begin{axis}[
  name=plot4,
  width=0.5*\textwidth, height=5 cm,
 at=(plot3.right of north east), anchor=left of north west,
    y label style={at={(axis description cs:-0.065,.5)},rotate=90,anchor=south},
  xmin=1945,
  xmax=2019,
  ymin=0.065,
  ymax=0.23,
         axis x line*=bottom,
axis y line*=left,
  tick label style={font=\footnotesize,/pgf/number format/fixed},
  title={\footnotesize Panel D. Investment-output ratio:  $X/Y$ },
      title style={at={(axis description cs:0.325,0.95)}, anchor=south} , 
  xticklabel style={/pgf/number format/set thousands separator={}},
    ytick={0.1,0.13,...,0.22},
  xtick={1950,1960,1970,...,2020}  ]

     \addplot [mark=o, mark size = 1.5pt,mark options={fill=gray!30}, draw=gray,line width=1, smooth,forget plot] table[x index = {0}, y index = {2},
  each nth point={2}]{\datatablew};
     
     \addplot [Red!70,thick,smooth,mark=square*,mark size=2pt, dash dot, line width=1.25,
    mark options={fill=Red!70,draw opacity=0}] coordinates{
    (1949,0.08201)
     (1962.5,0.0966) 
   };
            \addplot [Red!70,thick,smooth,mark=square*,mark size=2pt, dash dot, line width=1.25,
    mark options={fill=Red!70,draw opacity=0}] coordinates{
             (1962.5, 0.10034) 
             (1980, 0.1259)
   };
         \addplot [Red!70,thick,smooth,mark=square*,mark size=2pt, dash dot, line width=1.25,
    mark options={fill=Red!70,draw opacity=0}] coordinates{
             (1980,  0.13124) 
             (1998, 0.1486)
   };
         \addplot [Red!70,thick,smooth,mark=square*,mark size=2pt, dash dot, line width=1.25,
    mark options={fill=Red!70,draw opacity=0}] coordinates{
             (1998,  0.146028) 
             (2009, 0.16851)
   };
  
        \addplot [NavyBlue!70,thick,smooth,mark=triangle*,mark size=2.5pt, dashed, line width=1.25,
    mark options={fill=NavyBlue!70,draw opacity=0}] coordinates{
    (1949, 0.08201)
     (1962.5, 0.08436) 
   };
            \addplot [NavyBlue!70,thick,smooth,mark=triangle*,mark size=3pt, dashed, line width=1.25,
    mark options={fill=NavyBlue!70,draw opacity=0}] coordinates{
             (1962.5, 0.10034) 
             (1980, 0.1037)
   };
         \addplot [NavyBlue!70,thick,smooth,mark=triangle*,mark size=3pt, dash dot, line width=1.25,
    mark options={fill=NavyBlue!70,draw opacity=0}] coordinates{
             (1980,  0.13124) 
             (1998, 0.12880)
   };
         \addplot [NavyBlue!70,thick,smooth,mark=triangle*,mark size=3pt, dashed, line width=1.25,
    mark options={fill=NavyBlue!70,draw opacity=0}] coordinates{
             (1998,   0.146028) 
             (2009, 0.13959)
   };

  \addplot  [Green!70,smooth,mark=diamond*,mark size=4pt, dotted,line width=1.5,
    mark options={fill=Green!70,draw opacity=1}]  table[x index = {0}, y index = {6}]{\datatableN};

    \end{axis}

    \begin{axis}[
  name=plot5,
  width=0.5*\textwidth, height=5 cm,
   at=(plot3.below south east), anchor=above north east, 
    y label style={at={(axis description cs:-0.065,.5)},rotate=90,anchor=south},
 xmin=1945,
  xmax=2020,
  ymin=2,
  ymax=14.5,
         axis x line*=bottom,
axis y line*=left,
  tick label style={font=\footnotesize},
  title={\footnotesize Panel E. Rate of unemployment},
    title style={at={(axis description cs:0.27,0.95)}, anchor=south} , 
  xticklabel style={/pgf/number format/set thousands separator={}},
  xtick={1950,1960,1970,...,2010},
    ytick={4,6.5,...,14.5}  ]  

  \addplot [mark=o, mark size = 1.5pt,mark options={fill=myblue!30}, draw=gray,line width=1, smooth] table[x index = {0}, y index = {1},
  each nth point={4}]{\datatablee};
  
 %
 
    \addplot [Red!70,thick,smooth,mark=square*,mark size=2pt, dash dot, line width=1.25,
    mark options={fill=Red!70,draw opacity=0}] coordinates{
    (1950, 5.78369)
     (1962.5, 5.8725) 
   };
            \addplot [Red!70,thick,smooth,mark=square*,mark size=2pt, dash dot, line width=1.25,
    mark options={fill=Red!70,draw opacity=0}] coordinates{
             (1962.5, 7.0428) 
             (1980, 7.0739)
   };
         \addplot [Red!70,thick,smooth,mark=square*,mark size=2pt, dash dot, line width=1.25,
    mark options={fill=Red!70,draw opacity=0}] coordinates{
             (1980, 10.856) 
             (1998, 11.409)
   };
         \addplot [Red!70,thick,smooth,mark=square*,mark size=2pt, dash dot, line width=1.25,
    mark options={fill=Red!70,draw opacity=0}] coordinates{
             (1998,  8.9310345) 
             (2009, 9.00264)
   };
  
        \addplot [NavyBlue!70,thick,smooth,mark=triangle*,mark size=2.5pt, dashed, line width=1.25,
    mark options={fill=NavyBlue!70,draw opacity=0}] coordinates{
    (1950, 5.783691)
     (1962.5, 6.5261) 
   };
            \addplot [NavyBlue!70,thick,smooth,mark=triangle*,mark size=3pt, dashed, line width=1.25,
    mark options={fill=NavyBlue!70,draw opacity=0}] coordinates{
             (1962.5, 7.0428) 
             (1980, 9.7262)
   };
         \addplot [NavyBlue!70,thick,smooth,mark=triangle*,mark size=3pt, dash dot, line width=1.25,
    mark options={fill=NavyBlue!70,draw opacity=0}] coordinates{
             (1980, 10.856) 
             (1998, 8.48269)
   };
         \addplot [NavyBlue!70,thick,smooth,mark=triangle*,mark size=3pt, dashed, line width=1.25,
    mark options={fill=NavyBlue!70,draw opacity=0}] coordinates{
             (1998,   8.9310345) 
             (2009, 6.47234)
   };

    \addplot [Green!70,smooth,mark=diamond*,mark size=4pt, dotted,line width=1.5,
    mark options={fill=Green!70,draw opacity=1}]  table[x index = {0}, y index = {4}]{\datatableN};

    \end{axis}

      \begin{axis}[
  name=plot6,
  width=0.5*\textwidth, height=5 cm,
  at=(plot5.right of north east), anchor=left of north west,
    y label style={at={(axis description cs:-0.065,.5)},rotate=90,anchor=south},
  xmin=1945,
  xmax=2019,
  ymin=0.0,
  ymax=0.9,
           axis x line*=bottom,
axis y line*=left,
  tick label style={font=\footnotesize},
 legend pos=south west,
  legend style={draw=none,font=\tiny},
   legend cell align={left}, 
  title={\footnotesize Panel F.  Capitalist savings rate:  $s$ },
      title style={at={(axis description cs:0.3,0.95)}, anchor=south} , 
  xticklabel style={/pgf/number format/set thousands separator={}},
    ytick={0.0,0.2,...,0.9} ,
  xtick={1950,1960,1970,...,2020}  ]

      \addplot [mark=o, mark size = 1.5pt,
      mark options={fill=gray!50}, draw=gray,line width=1, smooth] table[x index = {0}, y index = {1},
  each nth point={6}]{\datatableS}; 
      \addlegendentry{ with IVA and CCAdj}; 

           \addplot [mark=square*, mark size = 1.5pt, mark options={fill=gray!90},  draw=black!50,line width=1, smooth] table[x index = {0}, y index = {2},
  each nth point={7}]{\datatableS};
      \addlegendentry{without IVA and CCAdj};


    \addplot [Red!70,thick,smooth,mark=square*,mark size=2pt, dash dot, line width=1.25,
    mark options={fill=Red!70,draw opacity=0}] coordinates{
    (1950, 0.4429949)
     (1962.5, 0.428735) 
   };
            \addplot [Red!70,thick,smooth,mark=square*,mark size=2pt, dash dot, line width=1.25,
    mark options={fill=Red!70,draw opacity=0}] coordinates{
             (1962.5,0.5682882) 
             (1980, 0.54285)
   };
         \addplot [Red!70,thick,smooth,mark=square*,mark size=2pt, dash dot, line width=1.25,
    mark options={fill=Red!70,draw opacity=0}] coordinates{
             (1980,0.740244) 
             (1998, 0.7802)
   };
         \addplot [Red!70,thick,smooth,mark=square*,mark size=2pt, dash dot, line width=1.25,
    mark options={fill=Red!70,draw opacity=0}] coordinates{
             (1998, 0.633710) 
             (2009, 0.63963)
   };
  
        \addplot [NavyBlue!70,thick,smooth,mark=triangle*,mark size=2.5pt, dashed, line width=1.25,
    mark options={fill=NavyBlue!70,draw opacity=0}] coordinates{
    (1950, 0.4429949)
     (1962.5, 0.556023) 
   };
            \addplot [NavyBlue!70,thick,smooth,mark=triangle*,mark size=3pt, dashed, line width=1.25,
    mark options={fill=NavyBlue!70,draw opacity=0}] coordinates{
           (1962.5, 0.5682882) 
             (1980, 0.76944)
   };
         \addplot [NavyBlue!70,thick,smooth,mark=triangle*,mark size=3pt, dash dot, line width=1.25,
    mark options={fill=NavyBlue!70,draw opacity=0}] coordinates{
             (1980,0.740244) 
             (1998, 0.681552)
   };
         \addplot [NavyBlue!70,thick,smooth,mark=triangle*,mark size=3pt, dashed, line width=1.25,
    mark options={fill=NavyBlue!70,draw opacity=0}] coordinates{
         (1998, 0.633710) 
             (2009, 0.500223)
   };

  \addplot  [Green!70,smooth,mark=diamond*,mark size=4pt, dotted,line width=1.25,
    mark options={fill=Green!70,draw opacity=0}]  table[x index = {0}, y index = {5}]{\datatableN};
  \end{axis}

    \end{tikzpicture}
  \end{center}
  \caption{Steady-state paths associated with the automation and labor institutions hypotheses. \emph{Notes---}  Panels A, B, C and D use  BLS-BEA  data   \protect  \cite{eldridge2020toward}. Panel E  uses the non-farming unemployment rate  data of  \protect  \citeA{petrosky2021unemployment}. Each  savings rate express the  ratio of  undistributed corporate profits after taxes  to Corporate Profits after taxes.    \label{fig:counterfactual_test}}
  \end{figure}
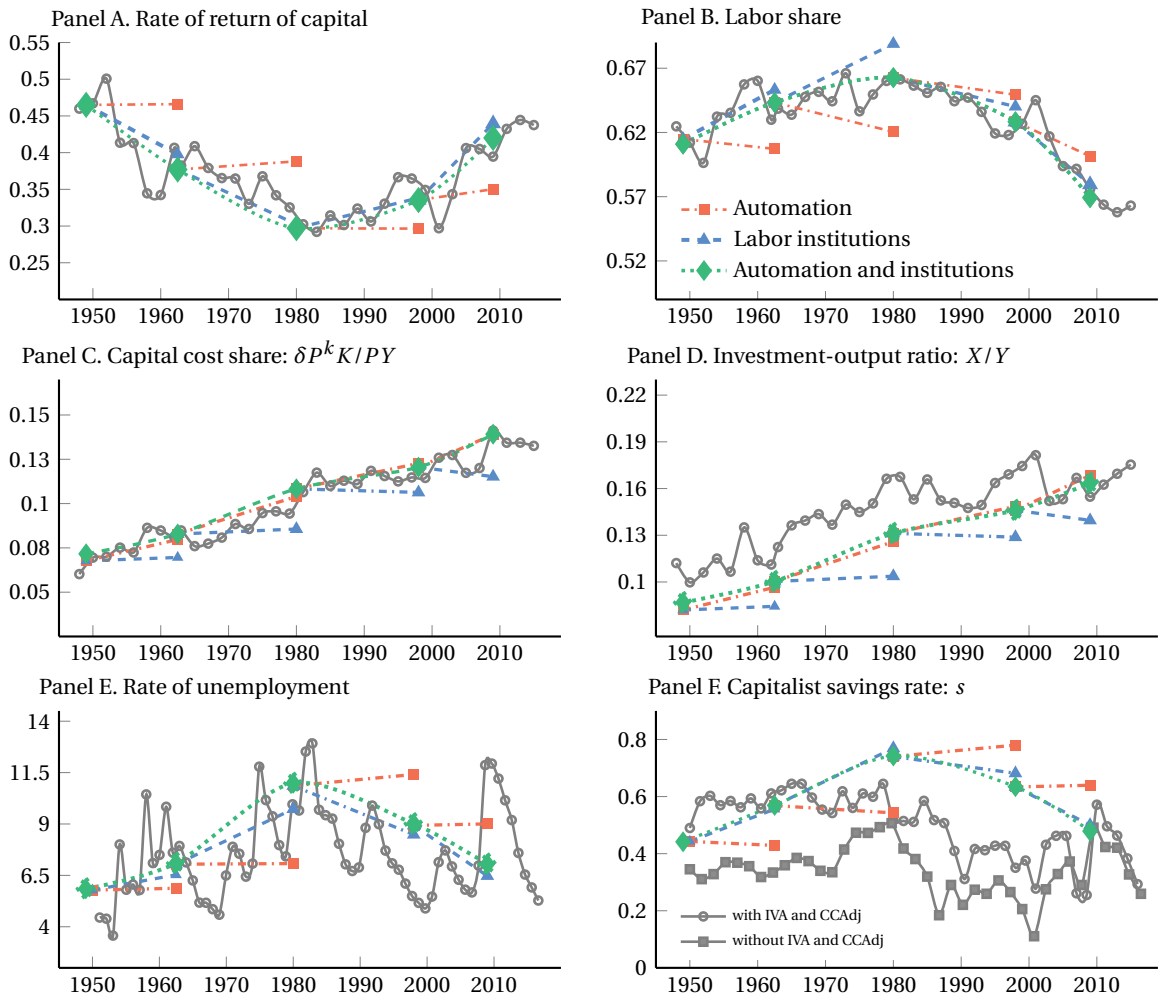

\textbf{Main Empirical Results.}  Figure \ref{fig:counterfactual_test} contains a summary of the main empirical results. The green diamonds depict the adjusted values of the model obtained by estimating $m_{t}$ and $\beta^{w}_{t}$ to target a five year average of the rate of return and the capital cost share around the years 1950, 1963, 1980,  1998, and 2010.  Correspondingly, to test if  changes in automation or  in labor institutions can individually describe  the behavior of the postwar US economy, I vary the parameters associated with each hypothesis and leave the remaining ones unaltered. For example, to evaluate the hypothesis that automation explains the decline of the labor share following the 1980s, I set $m$ to its adjusted value from 1996-2001 and leave  the remaining parameters equal to the calibrated values of  the period 1978-1983.

The results associated with the  institutions hypothesis  in Figure \ref{fig:counterfactual_test} (Panels A,  B, and E)  are broadly aligned with the empirical findings of \citeA{stansbury2020declining}, who show that changes in the bargaining power of workers offers a unified explanation for the behavior of the labor share, capital returns, and the rate of unemployment.  In Appendix \ref{appendix:data_c}, I show that the adjusted equilibrium values of the model are also consistent  with the changes of the vacancy-unemployment ratio.

The clear differences between the blue and green lines in Figure \ref{fig:counterfactual_test} (Panels C and D), however, weaken the predictive power of the labor institutions hypothesis. Contrary to what is reported by the data of the capital cost share and the investment-output ratio, the model with a constant $m$ undervalues the rise of  investment  since periods of declining labor shares should generally reduce, not rise, the participation of capital on aggregate output (see Proposition \ref{prop:comp_stat} (ii)). 

The rise in the rate of automation  presents a plausible explanation for the increase in  the  capital cost share  and the investment-output  ratio, together with  the fall in the labor share since the 1980s. A simple inspection of Panels C and D in Figure \ref{fig:counterfactual_test} shows that the red and green lines are almost perfectly aligned, meaning that the main variations in technology are described by changes in automation.  Appendix \ref{appendix:data_c} presents   the data of the net investment-output ratio, the capital-output ratio and the rate of profit,  and shows  that an adequate representation of these variables  requires a growing rate of automation consistent with the data in Figure \ref{fig:automation} below.\footnote{The model slightly underestimates the capital-output ratio, which implies  that the depreciation rate chosen in the calibration is probably too high. However, given that reducing $\delta$ also implies a higher rate of savings in equilibrium (see Theorem \ref{theorem:equil_harrod}), I preferred not to change the calibration and simply point out that the model can improve its predictive power of technical change by lowering  $\delta$. A more detailed explanation of this problem can be found in Appendix \ref{appendix:data_c}.} 

\citeA{moll2022uneven} reach similar conclusions in an exercise of transitional dynamics,  though they attribute most of the increase in the capital share to a surge in the return to wealth, rather than a rise in the capital-output ratio.   Like most models using a task-based framework (e.g., \citeNP{acemoglu2018race,hemous2022rise}), \citeA{moll2022uneven} work with the assumption that labor is inelastically supplied at full employment, and that factor prices are determined by their corresponding marginal products. This  is diametrically opposed to the approach of this paper, and it leads to predictions which are  at odds with the data of the rate of unemployment, on one hand, and with the data of the labor share and capital returns \emph{before} the 1980s, on the other.

The behavior of the savings rate in Figure \ref{fig:counterfactual_test} (Panel F) can be understood  using Theorem \ref{theorem:equil_harrod} (A).\footnote{ Let us remember  that the model works with the assumption that capitalists finance all capital investment from retained earnings, so it is not a surprise that the savings rate in the model is generally higher than in the data. The important point here is that---regardless of the level---all empirical measures of the savings rate show a positive trend before the 1980s, when the rate of return of capital was falling, and a negative trend after the 1980s when capital returns were rising.} Essentially,  if the long-run rate of growth of the economy is approximately constant, the savings rate will tend to move in the opposite direction to the rate of return of capital.  This presents a simple mechanism that can explain  the puzzle  of a rising wealth-to-GDP ratio and a decreasing  private savings rate since the  1980s found by authors like  \citeA{eggertsson2021kaldor}. 

The changes of the capitalist savings rate  have major   implications from  a political economy perspective. Particularly, it reveals that---because the growth rate of capitalist economies is bounded by the rate of profit\footnote{This is a well known result that can be traced back to \citeA{neumann1945model}.}---the system will probably  find bottlenecks which impede its reproduction when the two rates get closer together. This means that, though there may not exist a negative relation between growth and, say, a  widening institutional support to workers (see, e.g., Figures 11.12 and 11.13 in \citeA{piketty2020capital}), if these changes   lower the profitability of  capital it  is likely that the system will try to adjust itself through  economic crises or  political manifestations that end up   favoring capital over labor. The next section explores how this conclusion and the main empirical results outlined in Figure \ref{fig:counterfactual_test} are consistent with and provide an analytically foundation to some of the historical events of the postwar US economy. 

    \section{A Brief Historical Analysis of  the US Economy}\label{sec:hist_analysis}

   This section focuses on two key questions. First, I explore how the history and data of welfare and labor institutions is associated with changes in capital returns and the reproduction capacity of the system, and whether this historical evidence can be reconciled  with the predictions of bargaining power derived from the model.  Second, I evaluate to what extent is the data and history of technical change consistent with the model predictions of automation based on equation \eqref{subeq:steady_marg_prod_cap}, and how some of these changes may have been prompted by institutional changes in the US economy.

\subsection{Historical analysis of  bargaining power and profitability} 
\begin{figure}
\begin{center}
\pgfplotstableread[col sep=comma,]{dt_results_nash_SEP.csv}\datatable

  \begin{tikzpicture}
  \begin{axis}[
 axis x line*=bottom,
axis y line*=left,
  name=plot1,
    legend columns=2, 
  legend pos=south west,
  legend style={draw=none,font=\footnotesize},
   legend cell align={left}, 
         axis x line*=bottom,
axis y line*=left,
  name=plot1,
  width=0.95*\textwidth, height=8.5cm,
  xmin=1945,
  xmax=2015,
  ymin=0,
  ymax=0.091,
    legend pos=north west,
  legend style={draw=none,font=\footnotesize},
   legend cell align={left}, 
  xticklabel style={/pgf/number format/set thousands separator={}},
  xtick={1950,1955,1960,...,2015},
  ytick={0,0.02,0.04,...,0.08},
   extra y ticks={0.09}, extra y tick labels={$\eta_{w}$},
scaled ticks=false, tick label style={/pgf/number format/fixed} ]  
  
  \addplot [mark=diamond*, dotted, mark size = 4pt,mark options={solid,fill=Green!70}, draw=gray!80,line width=1.25, smooth] table[x index = {0}, y index = {1}]{\datatable};
          \addlegendentry{Relative bargaining power};

     \draw[
line width=1.25,
dotted,
->,
myblue!70
]
(1946,0.023)--node [pos=1, right,font=\tiny,align=center,text=myblue]{Employment Act \\
 (1946)}(1946,0.076);

     \draw[
line width=1.25,
dashed,
->,
BrickRed!70
]
(1947,0.024)--node [pos=1, right,font=\tiny,align=center,text=BrickRed]{Taft-Hartley\\
 (1947)}(1947,0.0165);
   
      \draw[
line width=1.25,
dotted,
->,
myblue!70
]
(1950,0.026)--node [pos=1,xshift=0.5em, above,font=\tiny,align=left,text=myblue]{Treaty of\\
 Detroit (1950)}(1950,0.034);
 
       \draw[
line width=1.25,
dashed,
->,
BrickRed!70
]
(1950,0.048)--node [pos=1,xshift=0.5em, above,font=\tiny,align=center,
text=BrickRed]{McCarran \\ Act (1950)}(1950,0.058);

   \draw[
line width=1.25,
dotted,
->,
myblue!70
]
(1961,0.035)--node [pos=1,xshift=0.25em, below,font=\tiny,align=center,
text=myblue]{AFDC-UP\\
 (1961)}(1961,0.025);

   \draw[
line width=1.25,
dotted,
->,
myblue!70
]
(1964,0.042)--node [pos=1,above,font=\tiny,align=center,text=myblue]{Economic\\ Opportunity Act  \\
 (1964)}(1972,0.042);

   \draw[
line width=1.25,
dashed,
->,
BrickRed!70
]
(1964,0.039)--node [pos=1,below,font=\tiny,align=center,text=BrickRed]{Revenue Act \\
 (1964)}(1968,0.039);

   \draw[
line width=1.25,
dotted,
->,
myblue!70
]
(1965,0.041)--(1965,0.024)node [pos=0.99,right, font=\tiny,align=center,text=myblue]{Medicaid \\
(1965)};

   \draw[
line width=1.25,
dotted,
->,
myblue!70
]
(1970,0.056)--(1970,0.0665)node [pos=0.7,above,font=\tiny,align=left,text=myblue]{ Occupational Safety\\
 and Health Act\\
  (1970)};

   \draw[
line width=1.25,
dashed,
->,
BrickRed!70
]
(1978,0.07)--(1978,0.035) node [pos=1,below,font=\tiny,align=center,text=BrickRed]{ Filibuster on \\
labor law \\
(1978)};

   \draw[
line width=1.25,
dashed,
->,
BrickRed!70
]
(1981,0.0824)--(1985,0.0824)node [pos=0.95,xshift=-0.15em,below,font=\tiny,align=center,text=BrickRed]{Volcker\\  (1981)};

   \draw[
line width=1.25,
dashed,
->,
BrickRed!70
]
(1981,0.074)--(1981,0.085)node [pos=0.95,above,font=\tiny,align=center,text=BrickRed]{OBRA (1981)};

   \draw[
line width=1.25,
dashed,
->,
BrickRed!70
]
(1981,0.073)--(1981, 0.05)node [pos=0.95,below,
 xshift=0.05em, font=\tiny,align=center,text=BrickRed]{ERTA \\
 (1981)};

   \draw[
line width=1.25,
dashed,
->,
BrickRed!70
]
(1986,0.07)--(1991,0.07)node [pos=1,above,font=\tiny,align=left,text=BrickRed]{Taxes on UI\\
 benefits\\
  (1986)};

   \draw[
line width=1.25,
dashed,
->,
BrickRed!70
]
(1996,0.0565)--(1996,0.036)node [pos=1,left,font=\tiny,align=center,text=BrickRed]{AFDC--TANF\\
 (1996)};

   \draw[
line width=1.25,
dotted,
->,
myblue!70
]
(2000,0.05)--(2005,0.05)node [pos=1,right,font=\tiny,align=center,text=myblue]{EITC\\
 (2000)};

   \draw[
line width=1.25,
dashed,
->,
BrickRed!70
]
(2001,0.0493)--(2001,0.055)node [pos=1,above,font=\tiny, align=center,text=BrickRed]{Economic Growth\\
 and Tax Relief \\
 Reconciliation Act \\
 (2001)};

   \draw[
line width=1.25,
dashed,
->,
BrickRed!70
]
(2003,0.044)--(2003,0.028) node [pos=0.9,below,xshift=-3em,font=\tiny,align=left,text=BrickRed]{Jobs and Growth Tax  Relief \\
Reconciliation Act (2003)};

\draw [thick,decoration={brace,amplitude=10pt,raise=1pt},decorate,
font=\tiny] 
  (axis cs:1948,0.001) --
    node[yshift=2 em]  {"Golden Age"} 
  (axis cs:1975,0.001);

\draw [thick,decoration={brace,amplitude=10pt,raise=1pt},decorate,
font=\tiny,align=center] 
  (axis cs:1979,0.001) --
    node[yshift=2.5em] {"Conservative \\
    retrenchment 1"} 
  (axis cs:1996,0.001);
  
\draw [thick,decoration={brace,amplitude=10pt,raise=1pt},decorate,
font=\tiny,align=center] 
  (axis cs:2001,0.001) --
    node[yshift=2.25 em] {"Conservative \\
    retrenchment 2"} 
  (axis cs:2009,0.001);

  \end{axis}

\begin{axis}[
  axis y line*=right,
  axis x line=none,
name=plot1,
  width=0.95*\textwidth, height=8.5 cm,
  xmin=1945,
  xmax=2015,
  ymin=0.88,
    ymax=1.00,
    legend columns=2, 
  legend pos=north east,
  legend style={draw=none,font=\footnotesize},
   legend cell align={left}, 
  xticklabel style={/pgf/number format/set thousands separator={}},
  xtick={},
  ytick={0.9,0.925,...,1.0},
   extra y ticks={1}, extra y tick labels={$\frac{\mu^{\text{min}}}{\mu}$},
scaled ticks=false, tick label style={/pgf/number format/fixed} ]

  \addplot [mark=*, dashed, mark size = 3pt,mark options={solid,fill=Black!70}, draw=gray!80,line width=1.25, smooth] table[x index = {0}, y index = {12}]{\datatable};
       \addlegendentry{Corridor of stability};
      
  \end{axis}
    \end{tikzpicture}
  \end{center}
  \caption{Relative wage bargaining power of workers and corridor of political and economic stability. \emph{Notes---} The dotted blue lines represent policies favoring the bargaining power of labor. The dashed red lines represent policies which may have contributed to weakening the power of workers.   \label{fig:bargaining_power}}
  \end{figure}
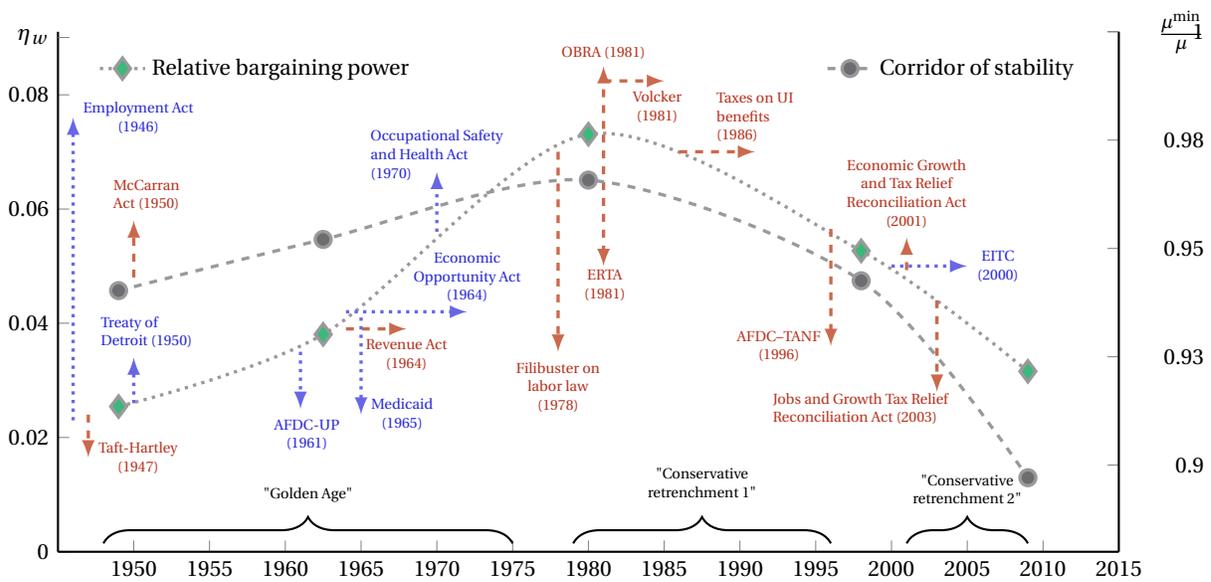

Figure \ref{fig:bargaining_power} tells a story about the balance of power between labor and capital in the postwar US economy.\footnote{The average value of $\eta_{w}$ in Figure \ref{fig:bargaining_power}  is close to the calibrations of  \citeA{hagedorn2008cyclical} and \citeA{petrosky2018endogenous}, but it deviates considerably from the values chosen by authors like \citeA{shimer2005cyclical}, \citeA{pissarides2009}, and \citeA{gertler2009unemployment}, who fix $\eta_{w} \geq 0.5$. Theorem \ref{theorem:equil_harrod} casts some doubts on  calibrations setting high values of $\eta_{w}$ given  that capitalist economies cannot grow without positive net aggregate profits. For instance, even if we set $\zeta=0$ in \eqref{eq:corridor} and use the model calibration of 1978-1982---which corresponds to the period with the highest  value of $\eta_{w}$---the maximum power of workers $\eta^{U}_{w}$ is about 0.12; way below 0.5. The bottom line in this respect is that  high values of $\eta_{w}$ are implausible, not because they cannot match specific features of the data, but because they fail to satisfy the minimum requirement of capitalist economies which is that the system can reproduce itself  at an increasing scale.  } The narrative of these events can be traced back to the Great Depression and the legacy of New Deal institutions, which paved the way for the introduction of federal cash and work relief programs in 1933, social insurance in 1935, a legal framework for collective bargaining in 1935, minimum wages in 1938,  federal regulation of working conditions in 1938, and tax hikes  on high income earners throughout the 1930s (\citeNP[p. 54]{noble1997welfare};  \citeNP {piketty2020capital}).\footnote{The crash of 1929  also opened a window for reforming the financial sector. It gave way, e.g.,  to the Glass-Steagall Act, which separated commercial banking from investment banking, and to the  Securities and Exchange Commission, which was intended  to reign over financial excesses \cite[p. 11]{eichengreen2014hall}. } 

By and large, these policies continued in the postwar era. The Employment Act of 1946, for example,  challenged the idea that the economy should be regulated by competitive forces alone and assigned  direct responsibility to the state in determining the level of employment \cite[p. 66]{bowles1982crisis}. In doing so, however, the government  drew limits on the capacity of labor in forming strikes, lock-downs or in organizing for disrupting production and investment decisions. The Taft-Hartley Act of 1947 and the McCarran Act of 1950 are specific examples   of these limits by providing a legal basis for the elimination of Communists and other leftist movements.

The agreement of workers in renouncing  all control over production and investment decisions meant that the role of labor organizations was largely centered on  securing  wage adjustments in relation to inflation and productivity improvements. The 1948 contract between GM and the United Auto Workers (UAW) presented an important step in this direction by introducing a Cost of Living Adjustment (COLA) clause and by allowing wages to be influenced by changes in the Consumer Price Index. By May of 1950, the UAW-GM contract---which came to be known as the ``Treaty of Detroit''--- guaranteed pensions, health insurance, and a 20 percent increase in the standard of living of auto worker under its provisions \cite[p. 123]{lichtenstein2002}. The COLA principle spread throughout the economy and even affected nonunion firms who approximated the conditions achieved by unions in the patterns of bargaining \cite[p. 368]{levy2011inequality}.

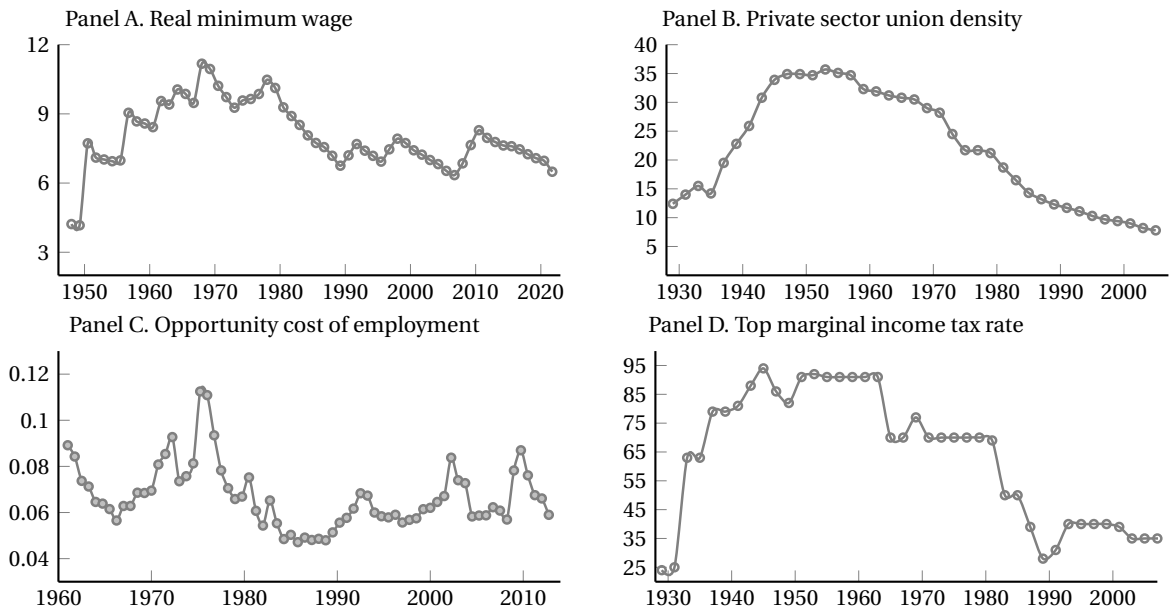
\begin{figure}
\begin{center}
\pgfplotstableread[col sep=comma,]{data_minwage.csv}\datatable
\pgfplotstableread[col sep=comma,]{data_unions.csv}\datatablee
\pgfplotstableread[col sep=comma,]{data_UI.csv}\datatableee
\pgfplotstableread[col sep=comma,]{data_mt.csv}\datatablex
  \begin{tikzpicture}

  \begin{axis}[
  name=plot1,
  width=0.5*\textwidth, height=4.65cm,
   at=(plot1.below south west), anchor=above north west,
    y label style={at={(axis description cs:-0.065,.5)},rotate=90,anchor=south},
  xmin=1946,
  xmax=2023,
  ymin=2,
  ymax=12,  
           axis x line*=bottom,
axis y line*=left,
  tick label style={font=\footnotesize},
  title={\footnotesize Panel A.  Real minimum wage},
    title style={at={(axis description cs:0.3,0.95)}, anchor=south} , 
  xticklabel style={/pgf/number format/set thousands separator={}},
   every y tick scale label/.style={at={(yticklabel cs:1)},anchor=south west},
  xtick={1950,1960,...,2010,2020},
      ytick={3,6,...,12}  ]

  \addplot [mark=o, mark size = 1.5pt,mark options={fill=gray!50}, draw=gray,line width=1, smooth] table[x index = {0}, y index = {1},,
  each nth point={5}]{\datatable};

  \end{axis}

   \begin{axis}[%
  name=plot2,
  width=0.5*\textwidth, height=4.65cm,
      at=(plot1.right of south east), anchor=left of south west,
    y label style={at={(axis description cs:-0.065,.5)},rotate=90,anchor=south},
  xmin=1928,
  xmax=2007,
  ymin=0,
  ymax=40,
           axis x line*=bottom,
axis y line*=left,
  tick label style={font=\footnotesize},
  legend pos=south west,
  legend style={draw=none,font=\footnotesize},
   legend cell align={left}, 
  title={\footnotesize Panel B. Private sector union density},
    title style={at={(axis description cs:0.35,0.95)}, anchor=south} , 
  xticklabel style={/pgf/number format/set thousands separator={}},
  xtick={1930,1940,...,2010},
    ytick={5,10,...,40}   ]

  \addplot [mark=o, mark size = 1.5 pt,mark options={fill=gray!50}, draw=gray,line width=1, smooth] table[x index = {0}, y index = {1},
  each nth point={2}]{\datatablee};
    \end{axis}

     \begin{axis}[%
 name=plot3,
  width=0.5*\textwidth, height=4.65cm,
  at=(plot1.below south west), anchor=above north west,
    y label style={at={(axis description cs:-0.065,.5)},rotate=90,anchor=south},
  xmin=1960,
  xmax=2014,
  ymin=0.03,
  ymax=0.13,  
           axis x line*=bottom,
axis y line*=left,
  tick label style={font=\footnotesize,/pgf/number format/fixed,
                                 /pgf/number format/precision=3},
  title={\footnotesize Panel C.  Opportunity cost of employment},
    title style={at={(axis description cs:0.43,0.95)}, anchor=south} , 
  xticklabel style={/pgf/number format/set thousands separator={}},
   every y tick scale label/.style={at={(yticklabel cs:1)},anchor=south west},
  xtick={1950,1960,...,2020}  ]

  \addplot [mark=*, mark size = 1.5pt,mark options={fill=gray!50}, draw=gray,line width=0.95, smooth] table[x index = {0}, y index = {1},
  each nth point={3}]{\datatableee};
      \end{axis}
  
   \begin{axis}[%
  name=plot4,
  width=0.5*\textwidth, height=4.65cm,
         at=(plot3.right of south east), anchor=left of south west,
    y label style={at={(axis description cs:-0.065,.5)},rotate=90,anchor=south},
  xmin=1928,
  xmax=2007,
  ymin=20,
  ymax=100,
           axis x line*=bottom,
axis y line*=left,
  tick label style={font=\footnotesize},
  legend pos=south west,
  legend style={draw=none,font=\footnotesize},
   legend cell align={left}, 
  title={\footnotesize Panel D. Top marginal income tax rate},
    title style={at={(axis description cs:0.36,0.95)}, anchor=south} , 
  xticklabel style={/pgf/number format/set thousands separator={}},
  xtick={1930,1940,...,2010},
    ytick={25,35,...,95}   ]

  \addplot [mark=o, mark size = 1.5pt,mark options={fill=gray!50}, draw=gray,line width=0.95, smooth] table[x index = {0}, y index = {1},
  each nth point={2}]{\datatablex};
    \end{axis}

    \end{tikzpicture}
  \end{center}
  \caption{Proxies of the institutional support to labor. \emph{Notes---} The minimum wage is the Federal Minimum Hourly Wage for Nonfarm Workers for the United States. Union density is the number of unionized workers as a share of the nonagricultural workforce and the data is from \protect  \citeA{hirsch2008sluggish}. Panel C reports the value of public benefits that unemployed forgo upon employed of \protect \citeA{chodorow2016cyclicality}. The data of top marginal income tax rate is from \protect \citeA{alvaredo2018world}.  \label{fig:unions_wages}}
  \end{figure}

The combination of wage adjustments in relation to inflation and labor productivity, together with the rise of fringe benefits resulting from wage bargaining agreements related to the Treaty of Detroit, provides a basis for understanding why the relative bargaining power of workers  increased during the 1950s as shown Figure \ref{fig:bargaining_power}.  This  is all the more   convincing when noting that the COLA principle was incorporated in more than 50 percent of union contracts by the early 1960s \cite[p. 123]{lichtenstein2002}, and that in this period  union density was close to its historical high (see Panel B in Figure \ref{fig:unions_wages}).\footnote{Recently, \citeA{taschereau2020union} presented a clear theoretical argument showing how the possibility of unionization  distorts the behavior of non-union firms, and can have significant effects in the economy even if the actual density of unions is not outstanding.}

The second expansion of the bargaining power of labor and the corridor of political and economic stability  can be explained by the rise of civil right movements and the rediscovery of poverty in the 1960s, which  opened a new window for social reforms leading to the War on Poverty (Economic Opportunity Act of 1964), the extension of medical care to the aged and the poor (Medicaid), the liberalization of public assistance, and the expansion of social security \cite[p. 79]{noble1997welfare}. The major changes of the decade started with president  Kennedy's extension of Aid to Families with Dependent Children (AFDC) for unemployed parents.\footnote{As a political compromise, Kennedy passed the Public Welfare Amendments of 1962 allowing  southern Democrats  some flexibility in the implementation of welfare in their states and enforcing stricter restrictions on benefits and eligibility \cite[p. 92]{noble1997welfare}. } Building on Kennedy's initiative, President Johnson implemented a more ambitious program with a focus on improving the living conditions of the worst-off members of society and expanded what \citeA{bowles1982crisis} and \citeA[p. 10]{lichtenstein2002} referred to as the citizen(social)-wage---representing ``that part of a person's consumption supplied by the state by virtue of his citizenship rather than directly acquired by the sale of labor power" \cite[p. 53]{bowles1982crisis}.  The rise of the citizen wage constituted a distributional victory of workers since it meant that they could finance a greater share of their consumption from social  expenditures. According to the estimates of  \citeA[p. 76]{bowles1982crisis}, the share of workers' consumption financed by welfare increased from about 13 percent in 1955 to 27 percent in 1977.\footnote{It is important not to overstate the role of welfare in the US economy in spite of the important results in poverty reduction during the 60s and 70s. By international standards,  the US assigned a relatively small share of GDP to social programs, such as unemployment, sickness, and maternity benefits \cite{rose1989exceptional}.}

The notion of citizen wages holds a close connection with the broad definition of UI  and non-UI benefits of \citeA{chodorow2016cyclicality} reported in Panel C of Figure \ref{fig:unions_wages}. The consistency of the two definitions not only establishes a bridge between the   policies reported in Figure \ref{fig:bargaining_power} with the data of unemployment benefits to labor productivity, but also provides a credible story explaining why the bargaining power of workers probably improved  from the early 1960s to the late 1970s---as predicted by the model. 

The extension of the welfare state and  the economic sphere of citizenship from the late 1940s  to the mid 1970s ultimately coincided  with the decrease in the profitability of capital and a rising rate of unemployment, revealing there probably exists  a juxtaposition between the principles of liberal democratic societies and those of capitalism. The basic problem, as shown formally in Theorem \ref{theorem:equil_harrod} and represented graphically in Figure \ref{fig:bargaining_power} with the corridor of political and economic stability, is that capitalism requires specific institutional arrangements  allowing  the expansion of capital at an increasing scale. Liberal democracy, by attaching rights on people rather than  property \cite{bowles1982crisis}, may confront the requirement of capital reproduction by improving the bargaining power of workers and by raising average wages through an increase in the citizen-wage. In this respect, societies may be incapable of reproducing the social relations forged by liberal policies \emph{and} the profits that maintain the accumulation of capital. 

 This contradictory nature of the welfare state with capitalism  was well understood by the conservative movement that finally materialized in the late 1970s. The extension of the COLA principles was filibustered under Carter's watch in 1978, making it harder   for workers to join unions and easier for employers to resits them \cite[p. 108]{noble1997welfare}. Reagan took these initiatives to a different level and proposed severe budget cuts in means-tested assistance and social service programs. Though many of these initiatives did not materialize to the degree they were  intended, the Omnibus Budget Reconciliation Act (OBRA) of 1981  managed to make significant reductions in food-stamp spending, AFDC assistance, and UI extensions by tightening the criteria for benefits eligibility \cite[pp. 116-119]{pierson1994dismantling}.\footnote{Reagan's initial cut proposal on social spending was  roughly twice as large as those ultimately accepted by Congress. Yet, according to \citeA[p. 206]{patterson2021america}, in two years OBRA increased poverty by 2 percent,  restricted eligibility to approximately 408,000 families, and eliminated benefits to other 300,000.}

 The political difficulty of a frontal reduction of welfare meant that the government had to search for indirect measures for cutting social  expenditures. The pinnacle of Reagan's reforms crystallized  with the Economic Recovery Tax  Act (ERTA)  of 1981; characterized for introducing substantial tax breaks to business and regressive cuts in personal income tax rates (see Panel D of Figure \ref{fig:unions_wages}). Combining income-tax cuts and increasing military spending, Reagan drove  deficits to record highs and managed to shift the concerns for social provisions for the poor and unemployed to a different one based on the need of achieving balanced budgets \cite[p. 123]{noble1997welfare}. A second indirect policy of social reform came with Volcker's tight-money shock. The high real interest rates of the early 1980s  increased the global demand for US securities, which ultimately crippled the recovery of employment in production sectors like manufacturing  and boosted the expansion of the financial sector following the recession of 1981 \cite{levy2011inequality}.

In addition to the  the aforementioned policies of the first conservative retrenchment, it should be noted---as shown in Figure \ref{fig:unions_wages} (Panels A and B)---that  the reduction of  federal real minimum wages and union density accelerated considerably in the wake of the 1980s. \citeA{DiNardo1996},  \citeA{card2002skill} and  \citeA{lemieux2008changing} present compelling evidence showing that much of the increase in wage inequality in the 1980s can be attributed  to the fall in minimum wages. The continuous fall in union density also helps explain the rising wage inequality after the 1990s, since unions not only protect the income of low paying jobs, but, as shown by \citeA{dinardo2000unions},  \citeA{rosenfeld2006widening}, and \citeA{lemieux2008changing}, also reduce the rents of management, executive and capital owners.

 It should come as no surprise then that Figure \ref{fig:bargaining_power} depicts a considerable decline in the bargaining power of labor following the 1980s. This decline was partly offset in the late  1990s by the Clinton government,  who---in spite of  replacing  AFDC   for the more restrictive program of Temporary Assistance for Needy Families (TANF)--- also took measures of redistribution by expanding the Earned Income Tax Credit (EITC), increasing the minimum wage, and rising the top income-tax rate \cite[p. 376]{levy2011inequality}. Some of this  was latter reversed by George W. Bush with the introduction of the Economic Growth and Tax Relief Reconciliation Act of 2001 and the Jobs and Growth Tax Relief Reconciliation Act of 2003, which adopted  a strategy of tax reduction for business and the wealthy (see Panel D of Figure \ref{fig:unions_wages}). 
 
 The  consequences of the conservative retrenchment just described are widespread and have been perfectly evident by the growing distrust of democratic institutions \cite{diamond2015facing}, the rise of deaths of despair \cite{case2021deaths}, the challenge against free trade \cite{dorn2020importing}, among other related manifestations. Though it would be far-fetched to put all the weight of these political reactions to the decline of the bargaining power of labor, it is a basis from which to  understand why people---especially those that have been negative affected by the turns of the economy over the past four decades---have reasons to demand radical changes in society.

  \subsection{Historical analysis of automation} One of the  hypothesis of this paper is that changes in technology and their lasting effects in the economy can be best  understood in the context of historic specific institutional environments. This view is well represented by the events of the US after the  1950s,  when a renewing concern over the  adverse effects of automation in the workplace resurfaced in the national debate. The government's approach at the time, in agreement with Taft-Hartley and  McCarran, was that  it should  provide public assistance  when needed for social dislocations generated by technological unemployment, but should not restrict the use of machines or even dispute the desirability of automation \cite[p. 180]{frey2019technology}.

                \begin{figure}
\begin{center}
\pgfplotstableread[col sep=comma,]{dt_results_nash_SEP.csv}\datatable
\pgfplotstableread[col sep=comma,]{dt_mech_model.csv}\datatablee
\pgfplotstableread[col sep=comma,]{dt_mech_Mann.csv}\datatablemech
\pgfplotstableread[col sep=comma,]{dt_mech_Dechezlepretre.csv}\datatablemechh

  \begin{tikzpicture}
  \begin{axis}[
           axis x line*=bottom,
axis y line*=left,
  name=plot1,
  width=0.87*\textwidth, height=7.5 cm,
  xmin=1945,
  xmax=2020,
  ymin=0.03,
    ymax=0.22,
    legend columns=1, 
  legend pos=south east,
  legend style={draw=none,font=\footnotesize},
   legend cell align={left}, 
  xticklabel style={/pgf/number format/set thousands separator={}},
  xtick={1950,1955,1960,...,2015},
  ytick={0.05,0.1,...,0.2},
   extra y ticks={0.215}, extra y tick labels={$1-m$},
scaled ticks=false, tick label style={/pgf/number format/fixed} ]

                  \addplot [mark=square* , mark size = 2 pt,mark options={solid,fill=myred!10},  draw=myred!50,line width=1, smooth] table[x index = {0}, y index = {1}]{\datatablemechh};
        \addlegendentry{\citeA{mann2021benign} };

          \addplot [mark=o, mark size = 2pt, draw=OliveGreen!70,line width=1.25, smooth] table[x index = {0}, y index = {1}]{\datatablee};
                       \addlegendentry{Steady-state automation measure};

          \addplot [mark=triangle*,
    mark options={fill=myblue!20}, mark size = 2.5 pt, draw=myblue!70,line width=1, smooth] table[x index = {0}, y index = {1}]{\datatablemech};
             \addlegendentry{\citeA{Dechezlepretre2019} };

\end{axis}

\begin{axis}[
  axis y line*=right,
  axis x line=none,
name=plot1,
  width=0.88*\textwidth, height=7.5 cm,
  xmin=1945,
  xmax=2020,
  ymin=0.1,
    ymax=0.35,
    legend columns=1, 
  legend pos=south west,
  legend style={draw=none,font=\footnotesize},
   legend cell align={left}, 
  xticklabel style={/pgf/number format/set thousands separator={}},
  xtick={},
  ytick={0.1,0.15,...,0.3},
   extra y ticks={0.345}, extra y tick labels={$1-\bar{m}$},
scaled ticks=false, tick label style={/pgf/number format/fixed} ]  
  
     \addplot [mark=square*, dashed, mark size = 3pt,mark options={solid,fill=orange!80}, draw=gray!70, line width=0.5, smooth] table[x index = {0}, y index = {3}]{\datatable};
       \addlegendentry{Cost minimizing automation measure };

                \draw[
thick,
|-|,
gray
]
(1947,      0.1649)--(1952,     0.1649) node [pos=0.5,above,yshift=1.25em, font=\footnotesize]{0.1677};

         \draw[
thick,
|-|,
gray
]
(1960, 0.2349)--(1965,0.2349) node [pos=0.5,above,yshift=0.25em, font=\footnotesize]{0.235};

         \draw[
thick,
|-|,
gray
]
(1978,0.3098) --(1982, 0.3098) node [pos=0.5,above,yshift=0.25em, font=\footnotesize]{0.309};

    \draw[
thick,
|-|,
gray
]
(1995, 0.2724)--(2001,0.2724)node [pos=0.5,above,yshift=0.5em, font=\footnotesize]{0.2724};

    \draw[
thick,
|-|,
gray
]
(2006, 0.199)--(2012,  0.199)node [pos=0.5,above,yshift=0.15em,font=\footnotesize]{0.199};

\end{axis}

    \end{tikzpicture}
  \end{center}
  \caption{Automation measures. \emph{Notes---} The steady-state measure of automation is obtained from \eqref{subeq:steady_profit_rate} by using the calibration in Table \ref{table:calibration_steady_state1} and the data of the  capital cost share and the rate of return of capital in Figure \ref{fig:counterfactual_test}. The cost-minimizing automation measure follows from Lemma \ref{lemma:lemma_2A_AR}.\label{fig:automation}}
  \end{figure}
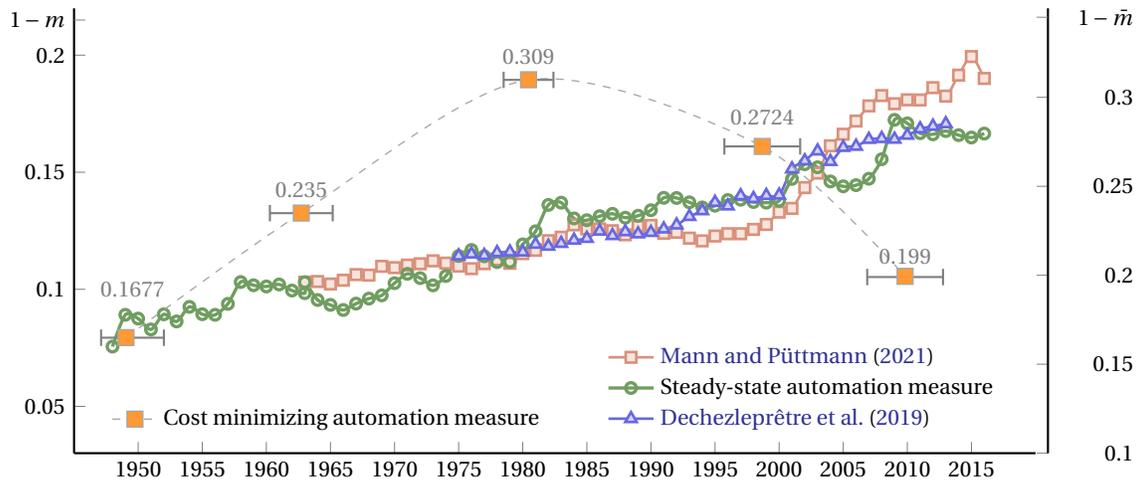
  
 This strategy was backed by a faith in Keynesianism--- crystallized in significant tax cuts for business and upper-income groups in 1964, the expansion of welfare throughout the 1960s and 70s, and the institutionalization of countercyclical economic policies---which ultimately  limited  the restoring effects that unemployment can have on profitability by setting a downward pressure on wage increases.\footnote{This conclusion  follows by increasing $Z_{t}$ and $\theta_{t}$ in equation  \eqref{eq:mark_up_nash}. Particularly, if $Z_{t}$ was increased by the government's concern over social welfare and if $\theta$ was kept relatively large as a result of demand driven policies, the result is a decreasing rate of return of capital, which is what is shown in  Figure \ref{fig:counterfactual_test}, Panel A. }  From this perspective it is expected that in the face of  mounting labor costs, especially after 1964, firms  accelerated the adoption of labor replacing technologies. This is precisely what the data of \citeA{mann2021benign} and the steady-state measure of automation derived from equation  \eqref{subeq:steady_profit_rate} show in Figure \ref{fig:automation}. The increasing relative  costs of labor is picked up by the  rise in $1-\bar{m}_{t}$ from 1962 to to 1980,  which provides theoretical  support to the view that the rise of welfare and the expansion of the citizen wage created the condition for automation to become a powerful player in reducing labor costs and increasing unemployment.
  
  By these standards, the computer revolution and similar technological improvements found  fertile grounds for creating potentially  disruptive effects on the labor market after the 1980s. According to the calibration of the model, by 1980 $1-\bar{m}_{t}$ was about twice as large as $1-m_{t}$, meaning that the system lied far above the boundary dividing  regions 1 and 2 in Figure \ref{fig:automation_reg} and that even small changes in automation could have generated significantly negative effects on employment and real wages. The expansion of labor replacing technologies  throughout the two decades from the early 1980s to the late 1990s was, however, relatively small  according to all measures in Figure \ref{fig:automation}. One of the reasons for this may have been the conservative retrenchment that followed after the late 1970s, which reduced the expansion of labor costs relative to the costs of capital. 
  
The 2000s saw a second revival of labor replacing technologies. Consistent with the findings of \citeA{autor2018automation}, Figure \ref{fig:automation} depicts a strong increase in the steady-state automation measure which moves along the data of \citeA{Dechezlepretre2019} and \citeA{mann2021benign}.   A peculiar characteristic of this period is that it coincided with a fall in the labor share, suggesting that the rise in automation was probably not prompted by rising labor costs but rather by technology improvements or tax cuts for capital. \citeA{acemoglu2020does} present evidence suggesting that the adoption of labor replacing technologies may have resulted from changes in the US tax system favoring the adoption of capital over labor. If  \citeA{acemoglu2020does} are correct, the fall in the cost-minimizing automation measure was likely  lower than was is reported in Figure \ref{fig:automation}, meaning that the disruptive effects of automation probably  continued to be important even after 2010.\footnote{ It is likely that COVID-19 exacerbated  the effects of automation on the labor market  since one of the defining characteristics of the  pandemic was the unprecedented increase in unemployment benefits and tight labor markets following the lock-downs. In this scenario it can be expected  that labor replacing technologies will be potentially effective  in reducing production costs of firms, which may  help explain the growing concerns in the media over the  effects automation.}

\section{Conclusions}\label{sec:conclusions}

This paper provides a framework that helps understand how institutions, automation, unemployment, and profitability  interact  in the dynamic setting of a general equilibrium analysis. At the center of the model is the notion that profits are  a surplus over costs of production, which brings back to light the importance of  political and institutional factors in the determination of income distribution. 

Among  the attractive  features of the paper is that it creates a link between task-based models, the surplus approach of the Classical economists, and  the literature on equilibrium unemployment. The merger of these  approaches introduces three  important theoretical contributions to the literature. First, it  establishes an endogenous theory of profits based on the relations of power between capitalists and workers. Second, it formalizes how unemployment and wage  dynamics are directly affected by the assignment of tasks between capital and labor. Third, it establishes how institutions can directly affect the task allocation of factors by intervening in the relative prices of labor and capital.

The empirical strength of the model is captured using two complementary strategies. First it is shown that  the behavior of income shares, capital returns, the rate of unemployment, and the big ratios of macroeconomics  in the American economy can be  explained by specific changes in labor institutions and technology. The second part  supports this conclusion by showing that the implied changes in the equilibrium measures of worker power and automation are consistent  with the history of welfare and technical change in the  US. Ultimately, this strategy highlights that there is much to be gained by locating the abstract reasoning of  economic theory into  the  historical context of a specific society.

From a political economy perspective, the article presents a framework to understand how capitalism interacts with the institutions of liberal democracy. For instance, one of the conclusions drawn from the analysis  is that there may exist a conflict between the social and economic sphere of citizenship with the profit-making capacity of the capitalist system.  The empowerment of workers can increase the outside option to employment and lead to a reduction of the rate of return  which may ultimately threaten the reproduction of capital at an increasing scale. In this respect, one of the key policy implications of the paper is that  the search for more equitable societies must consider the constraint imposed by profitability,  since a steady-state growth path cannot be sustainable  with declining rates of profit.   It is equally important  to note that   governments  play an active in protecting workers from the disruptive effects of unregulated markets---some of which are portrayed by the negative consequences that automation can have on employment and wages. The bottom line in this respect is that the balance between  more progressive societies and a more capital friendly system is, to a large extent, a political choice resulting from the particular time and place in the process of history.

The paper  sets the stage for at least four avenues of future research.  The first  is a study of the dynamic behavior of the model by introducing stochastic variables and evaluating how the economy reacts to shocks at the business cycle time scale. A  second avenue is to introduce credit in order  to understand how interest payments are determined when---consistent with business accounting---interest rates are not counted as part of costs of production but are rather paid from the surplus extracted from the process of production. This could serve the purpose of showing how ``capital costs" are separated from ``pure profits,"  and could consequently help understand the changes of capital, labor and profit shares in the past  50 years in the US; see, e.g., \citeA{karabarbounis2019accounting} and \citeA{barkai2020declining}. The model can also benefit from the introduction  of bargaining models offering a more accurate representation of the conflicting interests between capitalists and workers in wage negotiation processes.   Lastly, it is important  to introduce the government sector as an active agent so that institutional and political factors are themselves an endogenous response to the economic outcomes of the system.

\bibliographystyle{apacite}
\bibliography{references_ciep}

\newpage

\appendix
\numberwithin{equation}{section}
\renewcommand{\theequation}{\thesection\arabic{equation}}
\renewcommand{\thesubsection}{\thesection.\arabic{subsection}}

\section*{Appendix}

\section{Auxiliary Derivations}\label{appendix:AppendixA}

\subsection{Proof of Equation \eqref{eq:cap_exp_inv}}

Given the property of Poisson processes and the assumption of a large number of projects \cite[p. 5]{lucca2007resuscitating}: 

\begin{equation}\label{eq:poisson_prop}
\frac{\int_{i \in \mathcal{H}_{t}} \; x_{t}(i) di}{\pi_{I}}=\frac{\int_{i \notin \mathcal{H}_{t}} \; x_{t}(i) di}{1-\pi_{I}}=\int_{0}^{1} \; x_{t}(i)\; di \;\; \;\; \text{for }\;\; x_{t}(i)=\Big\{\mathcal{I}_{t}(i), \mathcal{I}^{\frac{\upsilon-1}{\upsilon}}_{t}(i) \Big\}.
\end{equation}

The objective of the investment   firms is to maximize $P^{I}_{t} I_{t} - P_{t} Y_{t}$ subject to \eqref{eq:tech_inv}. Using \eqref{eq:poisson_prop},  maximization problem yields  $P^{I}_{t}=P_{t} \Psi^{-1}_{t}$.  Similarly, using \eqref{eq:poisson_prop},   investment expenditure can be expressed as

\begin{equation*}
\begin{split}
\mathcal{X}_{t} &= P^{I}_{t} \int_{0}^{1} \mathcal{I}_{t}(i) \; di =  P^{I}_{t}  \Bigg[ \int_{i \in \mathcal{H}_{t-1}}  \mathcal{I}_{t}(i) \; di + \int_{i \notin \mathcal{H}_{t-1}}  \mathcal{I}_{t}(i) \; di \Bigg]\\
&=  P^{I}_{t}  \Bigg[  \int_{i \in \mathcal{H}_{t-1}}  \mathcal{I}_{t}(i) \; di + (1-\pi_{I}) \Bigg\{\int_{i \notin \mathcal{H}_{t-1}}  \mathcal{I}_{t}(i) di \Bigg\}/(1-\pi_{I})  \Bigg]\\
&=  P^{I}_{t}  \Bigg[ \int_{i \in \mathcal{H}_{t-1}}  \mathcal{I}_{t}(i) \; di + (1-\pi_{I}) \int_{0}^{1} \mathcal{I}_{t-1}  \Bigg]=  P^{I}_{t}  \Bigg[  \int_{i \in \mathcal{H}_{t-1}}  \mathcal{I}_{t}(i) \; di + (1-\pi_{I}) \frac{\mathcal{X}_{t-1}}{P^{I}_{t-1}}\Bigg]
\end{split}
\end{equation*}

A similar argument can be extended to express the amount of investment. Using \eqref{eq:tech_inv} and \eqref{eq:poisson_prop}:

\begin{equation*}
\begin{split}
I_{t} &= \Bigg(\int_{i \in \mathcal{H}_{t}} \mathcal{I}_{t}(i)^{\frac{\upsilon-1}{\upsilon}} di\Bigg)^{\frac{\upsilon}{\upsilon-1}}=\Bigg( \pi_{I} \int_{0}^{1}  \mathcal{I}_{t}(i)^{\frac{\upsilon-1}{\upsilon}} di\Bigg)^{\frac{\upsilon}{\upsilon-1}}\\
&=\Bigg[ \pi_{I}  \Bigg( \int_{i \in \mathcal{H}_{t-1}}  \mathcal{I}^{\frac{\upsilon-1}{\upsilon}}_{t}(i) \; di + \int_{i \notin \mathcal{H}_{t-1}}  \mathcal{I}^{\frac{\upsilon-1}{\upsilon}}_{t}(i) \; di \Bigg) \Bigg]^{\upsilon/(\upsilon-1)}\\
&= \Bigg[  \pi_{I}\Bigg(\int_{i \in \mathcal{H}_{t-1}}  \mathcal{I}^{\frac{\upsilon-1}{\upsilon}}_{t}(i) \; di + (1-\pi_{I}) \Bigg\{\int_{i \notin \mathcal{H}_{t-1}}  \mathcal{I}^{\frac{\upsilon-1}{\upsilon}}_{t}(i) di \Bigg\}/(1-\pi_{I})  \Bigg)\Bigg]^{\upsilon/(\upsilon-1)}\\
&= \Bigg[  \pi_{I} \int_{i \in \mathcal{H}_{t-1}}  \mathcal{I}^{\frac{\upsilon-1}{\upsilon}}_{t}(i) \; di + (1-\pi_{I}) I_{t-1}^{\frac{\upsilon-1}{\upsilon}} \Bigg]^{\upsilon/(\upsilon-1)}.
\end{split}
\end{equation*}

If  firms minimize investment expenditure subject to \eqref{eq:tech_inv}, they will all choose $\mathcal{I}_{t}(i)=\mathcal{I}_{t}$ for all $i \in \mathcal{H}_{t-1}$, so that

\begin{equation*}
\mathcal{I}_{t} = \pi_{I}^{\frac{2 \upsilon}{1-\upsilon}} \Big[I_{t}^{\frac{\upsilon-1}{\upsilon}} - (1-\pi_{I}) I_{t-1}^{\frac{\upsilon-1}{\upsilon}} \Big]^{\frac{\upsilon}{\upsilon-1}}.
\end{equation*}

Replacing the last equation on $\mathcal{X}_{t}$, it follows that

\begin{equation}
\begin{split}
\mathcal{X}_{t}&=  P^{I}_{t}  \Bigg[  \pi_{I}  \int_{0}^{1} \mathcal{I}_{t}(i) \; di + (1-\pi_{I}) \frac{\mathcal{X}_{t-1}}{P^{I}_{t-1}}\Bigg]\\
&= P^{I}_{t}  \Bigg[ \Omega(I_{t},I_{t-1}) + (1-\pi_{I}) \frac{\mathcal{X}_{t-1}}{P^{I}_{t-1}}\Bigg]
\end{split}
\end{equation}

where $\Omega(I_{t},I_{t-1}) \equiv \pi_{I}^{\frac{1+\upsilon}{1-\upsilon}} \Big[I_{t}^{\frac{\upsilon-1}{\upsilon}} - (1-\pi_{I}) I_{t-1}^{\frac{\upsilon-1}{\upsilon}} \Big]^{\frac{\upsilon}{\upsilon-1}}$.  This is precisely what is obtained from \eqref{eq:cap_exp_inv}.

\subsection{First-order Conditions of Workers and Capitalists}

The Lagrangian associated with the optimization problem of workers is

\begin{equation*}
\begin{split}
\mathcal{L}^{w}&=\sum_{t=0}^{\infty} \Big(\prod_{i=0}^{t} \beta^{w}_{i}\Big) \Bigg\{ \Big[  L_{t} \;\mathcal{U}^{we}(C^{we}_{t},h_{t}) +  U_{t}\; \mathcal{U}^{we}(C^{wu}_{t},0)\Big]+ \phi^{w}_{0,t}\Big(w_{t} N_{t} + U_{t} b_{t}- \\
& P_{t} C^{we}_{t} L_{t} -P_{t} C^{wu}_{t} U_{t} \Big)+ \phi^{w}_{1,t}\Big( (1-\lambda - U^{A}_{L_{t},t})L_{t} + f(\theta_{t})U_{t}-L_{t+1}\big) +\\  
 & \phi^{w}_{2,t}\Big(  U_{t} (1-f(\theta_{t}))   +  (\lambda + U^{A}_{L_{t},t} )L_{t} - U_{t+1} \big) \Bigg\}
\end{split}
\end{equation*}

Where $\phi^{w}_{i,t}$ are the Lagrange multipliers.  Using the  first order conditions it follows that:

\begin{equation}
\begin{split}
&C^{we}_{t}, C^{wu}_{t}:  \;\;\;   \mathcal{U}^{we}_{C^{we}_{t},t} - \phi^{w}_{0,t} P_{t} =0, \;\;   \;\;\;   \mathcal{U}^{wu}_{C^{wu}_{t},t} - \phi^{w}_{0,t} P_{t} =0\\
&L_{t}: \;  \beta^{w}_{t}\Big[ \mathcal{U}^{we}(C^{we}_{t},h_{t}) + \phi^{w}_{0,t} w_{t} h_{t}  -\phi^{w}_{0,t} P_{t} C^{we}_{t} + \phi^{w}_{1,t} (1-\hat{\lambda}_{t} ) + \phi^{w}_{2,t} \hat{\lambda}_{t}\Big] = \phi^{w}_{1,t-1}\\
&U_{t}: \;   \beta^{w}_{t}\Big[\mathcal{U}^{wu}(C^{wu}_{t},0) + \phi^{w}_{0,t} b_{t} -\phi^{w}_{0,t} P_{t} C^{wu}_{t} + \phi^{w}_{1,t} f(\theta_{t}) + \phi^{w}_{2,t}(1- f(\theta_{t}) )\Big]=\phi^{w}_{2,t-1}.
\end{split}
\end{equation}

Expressing all variables  in   consumption units by dividing by the marginal utility of consumption,  and expressing $\tilde{\beta}^{w}_{t} \equiv \beta^{w}_{t} \mathcal{U}^{w}_{C^{w}_{t},t}/ \mathcal{U}^{w}_{C^{w}_{t-1},t-1}$ and $\tilde{\phi}^{w}_{i,t}=\phi^{w}_{i,t}/\mathcal{U}^{w}_{C^{w}_{t},t}$,  the marginal value of an employed and unemployed worker satisfy 

\begin{equation*}
\Phi_{L_{t+1},t+1} = \frac{\tilde{\phi}^{w}_{1,t}}{\tilde{\beta}^{w}_{t+1}} \;\; \text{and}\;\;\; \Phi_{U_{t+1},t+1} = \frac{\tilde{\phi}^{w}_{2,t}}{\tilde{\beta}^{w}_{t+1}},
\end{equation*}

which completes the result in \eqref{eq:foc_worker}. 

 The Lagrangian associated with the capitalists optimization problem is

\begin{equation*}
\begin{split}
\mathcal{L}^{c}&= \sum_{t=0}^{\infty} {\beta^{c}}^{t}\;  \Bigg\{ \mathcal{U}^{c}_{t}(C^{c}_{t})+ \phi^{c}_{0,t}\Big( P_{t} Y_{t} - P^{k}_{t} I_{t} -w_{t} N_{t} - \kappa_{t} V_{t} -  P_{t} C^{c}_{t}-T_{t}\Big)\\
&+ \phi^{c}_{1,t}\Big( (1-\hat{\lambda}_{t} )L_{t} + q(\theta_{t})V_{t}-L_{t+1}\big)
\end{split}
\end{equation*}

As usual,  $\phi^{c}_{i,t}$ are the Lagrange multipliers. Expressing the first order conditions in consumption units using $\phi^{c}_{0,t} = \mathcal{U}^{c}_{C^{c}_{t},t}/P_{t}$, it follows that:

\begin{equation}\label{eq:foc_capitalist_app}
V_{t}: \; -\frac{\kappa_{t}}{P_{t}}  +\tilde{\phi}^{c}_{1,t} q(\theta_{t})=0; \;\;\;\;  L_{t} : \;  h_{t} Y_{N_{t},t} - \frac{w_{t}h_{t}}{P_{t}} + \tilde{\phi}^{c}_{1,t} (1-\hat{\lambda}_{t})=\frac{ \tilde{\phi}^{c}_{1,t-1}}{\tilde{\beta}^{c}_{t}}
\end{equation}

Where $\tilde{\phi}^{c}_{i,t} \equiv \phi^{c}_{i,t}/\mathcal{U}^{c}_{C^{c}_{t},t}$ and the value of an additional employed labor for the capitalist is $\Lambda_{L_{t+1},t+1}= \big(\tilde{\phi}^{c}_{1,t}/\tilde{\beta}^{c}_{t+1} \big)$.

 The shadow cost of capital in consumption units can be obtained from the  maximization problem of capital good producers. The Lagrangian in this problem can be formulated as

\begin{equation*}
\mathcal{L}^{cc}=\sum_{t=0}^{\infty} \Big(\prod_{i=0}^{t} \tilde{\beta}^{c}_{i}\Big) \Bigg\{ P^{k}_{t} I_{t}- P^{I}_{t} X_{t} - \tilde{\phi}^{c}_{3,t} \Big(  \Omega(I_{t},I_{t-1}) +(1-\pi_{I})X_{t-1} -X_{t}\Big) \Bigg\}
\end{equation*}

Using the first order conditions:

\begin{equation}\label{eq:foc_capital_app}
\begin{split}
&I_{t}:\; P^{k}_{t}=\tilde{\phi}^{c}_{3,t} \Omega_{I_{t},t}  +  \tilde{\beta}^{c}_{t+1}  \tilde{\phi}^{c}_{3,t+1} \Omega_{I_{t},t+1} \\
& X_{t}: \;  P^{I}_{t}= \tilde{\phi}^{c}_{3,t} -   \tilde{\beta}^{c}_{t+1} \tilde{\phi}^{c}_{3,t+1} (1-\pi_{I}) 
\end{split}
\end{equation}

Using the marginal productivity equation of capital, it follows that the demand for capital satisfies

\begin{equation}\label{eq:demand_capita_euler}
Y_{K_{t}} = \frac{\delta P^{k}_{t}}{P^{c}_{t}} = \delta P^{I}_{t} \Omega_{I_{t},t} + \delta \Big[\Omega_{I_{t},t} + \frac{\Omega_{I_{t},t+1}}{1-\pi_{I}} \Big] \sum_{i=1}^{\infty} \Big(\prod_{j=1}^{i} \tilde{\beta}^{c}_{t+j}\Big) (1-\pi_{I})^{i} P^{I}_{t+i}.
\end{equation}

An important implication of \eqref{eq:demand_capita_euler} is that, because $P^{I}_{t}=(1+\mu_{t})P^{c}_{t} \Psi_{t}^{-1}$, capital demand is a positive function of current and future rates of return.  If, for instance, capitalists receive news that in the future labor institutions will provide further support to workers, it can be expected that they will reduce the current demand for capital.

\section{Accounting Structure of the Model}\label{appendix:accounting_app}

The analysis in the text is carried out on the basis of an accounting structure where capitalists finance all capital investments and labor costs using retained earnings. This  assumption is made in order to highlight the nature of profits as a surplus before complicating the model with the introduction of interest payments and rents. Essentially, given that in the Classical tradition  interests and rents are paid from the aggregate surplus of society in the circulation process of capital, the first logical step is to understand how  profits are reproduced  before introducing exogenous sources of liquidity to  the model. 

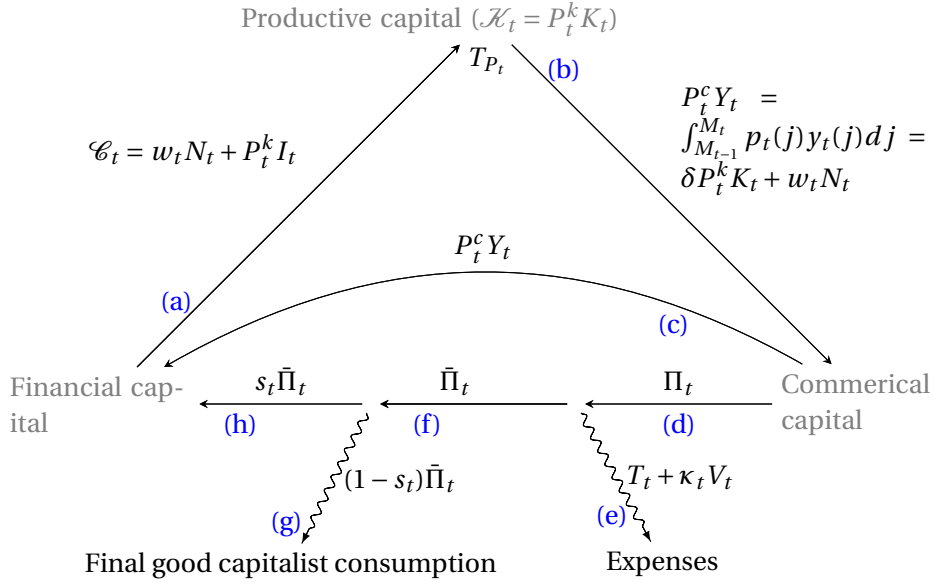
\begin{figure}
\begin{center}
\begin{tikzpicture} [node distance = 7.25 cm, on grid]
\node (q0) [gray,text width=2.4cm] {Financial capital};
\node (q1) [above right = of q0,gray, text width=6.5 cm] {Productive capital $(\mathcal{K}_{t}=P^{k}_{t}K_{t})$};
\node (q2) [below right = of q1,gray,text width=2.4cm] {Commerical capital};
\node (q5) [below  = of q0,gray] {};
\node (q6) [below  = of q2,gray] {};
\path (q0) -- (q2) node[pos=0.3,gray] (q4) {};
\path (q4) -- (q5) node[pos=0.29,black] (q3) {Final good capitalist consumption};
\path (q4) -- (q2) node[pos=0.5,gray] (q9) {};
\path (q9) -- (q6) node[pos=0.285,black] (q8) {Expenses};
\path [-stealth,black]
    (q0) edge node [left,pos=0.67,text width=3.4cm] {$\mathcal{C}_{t}=w_{t} N_{t}+P^{k}_{t} I_{t}$} (q1)
    (q1) edge  node[above right,pos=0.5, text width=4.3cm] {$P^{c}_{t}Y_{t}=\int_{M_{t-1}}^{M_{t}}p_{t}(j) y_{t}(j) dj=\delta P^{k}_{t} K_{t}+w_{t} N_{t}$} (q2)
       (q1) edge  node[right,pos=0.08,blue] {(b)}  (q2)
            (q2) edge   [bend right] node[below,pos=0.2,blue] {(c)}  (q0)
            (q2) edge  node[above,pos=0.5,black] {$\Pi_{t}$}  (q9)
             (q9) edge  node[below,pos=0.2,blue]{}  (q4)
        (q9) edge  node[above,pos=0.6] {$\bar{\Pi}_{t}$}  (q4)
     (q0) edge  node[left,pos=0.2,blue] {(a)}  (q1)
       (q0) edge  node[right,pos=0.95,black] {\;\;$ T_{P_{t}} $}  (q1)
          (q2) edge  node[below,pos=0.5,blue] {(d)}  (q9)
       (q4) edge  node[below,pos=0.75,blue] {(h)}  (q0)
        (q9) edge  node[below,pos=0.75,blue] {(f)}  (q4)
      (q4) edge  node[above] {$s_{t} \bar{\Pi}_{t}$}  (q0)
    (q2) edge [bend right] node[above] {$P^{c}_{t}Y_{t}$} (q0);
    \draw [->,snake=snake,segment amplitude=.5mm,
         segment length=2mm,
         line after snake=2mm,black] (q4) -- (q3)
    node[ right, midway,black]{$(1-s_{t}) \bar{\Pi}_{t}$}
    (q4) -- (q3)
    node[ left, pos=0.85,blue]{(g)};
        \draw [->,snake=snake,segment amplitude=.5mm,
         segment length=2mm,
         line after snake=2mm,black] (q9) -- (q8)
           node[ right, midway,black]{$T_{t} + \kappa_{t} V_{t}$}
    (q9) -- (q8)
        node[ left, pos=0.8,blue]{(e)};
\end{tikzpicture}
\vspace*{-45 mm}
\caption{Reproduction process of capital. \emph{Notes---} The  transitions of the flows of capital are described with  blue letters in parenthesis and the different forms of capital are represented in gray. \label{fig:circ_capital}}
\end{center}
\end{figure}

The underlying accounting structure of the model is well represented in Figure \ref{fig:circ_capital} using Marx's circuits of capital as formalized by \citeA{foley1986understanding}. At stage (a), the chain of intermediate and final good producers use a share of their money funds to hire labor services from workers and pay for new units of capital to capital good producers. 

The aggregate flow of capital outlays is committed to a production process which results in a flow value of unsold finished output ($P^{c}_{t} Y_{t}$).  As noted by numerous authors (e.g., \citeNP{haavelmo1960}),   not all components of capital outlays are transferred as   finished output after the same length of time in the production process, so the relation between $\mathcal{C}_{t}=w_{t}N_{t} + P^{k}_{t} I_{t} $ and $P^{c}_{t} Y_{t}$ can be accounted for by a convolution \cite{foley1982realization}:

\begin{equation}\label{eq:conserv_p1}
\sum_{t'=-\infty}^{t} \mathcal{C}_{t'} \mathcal{A}(t-t';t')=P^{c}_{t} Y_{t},
\end{equation}

where $\mathcal{A}(t-t';t') \geq 0$ for $ t-t' \geq 0$ represents the distributed shares in the value of capital outlays and $\sum_{t'=-\infty}^{t} \mathcal{A}(t-t';t')=1$. Intuitively, \eqref{eq:conserv_p1} says that finished goods   valued at costs (since it is unsold output) is equal to the weighted sum of the value of all previous capital outlays committed to the process of production.

To simplify the mathematical analysis, I  assume that each  unit of capital  stays in the production process for a time period $T_{P_{t}}$ and then emerges all at  once as a finished good. This is represented, similar to \citeA[p. 70]{foley1986understanding}, as

\begin{equation*}
P^{c}_{t} Y_{t} = \mathcal{C}_{t-T_{P_{t}}}.
\end{equation*}

That is, the value of the final good  measured at cost prices must be equal to the value of capital outlays entering the production process in period $t-T_{P_{t}}$. Given Assumption \ref{ass:steady_state_det}, the current value of capital outlays is discounted exponentially with a discount rate  $g_{t}$, denoting  the growth rate of the value of the capital stock,  such that

\begin{equation}\label{eq:prod_time}
P^{c}_{t} Y_{t} = e^{-g_{t} T_{P_{t}}} \mathcal{C}_{t}= \mathcal{C}_{t-T_{P_{t}}}.
\end{equation}

  The value of the final good valued at costs $(P^{c}_{t} Y_{t})$  subtracts from the value of the stock of productive capital in the transition from (a) to (b). The value of sales results from the sum of transition flows (c) and (d), such that   $P_{t}Y_{t}=P^{c}_{t} Y_{t} +\Pi_{t}$. In the transition from (d) to  (f), capitalists pay  taxes and vacancy expenses out of their realized profits. From this, final good producers decide   how much to retain in the circuit of capital with a value of $s_{t} \bar{\Pi}_{t}$, $s_{t}$ being the rate of savings, and how much to consume, with a value equal to $(1-s_{t})\bar{\Pi}_{t}$.

The two main assumption in the description of the circuit of capital in Figure \ref{fig:circ_capital} are that: (i)  commodities are sold immediately after they are produced; and (ii) that there is a zero time-lag between the moment the final good is sold and the moment firms demand new capital outlays. Formally, assumption (i) implies that there is no variation in commercial capital (i.e., there are no variations in final good inventories). Similarly, assumption (ii) states that the variations of financial capital ($\mathcal{K}^{F}_{t}$) are equal to zero, that is

\begin{equation}\label{eq:equil_finc_capital}
\frac{d \mathcal{K}^{F}_{t}}{dt} = P^{c}_{t} Y_{t} + s_{t} \tilde{\Pi}_{t} -  \mathcal{C}_{t} =0.
\end{equation}

This last equation states that savings $(s_{t} \tilde{\Pi}_{t})$ are equal to net investment $(P^{k}_{t} (I_{t}-\delta K_{t}))$. This relation is also used to derive the financial constraint of capitalist households \eqref{eq:cap_finac_constraint}. Using flows (f), (g) and (h) in Figure \ref{fig:circ_capital}, it follows that 

\begin{equation*}
\begin{split}
(1-s_{t}) \bar{\Pi}_{t} &= P_{t} C^{c}_{t} = \tilde{\Pi}_{t}-s_{t}\tilde{\Pi}_{t}\\
&= \tilde{\Pi}_{t}- \mathcal{C}_{t}  +  P^{c}_{t} Y_{t} \\
&= \Pi_{t} - T_{t} - \kappa_{t} V_{t} - \mathcal{C}_{t}  +  P^{c}_{t} Y_{t} \\
&= P_{t} Y_{t} - P^{c}_{t} Y_{t}  - T_{t} - \kappa_{t} V_{t} - \mathcal{C}_{t}  +  P^{c}_{t} Y_{t} \\
&= P_{t} Y_{t}- T_{t} - \kappa_{t} V_{t}- w_{t} N_{t} - P^{k}_{t} I_{t}.
\end{split}
\end{equation*}

Using the financial constraint of workers and the definition of profits of capital good producers in \eqref{eq:cap_prod_problem}, we can derive aggregate consumption in \eqref{eq:ag_consumption} and the aggregate resource constraint as

\begin{equation*}
P_{t} C_{t} = P_{t} C^{w}_{t}  + P_{t} C^{c}_{t}  + \Pi^{c}_{t}=P_{t}Y_{t} - \kappa_{t} V_{t} -P^{I}_{t} X_{t}.
\end{equation*}

Ultimately, though I preferred to omit all discussions about the circuit of capital in the main text, it is clear that all relevant accounting relations and financial constraints are obtained from Figure \ref{fig:circ_capital}.

\section{Aggregate Production Function and the Accounting Identity}\label{appendix:ces}

The aggregate production function in the main text can be written as

\begin{equation*}
Y_{t} = B\Big[(1-m_{t})^{1/\sigma} \Big(\Gamma^{K}_{t} K_{t}\Big)^{\frac{\sigma-1}{\sigma}} + \Big( \int_{0}^{m_{t}}{\Gamma^{N}_{t}}^{\sigma-1}(j) dj\Big)^{1/\sigma}\;  \Big(\Gamma^{N}_{t}(J^*_{t}) N_{t}\Big)^{\frac{\sigma-1}{\sigma}} \Big]^{\frac{\sigma}{\sigma-1}}
\end{equation*}

To simplify the mathematics I will assume that $\Gamma^{K}_{t}=\Psi_{t}=1$ and work with the fact that $ \int_{0}^{m_{t}}{\Gamma^{N}_{t}}^{\sigma-1}(j) dj \approx m_{t}$, so given that $\Gamma^{N}_{t}(j)=e^{\alpha j}$, then:

\begin{equation}\label{eq:ces_app}
y_{t} = B \Big[(1-m_{t})^{1/\sigma}  k_{t}^{\frac{\sigma-1}{\sigma}} + {m_{t}}^{1/\sigma}\; N_{t}^{\frac{\sigma-1}{\sigma}} \Big]^{\frac{\sigma}{\sigma-1}}
\end{equation}

Where $k_{t} \equiv K_{t} e^{-\alpha J^*_{t}}$ and $y_{t} \equiv Y_{t}  e^{-\alpha J^*_{t}}$. Using the properties of the aggregate production function, we know that $Y_{t} = \delta P^{k}_{t} K_{t} + w_{t} N_{t}=$ and $y_{t} = \delta P^{k}_{t} k_{t} + \hat{w}_{t} N_{t}$, where $\hat{w}_{t} = w_{t} e^{-\alpha J^{*}_{t}}$.  Working in continuous time (for simplicity), denoting $\dot{x} \equiv dx/dt$ and $g_{x} \equiv \dot{x}/x$, then:

\begin{equation*}
g_{y} =(1-\Omega^{c}_{t})g_{P^{k}} + \Omega^{c}_{t} g_{\hat{w}} + (1-\Omega^{c}_{t})g_{k} + \Omega^{c}_{t} g_{N} = \psi_{t} +  (1-\Omega^{c}_{t})g_{k} + \Omega^{c}_{t} g_{N}.
\end{equation*}

Here $\Omega^{c} = wN/Y$ is the labor share on \emph{costs} and $(1-\Omega^{c})=\delta P^{k}K/Y$. We are interested in finding some functional form that will track the behavior of $\Omega^{c}_{t}$ such that it will always provide a good prediction of $y_{t}$ at any moment in time. Our purpose is to show that the CES production function \eqref{eq:ces_app} solves this problem. 

Start with the partial derivative of $y_{t}$ in \eqref{eq:ces_app} with respect to time, from which it follows that

\begin{equation}\label{eq:ces_growth_app}
g_{y} = \frac{\dot{m}_{t}}{(1-\sigma)D_{t}} \Big[(1-m_{t})^{\frac{1-\sigma}{\sigma}}k_{t}^{\frac{\sigma-1}{\sigma}} - m_{t}^{\frac{1-\sigma}{\sigma}}N_{t}^{\frac{\sigma-1}{\sigma}}\Big] + \frac{(1-m_{t})^{1/\sigma} k_{t}^{\frac{\sigma-1}{\sigma}}}{D_{t}} g_{k} + \frac{m_{t}^{1/\sigma} N_{t}^{\frac{\sigma-1}{\sigma}}}{D_{t}} g_{N}.
\end{equation}

Where $D_{t} \equiv \Big[(1-m_{t})^{1/\sigma}  k_{t}^{\frac{\sigma-1}{\sigma}} + {m_{t}}^{1/\sigma}\; N_{t}^{\frac{\sigma-1}{\sigma}} \Big]$. Using the marginal productivity conditions in \eqref{eq:marg_prods} (and the assumption that $\Gamma^{K}=1$):

\begin{equation*}
\begin{split}
\delta P^{k}_{t} &= B^{\frac{\sigma-1}{\sigma}} (1-m_{t})^{1/\sigma} (Y_{t}/K_{t})^{1/\sigma}\\
w_{t}  & \approx m_{t}^{1/\sigma} B^{\frac{\sigma-1}{\sigma}} e^{\alpha(\sigma-1)J^*_{t}/\sigma} (Y_{t}/N_{t})^{1/\sigma},
\end{split}
\end{equation*}

it implies that $\Omega^{c}_{t}  \equiv w_{t}N_{t}/Y_{t} \approx  m_{t}^{1/\sigma} N^{\frac{\sigma-1}{\sigma}}_{t}/D_{t}$ and $1-\Omega^{c}_{t}  \equiv \delta P^{k}_{t}K_{t}/Y_{t} = (1-m_{t})^{1/\sigma} k^{\frac{\sigma-1}{\sigma}}_{t}/D_{t}$. Replacing these equations in \eqref{eq:ces_growth_app}, we get:

\begin{equation}\label{eq:ces_growth_app}
g_{y} = (1-\Omega^{c}_{t})g_{k} + \Omega^{c}_{t} g_{N}
\end{equation}

That is, the aggregate production function will hold exactly if---\emph{after correcting for the time trend of labor productivity}---$\psi_{t} \approx 0$, which is what is expected on average out of the stationary rates of growth of wages and capital prices. 

This close connection with the accounting identity can be used to reveal  the possible illusion of  forming \emph{causal} interpretations from  marginal productivity equations.  For example, one may hope to test the validity of marginal productivity equations by using $\Omega^{c}_{t}  \approx  m_{t}^{1/\sigma} N^{\frac{\sigma-1}{\sigma}}_{t}/D_{t} $ and estimating

\begin{equation*}
g_{\Omega^{c}} = \frac{1}{\sigma} g_{m} - \frac{\sigma-1}{\sigma}(1-\Omega^{c}_{t})\Big[g_{k} - g_{N}\Big],
\end{equation*}

not knowing that this equation must necessarily hold if $\psi_{t} \approx 0$. The equation above presents two alternatives to interpret a fall in the wage share.  First, as noted by \citeA[p. 85]{felipe2013aggregate}, it may decline if capital and labor are gross substitutes and $K_{t}$ grows faster than $N_{t}$. This is the approach used by  \citeA{piketty2014capital} and others to ``explain" the fall in the wage share. Second, the share of labor may fall if $m_{t}$ declines---regardless of the value of $\sigma$. This is one of the innovations of the production function with automation, since it presents an alternative that can account for the falling wage share without assuming gross substitution between labor and capital.

One of the problems here is that $g_{\Omega^{c}}$ may decrease as a result of, say, a lower bargaining power of workers, but by interpreting this fall using marginal productivity equations one may wrongly associate it with, say, increasing automation. Furthermore, one may confuse the direction of causality by thinking that changes in the distribution of income are caused by the components of the production function when these may go the other way: the fall in the wage share may explain the changes in the parameters of the production function.

The bottom line here is that it is generally questionable to assign claims of causality to aggregate production functions when these may just be representing an accounting relation. The model in the paper partly addresses this issue in two ways. First, by creating a relation between the aggregate production function and the aggregate costs of production, rather than aggregate income,  the model is capable of introducing institutional and political factors as determinants of the labor share. Secondly, by restricting $\sigma<1$, the model can identify changes in technology with automation if these are also associated with increasing unemployment, lower steady-state real wages, and a higher capital-output ratio.

\section{ Proofs}

\subsection{Proof of Theorem \ref{theorem:equil_harrod} }\label{appendix:proof_theorem}

\textbf{Part A.} Let the stationary values be defined   $\hat{Y}_{t}=Y_{t} e^{-\alpha J^{*}_{t}}$,  $\hat{C}^{c}_{t}=C^{c}_{t} e^{-\alpha J^{*}_{t}}$, $\hat{C}^{w}_{t}=C^{w}_{t} e^{-\alpha J^{*}_{t}}$, $\hat{w}_{t}=w_{t} e^{-\alpha J^{*}_{t}}$, $\hat{K}_{t} = K_{t}  e^{-\alpha J^{*}_{t}} \Psi^{-1}_{t}$, $\hat{I}_{t} = I_{t}  e^{-\alpha J^{*}_{t}} \Psi^{-1}_{t}$, $\hat{P}^{I}_{t}=P^{I}_{t}  \Psi_{t}$, $\hat{P}^{k}_{t}=P^{k}_{t}  \Psi_{t}$, $\hat{\mathcal{X}}_{t}=\mathcal{X}_{t} e^{-\alpha J^{*}_{t}}$,  and $\hat{\mathcal{K}}_{t}=\mathcal{K}_{t} e^{-\alpha J^{*}_{t}}$. Given the accounting structure defined in online Appendix \ref{appendix:accounting_app}, the evolution of capital can be described as

\begin{equation*}
\mathcal{K}_{t+1} = \mathcal{K}_{t} + s_{t}(r_{t}-\tau_{t}-\zeta_{t})\mathcal{K}_{t}.
\end{equation*}

Where $r_{t} \equiv  \Pi_{t}/\mathcal{K}_{t}$ is the rate of profit,  $\tau_{t} \equiv T_{t}/\mathcal{K}_{t}$ is the  share of taxes on capital value,  $\zeta_{t} = \kappa_{t} V_{t}/\mathcal{K}_{t}$ is the share of vacancy costs to capital, and $\mathcal{K}_{t} = P^{k}_{t} K_{t}$ is the current value of the capital stock.  Defining $\mathcal{K}_{t}=\hat{\mathcal{K}} e^{\alpha J^{*}_{t}}$  and denoting the current rate of growth of productive capital as $g_{t} = (\mathcal{K}_{t+1}-\mathcal{K}_{t})/\mathcal{K}_{t}$, then

\begin{equation*}
 e^{\alpha (J^{*}_{t+1}-J^{*}_{t})}= 1 + g_{t}.
\end{equation*}

Given Assumption \ref{ass:steady_state_det} (ii), it follows that $g_{t} \approx g$. Now, the existence of an equilibrium where $g_{t} \rightarrow g$ depends on the existence and stability of the equilibrium in the labor market.  This equilibrium is guaranteed if $b_{1}$ is sufficiently small. To verify this claim we can evaluate the partial derivatives of $(1+\mu^{S})$ and $\mu^{D}$ near the equilibrium and check the conditions by which $\partial \mu^{D}/\partial \theta >0$ and $\partial (1+\mu^{S})/\partial \theta <0$.

Starting with $\partial (1+\mu^{S})/\partial \theta <0$, it is useful to note that $\hat{Z} = \hat{b} + (\epsilon_{1}/(1+\epsilon_{1}) h \hat{Y}_{N}$ and $\hat{b} = b_{0}^{\beta} + b_{1}^{\beta} (1-L)$. Thus, the Nash solution in steady-state can be written as

\begin{equation*}
(1+\mu^{S})= \frac{h \hat{Y}_{N}}{ (1-\eta_{w})\big( b_{0}^{\beta} + b_{1}^{\beta} (1-L) \big) + (\epsilon_{1} + \eta_{w})/(1+\epsilon_{1}) \;  h \hat{Y}_{N} + \eta_{w} \hat{\kappa} \Big[ (1-\tilde{\beta})(1-\hat{\lambda})q(\theta)^{-1} + \tilde{\beta} \theta\Big] }
\end{equation*}

where $\tilde{\beta} \equiv \beta^{w}/\beta^{c}$. Expressing $\bar{\omega} = (1-\eta_{w})\big( b_{0}^{\beta} + b_{1}^{\beta} (1-L) \big) + (\epsilon_{1} + \eta_{w})/(1+\epsilon_{1}) \;  h \hat{Y}_{N} + \eta_{w} \hat{\kappa} \Big[ (1-\tilde{\beta})(1-\hat{\lambda})q(\theta)^{-1} + \tilde{\beta} \theta\Big] $, it follows that

\begin{equation}\label{eq:derivative_nash}
\frac{\partial (1+\mu^{S})}{\partial \theta} = \frac{h_{\theta} \hat{Y}_{N}}{\bar{\omega}} + \frac{h \hat{Y}_{N \theta}}{\bar{\omega}} - \frac{h \hat{Y}_{N}}{\bar{\omega}^2} \Big[  \hat{\kappa} \eta_{w} \big(\tilde{\beta} - (1-\tilde{\beta})(1-\hat{\lambda}) \frac{ q'(\theta)}{q(\theta)^{2} } \big)+ \frac{\epsilon_{1}+\eta_{w}}{1+\epsilon_{1}} \big( h_{\theta} \hat{Y}_{N} + h \hat{Y}_{N \theta}\big) - (1-\eta_{w}) b_{1}^{\beta} L_{\theta} \Big] 
\end{equation}

If $b^{\beta}_{1} =0$, we would get an equation close to the usual Nash solution, which is known to have a negative slope. However, if  $b^{\beta}_{1}  >0$, we can have $\frac{\partial (1+\mu^{S})}{\partial \theta}  >0$, specially for small values of $\theta$ given that  $L_{\theta}$ increases as $\theta \rightarrow 0$.  This explains why the slope of $\mu^{S}$ is initially  positive in Figure \ref{fig:labor_market_comp_stat}. 

Correspondingly, the partial derivative of $\mu^{D}$ with respect to $\theta$ is 

\begin{equation*}
\frac{\partial \mu^{D}}{\partial \theta} = -{\mu^{D}}^{2} \Big[ \frac{h_{\theta} \hat{Y}_{N} q(\theta) }{\hat{\lambda} \hat{\kappa}} + \frac{h\hat{Y}_{N \theta}}{\hat{\lambda} \hat{\kappa}} + \frac{h \hat{Y}_{N}q'(\theta)}{\hat{\lambda} \hat{\kappa}} \Big],
\end{equation*}

which is  positive given the assumption of decreasing marginal returns and $q'(\theta) <0$. Thus, a stable equilibrium in the labor market can be obtained if $b^{\beta}_{1}$ is sufficiently low.

\subsection*{Part B}

\textbf{Marginal productivity of capital.}  Using \eqref{eq:marg_prods}, we know that the marginal productivity of capital satisfies $Y_{K_{t}} = \delta P^{k}_{t}/P^{c}_{t}$. Expressing this equation in its stationary form it follows that $\hat{Y}_{K_{t}} = Y_{K_{t}} \Psi_{t} = \delta P^{k}_{t} \Psi_{t}/P^{c}_{t}$. Now, given the first order conditions of capitalists in  Appendix \ref{appendix:AppendixA}:

\begin{equation*}
\hat{Y}_{K_{t}} = \frac{\delta \hat{P}^{k}_{t}}{P^{c}_{t}} = \delta (1+\mu_{t}) \Omega_{I_{t},t} + \delta \Big[\Omega_{I_{t},t} + \frac{\Omega_{I_{t},t+1}}{1-\pi_{I}} \Big] \sum_{i=1}^{\infty}  (1-\pi_{I})^{i} e^{-i\bar{z}}(1+ \mu_{t+i}).
\end{equation*}

where I assume (without any loss of generality) that $\beta^{c}=1$. In the limit, since 

\begin{equation*}
\begin{split}
\Omega_{I_{t},t} \rightarrow \Omega_{I} = \pi^{\frac{1+\upsilon}{1-\upsilon}}_{I} \Big[1 -(1-\pi_{I}) e^{-\frac{\bar{z}(\upsilon-1)}{\upsilon}}\Big]^{\frac{1}{\upsilon-1}}\\
\Omega_{I_{t},t+1} \rightarrow \Omega^{+}_{I} = -(1-\pi_{I}) \pi^{\frac{1+\upsilon}{1-\upsilon}}_{I} \Big[1 -(1-\pi_{I}) e^{-\frac{\bar{z}(\upsilon-1)}{\upsilon}}\Big]^{\frac{1}{\upsilon-1}}e^{\bar{z}/\upsilon},
\end{split}
\end{equation*}

it follows that 

\begin{equation}
\hat{Y}_{K_{t}} = \frac{\delta \Omega_{I} (1+\mu)}{1-(1-\pi_{I})e^{-\bar{z}}} \Big[1-(1-\pi_{I})e^{-\bar{z}(\upsilon-1))/\upsilon}\Big]
\end{equation}

which is the result in \eqref{subeq:steady_marg_prod_cap}. 

\textbf{Rate of profit.} The rate of profit, unlike the rate of return of capital, is the ratio of a flow over a stock. By definition, $r_{t} \equiv \Pi_{t}/(P^{k}_{t}K_{t})$. Making use of equation \eqref{eq:prod_time} in Appendix \ref{appendix:accounting_app}: 

\begin{equation*}
\begin{split}
g_{t} &= \frac{(w_{t} N_{t} + P^{k}_{t} I_{t})\big(1- e^{g_{t} T_{P_{t}}}\big)}{\mathcal{K}_{t} }= \frac{\Big[s_{t}\big(	\Pi_{t} - T_{t} - \kappa_{t} V_{t}\big) + P^{c}_{t} Y_{t} \Big] \big(1- e^{g_{t} T_{P_{t}}}\big)}{\mathcal{K}_{t} }\\
g_{t}&=\Big[s_{t}(r_{t}-\tau_{t}-\zeta_{t}) + \frac{P^{c}_{t}Y_{t}}{\mathcal{K}_{t}}\Big](1-e^{-g_{t}T_{P_{t}}})\\
g_{t}& = \Big[s_{t}(r_{t}-\tau_{t}-\zeta_{t} ) + \frac{r_{t}}{\mu_{t}}\Big](1-e^{-g_{t}T_{P_{t}}}).
\end{split}
\end{equation*}

Where $\tau_{t} \equiv T_{t}/\mathcal{K}_{t}$ is the  share of taxes on capital value,   $\zeta_{t} = \kappa_{t} V_{t}/\mathcal{K}_{t}$ is the share of vacancy costs to capital, and $T_{P_{t}}$ is the average production time of the final good defined in Appendix \ref{appendix:accounting_app}. Given that in equilibrium $s_{t}(r_{t}-\tau_{t}-\zeta_{t})=g_{t}$, it follows that

\begin{equation}\label{eq:prof_rate_approx}
\begin{split}
e^{g_{t}T_{P_{t}}} &= 1+ \frac{g_{t}\mu_{t}}{r_{t}}\\
\Rightarrow r_{t} &\approx \frac{\mu_{t}}{T_{P_{t}}}.
\end{split}
\end{equation}

Given Assumption \ref{ass:steady_state_det} (ii) and the assumption that the final good is produced after an average time lag $T_{P_{t}}$, it must hold that $P^{k}_{t-T_{P_{t}}} I_{t-T_{P_{t}}}  + w_{t-T_{P_{t}}}  N_{t-T_{P_{t}}}  = e^{-g_{t} T_{P_{t}}} \big[ P^{k}_{t}I_{t} +w_{t} N_{t}\big] = P^{c}_{t} Y_{t} = \delta P^{k}_{t} K_{t} + w_{t} N_{t}$, such that

\begin{equation*}
T_{P_{t}} = \frac{1}{g_{t}} \mathrm{log} \Bigg(\frac{P^{k}_{t}I_{t} +w_{t} N_{t}}{\delta P^{k}_{t} K_{t} + w_{t} N_{t}}\Bigg) = \frac{1}{g_{t}} \mathrm{log} \Bigg(\frac{\hat{P}^{k}_{t}\hat{I}_{t} +\hat{w}_{t} N_{t}}{\delta \hat{P}^{k}_{t} \hat{K}_{t} + \hat{w}_{t} N_{t}}\Bigg) 
\end{equation*}

Now, since $\hat{I}_{t} \approx (\delta + g_{t}) \hat{K}_{t}$, the last equation becomes

\begin{equation*}
T_{P_{t}} = \frac{1}{g_{t}} \; \mathrm{log}\Big(1 + g_{t} \frac{1}{\delta + (\hat{w}_{t} N_{t}/\hat{P}^{k}_{t}\hat{K}_{t})}\Big).
\end{equation*}

Using equation \eqref{eq:labor_share_inc}, $(\hat{w}_{t}N_{t}/\hat{P}^{k}_{t}\hat{K}_{t})= (\hat{w}_{t}N_{t}/P_{t}\hat{Y}_{t}) \times (P_{t}\hat{Y}_{t}/P^{k}_{t}\hat{K}_{t})$ and  $(P
_{t}\hat{Y}_{t}/P^{k}_{t}\hat{K}_{t})=(1+\mu_{t})\delta/(1-(\hat{w}_{t}N_{t}/P^{c}_{t}\hat{Y}_{t}))$, such that $(\hat{w}_{t}N_{t}/\hat{P}^{k}_{t}\hat{K}_{t})=\delta \Omega^{c}_{t}/(1-\Omega^{c}_{t})$ and 

\begin{equation}\label{eq:stat_prod_time_app}
T_{P_{t}} \rightarrow  \frac{1}{g} \; \mathrm{log} \Big(1+ g \frac{1-\Omega^{c}}{\delta}\Big) \approx \frac{1-\Omega^{c}}{\delta}.
\end{equation}

Replacing \eqref{eq:stat_prod_time_app} in \eqref{eq:prof_rate_approx}, it follows that

\begin{equation}
r_{t} \rightarrow \frac{\delta \mu}{1-\Omega^{c}}
\end{equation}

which is the result in \eqref{subeq:steady_profit_rate}.

\textbf{Investment-output ratio} The function $\Omega_{t}$ can be written as $\Omega_{t} = \pi^{\frac{1+\upsilon}{1-\upsilon}}_{I} I_{t} \Big[ 1- (1-\pi_{I})(\frac{\hat{I}_{t-1}}{\hat{I}_{t}})^{\frac{\upsilon -1}{\upsilon}} e^{-\bar{z}(\upsilon-1)/\upsilon}\Big]^{\frac{\upsilon}{\upsilon-1}}$. Given that $X_{t}=\hat{X}_{t} \Psi_{t} e^{\alpha J^{*}_{t}}$, in equilibrium it holds that

\begin{equation*}
\hat{X} = \frac{\hat{I} \pi^{\frac{1+\upsilon}{1-\upsilon}}_{I}  \Big[ 1- (1-\pi_{I})e^{-\bar{z}(\upsilon-1)/\upsilon}\Big]^{\frac{\upsilon}{\upsilon-1}}}{1-(1-\pi_{I})e^{-\bar{z}}}
\end{equation*} 

Using $\hat{I} \approx (\delta +g) \hat{K}$ and $\hat{P}^{I}_{t}=P^{I}_{t} \Psi_{t}$: 

\begin{equation*}
\frac{\hat{X}}{\hat{Y}} = \frac{(\delta +g) \hat{K}\pi^{\frac{1+\upsilon}{1-\upsilon}}_{I}  \Big[ 1- (1-\pi_{I})e^{-\bar{z}(\upsilon-1)/\upsilon}\Big]^{\frac{\upsilon}{\upsilon-1}}}{\hat{Y}\big(1-(1-\pi_{I})e^{-\bar{z}}\big)}.
\end{equation*}

Solving $\hat{K}/\hat{Y}$ from \eqref{eq:marg_prods} we obtain \eqref{subeq:steady_inv_output}.

\subsection{Proof of  Lemma \ref{lemma:lemma_2A_AR}}\label{appendix:lemma}

This part of the proof follows from Lemma A2 of \citeA{acemoglu2018race}. The main difference here is that the automation measure function depends on the rate of return of capital and is bounded in regions strictly inside $(0,1)$. 

We can begin by noting that  at the boundary of region 2 in Figure \ref{fig:automation_reg}, it must be true that $\delta \hat{P}^{k}_{t}/\gamma_{k} = w_{t} e^{-\alpha J^{*}_{t}} = \hat{w}_{t}$. Using this condition in the ideal price index, it follows that:

\begin{equation*}
1 = \Bigg(\frac{\delta \hat{P}^{k}_{t}}{\gamma_{k}}\Bigg)^{1-\sigma} \; \Bigg[1 -\bar{m}(\mu_{t}) + \frac{e^{\alpha(\sigma-1)\bar{m}(\mu_{t})}-1}{\alpha(\sigma-1)}\Bigg]
\end{equation*} 

The sign of $\bar{m}(\mu_{t})$ can be deduced using a Taylor expansion on $\mathrm{exp}(x) = 1 + x +x^2/2 + \mathcal{O}(x^3)$, with $x = \alpha(\sigma-1)\bar{m}(\mu_{t})$, such that\footnote{Using the calibration in Table \ref{table:calibration_steady_state1}, the errors of this approximation are at most 5$\%$.}

\begin{equation*}
\bar{m}(\mu_{t})   \approx \sqrt{\frac{2}{\alpha(1-\sigma)} \Bigg( 1 - \Big(\frac{\delta \hat{P}^{k}_{t}}{\gamma_{k}}\Big)^{\sigma-1}\Bigg)} = \sqrt{\frac{2}{\alpha(1-\sigma)} \Bigg( 1 - \Big(\frac{\hat{Y}_{k_{t}}}{\gamma_{k}}\Big)^{\sigma-1}\Bigg)}
\end{equation*}

The assumption that $\gamma_{k} = \delta P^{k}_{t}(0)$ guarantees that $\bar{m}(\mu_{t}) >0$ since $\gamma_{k} < \delta P^{k}_{t}(\mu_{t})$ for all $\mu_{t} >0$. Furthermore, given that $\hat{Y}_{k}$ is an increasing function of $\mu$, then $\bar{m}(\mu)$ is also an increasing function of the rate of return.

\subsection{Proof of Proposition \ref{prop:comp_stat}}\label{appendix:prop}

\subsubsection{Automation} 

\textbf{Wages.} The first part is analogous to  the proof of Lemma A2 of \citeA{acemoglu2018race}. Defining $\hat{w}_{t} = w_{t} e^{-\alpha J^{*}_{t}}$ and using the ideal price index condition

\begin{equation*}
\hat{w}^{1-\sigma}_{t} = \frac{B^{1-\sigma} -(1-m^*_{t})\Big(\delta P^{k}_{t}/\Gamma^{k}_{t}\Big)^{1-\sigma}}{\int_{0}^{m^{*}_{t}} \Gamma^{N}_{t}(j) dj}.
\end{equation*}

By implicit differentiation, it follows that

\begin{equation*}
\hat{w}'_{t}/\hat{w}_{t} = \frac{1}{1-\sigma} \Bigg( \frac{\Big(\delta P^{k}_{t}/\Gamma^{k}_{t}\Big)^{1-\sigma}}{\Big(\int_{0}^{m^{*}_{t}} \Gamma^{N}_{t}(j) dj\Big)\hat{w}^{1-\sigma}_{t}} - \frac{e^{\alpha(\sigma-1)m^{*}_{t}}}{\int_{0}^{m^{*}_{t}} \Gamma^{N}_{t}(j) dj} \Bigg)
\end{equation*}

Using  $\tilde{w}_{t} = w_{t} e^{-\alpha M_{t}}$ and  $\int_{0}^{m^{*}_{t}}  \tilde{w}^{1-\sigma}_{t}  {\Gamma^{N}_{t}}^{(1-\sigma)} (j) dj = \hat{w}^{1-\sigma}_{t} \int_{0}^{m^{*}_{t}}  \hat{w}^{1-\sigma}_{t}  {\Gamma^{N}_{t}}^{(\sigma-1)} (j) dj$

\begin{equation*}
\hat{w}'_{t}/\hat{w}_{t} = \frac{1}{1-\sigma} \Bigg( \frac{\Big(\delta P^{k}_{t}/\Gamma^{k}_{t}\Big)^{1-\sigma} - \tilde{w}^{1-\sigma}_{t}}{ \tilde{w}^{1-\sigma}_{t} \; \int_{0}^{m^{*}_{t}} \Gamma^{N}_{t} (j) dj } \Bigg)
\end{equation*}

Meaning that $\hat{w}'_{t} >0$ if $\tilde{w}_{t} < \delta \hat{P}^{k}_{t}/\gamma_{k}$, which is the condition to be in region 1 in Figure \ref{fig:automation_reg}. 

\textbf{Hours.} Using the Nash solution with SEP preferences, it follows that

\begin{equation*}
\epsilon_{0} h^{1/\epsilon_{1}}_{t} =\frac{\hat{Y}_{N_{t}}}{\hat{C}^{we}_{t} P_{t}} = \frac{1}{L_{t}h_{t} + (1-L_{t})\hat{b}_{t}/\hat{w}_{t}},
\end{equation*}

since $C^{we}_{t}=C^{wu}_{t}=C^{w}_{t}$ and $w_{t}=Y_{N_{t}}$. By implicit differentiation:

\begin{equation*}
\Big[\epsilon_{0}/\epsilon_{1} h^{1/\epsilon_{1} -1}_{t}  + \frac{L_{t}w^2_{t}}{(P_{t}C^{w}_{t})^2}\Big] dh =  -\frac{dm}{(P_{t}C^{w}_{t}/w_{t})^2} \Bigg[ L_{\theta} \theta_{m} \big(h_{t} - \hat{b}_{t}/\hat{w}_{t}\big) - (1-L_{t})\frac{\hat{w}'_{t}}{\hat{w}_{t}} \frac{\hat{b}_{t}}{\hat{w}_{t}} \Bigg] 
\end{equation*}

The product $(1-L_{t})\frac{\hat{w}'_{t}}{\hat{w}_{t}} \frac{\hat{b}_{t}}{\hat{w}_{t}} \approx 0$, so 

\begin{equation*}
\frac{dh}{dm} \propto - L_{\theta} \theta_{m} \big(h_{t} - \hat{b}_{t}/\hat{w}_{t}\big)
\end{equation*}

where $L_{\theta} \equiv \partial L_{t}/\partial \theta_{t} >0$ and $\theta_{m} \equiv \partial \theta_{t}/ \partial m_{t}$. Since $\hat{w}_{t} > \hat{b}_{t}$ and $h_{t} \approx 1$ by assumption, the sign of $dh/dm$ is determined by $\theta_{m}$. 

\textbf{Rate of return and vacancy/unemployment ratio.} This is the main part of the proof of Proposition \ref{prop:comp_stat}. Using the equations in \eqref{eq:steady_mu}, we can explore the changes in the labor market to a variation in $m_{t}$ near the equilibrium $\theta=\theta^*$. Starting with the Nash solution for wages:

\begin{equation*}
\left.{\frac{\partial (1+\mu^{S})}{\partial m}}\right\vert_{\theta=\theta^*}= \frac{h_{m} \hat{w}}{\bar{\omega}} + \frac{\hat{w}' h}{\bar{\omega}} - \frac{h \hat{w}}{\bar{\omega}^2} \Big[ (1-\eta_{w}) \hat{Z}_{m} +\eta_{w} h_{m} \hat{w} + \eta_{w} h \hat{w}'\Big].
\end{equation*}

Where $ \bar{\omega} \equiv (1-\eta_{w})\hat{Z} + \eta_{w}\Big( h \hat{w} + \hat{\kappa}\Big( (1-\beta^{w}/\beta^{c})(1-\hat{\lambda})/q(\theta) + (\beta^{w}/\beta^{c}) \theta \Big)$. Given SEP preferences, the steady-state value of the opportunity cost of employment is $\hat{Z}_{t} = \hat{b}_{t} + \epsilon_{1}/(1+\epsilon_{1}) h_{t} \hat{w}_{t}$. From this it follows that $\hat{Z}_{m} = \epsilon_{1}/(1+\epsilon_{1}) h_{m} \hat{w} + \epsilon_{1}/(1+\epsilon_{1}) h \hat{w}'$. Now, given that we are fixing $\theta=\theta^*$, then $h_{m} \approx 0$, such that

\begin{equation*}
\begin{split}
\left.{\frac{\partial (1+\mu^{S})}{\partial m}}\right\vert_{\theta=\theta^*} \approx  &  \frac{\hat{w}' h}{\bar{\omega}} - \frac{h \hat{w}}{\bar{\omega}^2} \Big[ (1-\eta_{w}) \frac{\epsilon_{1}}{1+\epsilon_{1}} h \hat{w}' + \eta_{w} h \hat{w}' \Big]\\
=&  \frac{\hat{w}'h}{\bar{\omega}} \Bigg[1 - \frac{\hat{w}h}{\bar{\omega}} \Big(\eta_{w} + (1-\eta_{w})\frac{\epsilon_{1}}{1+\epsilon_{1}} \Big) \Bigg].
\end{split}
\end{equation*}

Thus, $\left.{\frac{\partial (1+\mu^{S})}{\partial m}}\right\vert_{\theta=\theta^*} >0$ if $\hat{w}'_{t} >0$, which occurs in region 1 of Figure \ref{fig:automation_reg}. In this case,  the Nash solution moves to the right with an increase of $m$ (or to the left when $m$ declines, as shown in Figure \ref{fig:labor_market_comp_stat}). 

Repeating the same exercise with the labor demand equation:

\begin{equation*}
\left.{\frac{\partial \mu^{D}}{\partial m}}\right\vert_{\theta=\theta^*}  =- \frac{{\mu^{D}}^{2} \beta^{c} \hat{w}'q(\theta)}{(1-\beta^{c}(1-\hat{\lambda}))\hat{\kappa}}.
\end{equation*}

That is, $\left.{\frac{\partial \mu^{D}}{\partial m}}\right\vert_{\theta=\theta^*}  <0$ if $\hat{w}'_{t} >0$. Given that $\mu^{S}$ is a decreasing function and $\mu^{D}$ is increasing close to the initial equilibrium, it follows that the labor market reaches a new equilibrium $\theta^{**}>\theta^{*}$ following an increase in $m$ if $m>\bar{m}(\mu)$. The increase in the vacancy/unemployment ratio it follows that $L_{t}$ rises with a higher $m$. 

The final effect on $\mu_{t}$ depends on the model parameters and cannot be determined  a priori. It is most likely, however,  that $\partial \mu/\partial m \approx 0$ given that the labor supply and demand equations move in opposite directions. In what follows I will work with this assumption.

\textbf{Labor Share on Costs of Production.} Using \ref{subeq:steady_marg_prod_cap}, the share of labor on costs of production satisfies $\Omega^{c} = 1- (B\gamma_{k})^{\sigma-1} (1-m^*) \Big(\frac{\delta (1+\mu)\Omega_{I}\big(1-(1-\pi_{i})e^{-\bar{z}(\upsilon-1)/\upsilon}\big) }{1-(1-\pi_{I})e^{-\bar{z}}}\Big)^{1-\sigma}$, such that

\begin{equation*}
\left.{\frac{\partial \Omega^{c}}{\partial m}}\right\vert_{\mu=\mu^*}  =  (B\gamma_{k})^{\sigma-1}  \Bigg[\frac{   \delta (1+\mu)\Omega_{I}\big(1-(1-\pi_{I})e^{-\bar{z}(\upsilon-1)/\upsilon}\big)}{1-(1-\pi_{I})e^{-\bar{z}}}\Bigg]^{1-\sigma}
\end{equation*}

The sign of $\left.{\frac{\partial \Omega^{c}}{\partial m}}\right\vert_{\mu=\mu^*} $ is positive if $m^*=m$, which is the case when $m > \bar{m}(\mu)$. 

\textbf{Investment Expenditure to Output ratio.}  Starting with the steady-state investment-output ratio in \eqref{subeq:steady_inv_output}, it readily follows that $\left.{\frac{\partial \hat{X}_{t}/\hat{Y}_{t}}{\partial m}}\right\vert_{\mu=\mu^*} $ is negative if $m^*=m$. 

\subsubsection{Unemployment Benefits}  This part of the proof is equivalent for unemployment benefits and the discount factor of workers.\footnote{This is because we are working with SEP preferences. Other preference functions might lead to qualitatively different results for changes in $\hat{b}$ and $\beta^w$. }  Similar to the previous part of the proof, it is convenient to start with the steady-state  equation of wages.

\textbf{Wages.} Using the ideal price condition, we can express the steady-state value of wages as

\begin{equation*}
\hat{w}^{1-\sigma} = \frac{B^{1-\sigma} - (1-m^*)\Big(\delta\hat{P}^{k}/\gamma_{k}\Big)^{1-\sigma} }{\int_{0}^{m^*} {\Gamma^{N}}^{\sigma-1}_{t}(j) dj} = \frac{B^{1-\sigma} - (1-m^*)\Big(\hat{Y}_{k}\gamma_{k}\Big)^{1-\sigma} }{\int_{0}^{m^*} {\Gamma^{N}}^{\sigma-1}_{t}(j) dj}
\end{equation*}

The partial derivative of $\hat{w}$ with respect to is then determined by $\partial \mu/\partial \hat{b}$ given that $\partial \hat{Y}_{k}/\partial \mu >0$. 

\textbf{Hours.} Using again the Nash solution with SEP preferences:

\begin{equation*}
\Bigg[\frac{\epsilon_{0}}{\epsilon_{1}} h^{1/\epsilon_{1} - 1} + \frac{L \hat{w}^2}{(P \hat{C}^{w})^2}\Bigg] dh = \frac{d \hat{b}}{(P \hat{C}^{w}/\hat{w})^2} \Bigg[ L_{\theta} \theta_{\hat{b}} \big[ h - \hat{b}/\hat{w} \big] -(1-L) \frac{\hat{b}}{\hat{w}} \frac{\hat{w}_{\hat{b}}}{\hat{w}} + (1-L)/\hat{w}\Bigg].
\end{equation*}

As before, given that $(1-L)\hat{b}/\hat{w} \approx 0$, then 

\begin{equation*}
\frac{dh}{d\hat{b}} \approx - \text{contant} \; \times \; L_{\theta} \theta_{\hat{b}} \big[ h - \hat{b}/\hat{w} \big].
\end{equation*}

The sign of $\frac{dh}{d\hat{b}} $ is consequently determined, for the most part, by the sign of $\theta_{\hat{b}}$.

\textbf{Rate of return and vacancy/unemployment ratio.} Given that the sign of $\hat{w}_{\hat{b}}$ and $\frac{dh}{d\hat{b}} $ are determined by $\partial \mu/\partial \hat{b}$ and $\theta_{\hat{b}}$, we can work this part with the initial  assumption that $\frac{dh}{d\hat{b}} = \frac{dh}{d\hat{b}}  = 0$. Starting with the Nash solution near the initial equilibrium $\theta=\theta^*$:

\begin{equation*}
\left.{\frac{\partial (1+\mu^{S})}{\partial \hat{b}}}\right\vert_{\theta=\theta^*} = -(1-\eta_{w}) \frac{h \hat{w}}{\bar{\omega}^2} < 0
\end{equation*}

Similarly, given the same assumptions:

\begin{equation*}
\left.{\frac{\partial \mu^{D}}{\partial \hat{b}}}\right\vert_{\theta=\theta^*} = 0.
\end{equation*}

Joining the effects on the labor supply and labor demand equations, it follows that $\mu \downarrow$, $\theta \downarrow$, $L \downarrow$, $\hat{w} \uparrow $ and $h \uparrow$. 

\textbf{Labor share on costs of production.} Using \eqref{eq:marg_prods} it is straightforward to show that

\begin{equation*}
\frac{\partial \Omega^{c}}{\partial \hat{b}} = -(1-\sigma) (1-m^*)(B\gamma_{k})^{\sigma-1} {\hat{Y}_{k}}^{-\sigma} \frac{\partial \hat{Y}_{k}}{\partial \mu} \frac{\partial \mu}{\partial \hat{b}} > 0
\end{equation*}

if $\frac{\partial \mu}{\partial \hat{b}} < 0$ and $\sigma<1$. 

\textbf{Investment-Output and Capital-Output ratio.} Given that $\partial \hat{Y}_{k}/\partial \hat{b} > 0$, it is obvious from \eqref{eq:marg_prods} and \eqref{subeq:steady_inv_output} that $\partial (\hat{K}/\hat{Y})/\partial \hat{b} >0$ and $\partial (\hat{X}/\hat{Y})/\partial \hat{b} >0$ when $\partial \mu/\partial \hat{b}<0$. 

\section{Data Description and Analysis}\label{AppendixD}

\subsection{Data Description}\label{appendix:data_c} I use data from the BEA-BLS integrated industry-level production account from 1947 to 2016 with the intention of creating a  mapping between the model in the paper and the sectors of the economy that is meant to represent. Particularly, given that the paper focuses on the profit-making capacity of the economy, it makes sense that it concentrates on the specific sectors that are  contributing to the direct creation of aggregate profits. Though this is a controversial topic that was largely submerged with the rise of neoclassical economics---which interpreted all potentially marketable activities to be production activities--- it is instrumental for understanding the conditions allowing the reproduction of capital at an increasing scale. Here I take on the view that new wealth results from the creation of aggregate profits, and that these are the outcome of a social relation joining capitalists and workers in the production process of tangible output (goods or services). Taking a practical approach to this problem I concentrate exclusively on what \citeA{basu2013dynamics} denoted as ``value-adding sectors", which consist on the following industries:\footnote{Sectors not included in the list,  like the financial industry ( Finance, Insurance,, and Real Estate---FIRE), Education and Health Services, and Professional and Business Services, share the characteristic that national accounts impute value added onto them to make it equal to the incomes generated. The BEA, for example, calculates the value added of the banking sector from interest rate spreads between lending and deposits rates, which has direct relation with the production of goods and services in the economy.}

\begin{itemize}
\item Utilities
\item Construction
\item Manufacturing
\item Wholesale trade
\item Retail trade
\item Transportation and warehousing
\item Information
\item Administrative and waste management
\item Arts, entertainment, and recreation
\item Accommodation and food services
\item Other services, expect government
\end{itemize}

Basing the analysis on these sectors and the theoretical argument of the model,  the rate of return of capital is measured 

\begin{equation*}
\mu_{t} = \frac{P_{t}Y_{t}-P^{c}Y_{t}}{P^{c}Y_{t}}= \frac{P_{t}Y_{t} - \delta P^{k}_{t} K_{t}-w_{t} N_{t}}{\delta P^{k}_{t} K_{t}+w_{t} N_{t}},
\end{equation*}

where $P_{t}Y_{t}$ is nominal gross output minus nominal intermediate input. To obtain the value of depreciation I use Table 3.4ESI from the Fixed Assets Tables. The value of labor costs is the sum of nominal college labor input and nominal non-college labor input in the BEA-BLS data. 

The labor share is estimated as nominal college labor input plus nominal non-college labor input over net output (nominal gross output minus nominal intermediate input). The capital cost share is the nominal value of depreciation over nominal net output. The investment-output ratio, in turn, is the nominal value of fixed investment obtained from Table 3.7ESI of the Fixed Assets Tables over nominal net output.

To estimate the equilibrium capitalist savings rate I used Table 5.1. Saving and Investment by Sector and estimated        Undistributed corporate profits -      Capital consumption adjustment, corporate as the measure of retained profits for the demand of capital outlays. Net aggregate profits are measured by Corporate Profits After Tax with Inventory Valuation Adjustment (IVA) and Capital Consumption Adjustment (CCAdj). The rate of savings is ( Undistributed corporate profits -      Capital consumption adjustment) divided by net aggregate profits. Lastly, to avoid double y-axis I added 0.25 to the savings rate.

\subsection{Additional Data and Figures}

\begin{figure}[H]
    \begin{center}
    \includegraphics[width=16.5 cm,height=17.5 cm]{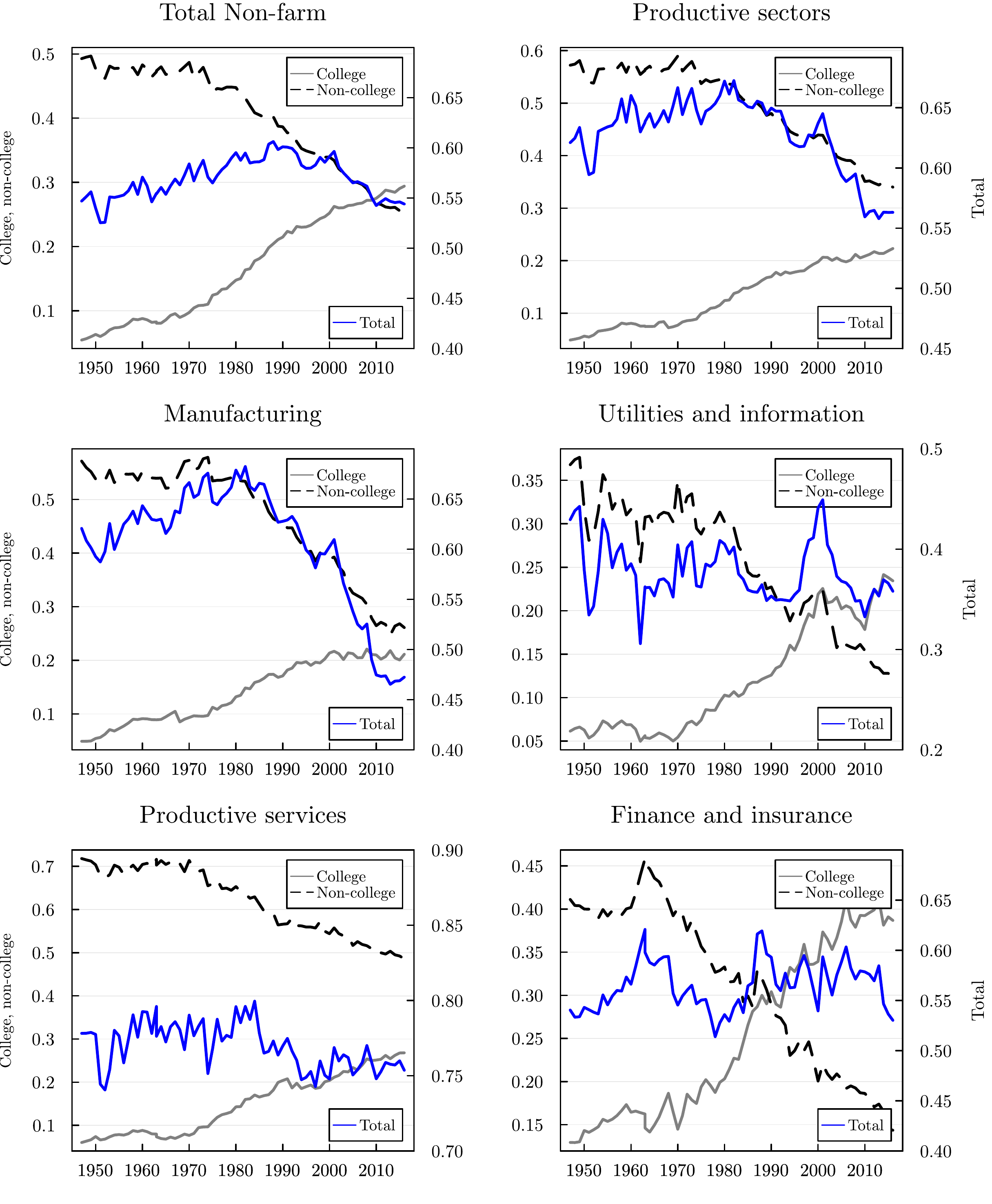}
    \end{center}
          \caption{Labor shares. \emph{Notes---} The productive service sector in the sum of Administrative and waste management, arts, Entertainment, and recreation, Accommodation and food services, and Other services, expect government. All the data is from the BEA-BLS integrated industry-level production account.  \label{fig:lab_shares1}}
  \end{figure}

This subsection compares the labor shares by sectors in the US economy using the BEA-BLS integrated industry-level production account \cite{eldridge2020toward}. In the first row of Figure \ref{fig:lab_shares1} I compare the labor shares of the total non-farm economy with the measure proposed of productive sectors.  As noted above, the sectors excluded from the analysis are those with questionable assignments of value added. By employing this modification we find two important changes: the labor share is generally higher than the total non-farm sector; and the fall in the labor share is much clearer in the productive sectors right after the early 1980s. This is an important difference given that a variety of measures of the share of labor only exhibit a clear fall after the 2000s. 

\begin{figure}[H]
    \begin{center}
    \includegraphics[width=16.5 cm,height=17.5 cm]{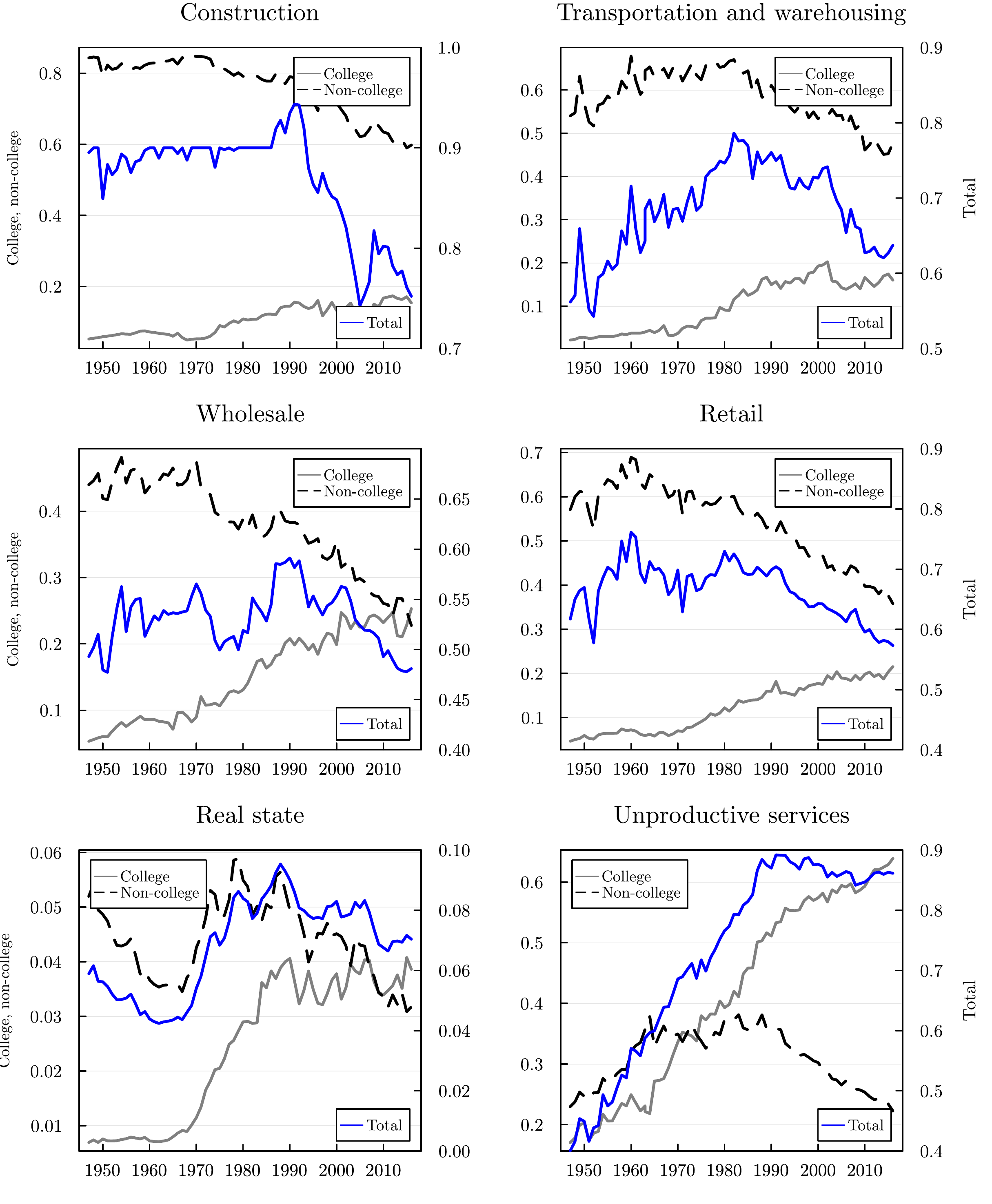}
    \end{center}
          \caption{Labor shares. \emph{Notes---} The unproductive service sector is the sum of Professional, scientific, and technical services, Management of companies and enterprises, Educational services, and Health care and social assistance . All the data is from the BEA-BLS integrated industry-level production account. \label{fig:lab_shares2}}
  \end{figure}

The data in Figures \ref{fig:lab_shares1} and \ref{fig:lab_shares2} offer a way of interpreting the behavior of the labor shares. First, it quite clear that---with the exception of unproductive service sectors (defined below)---the share of wages of non-college decreased significantly in the decade between 1970 and 1980.  That is, dividing the share of labor between college and non-college workers, it is clear that the entire fall of the aggregate wage share can be accounted for the declining participation of non-college workers on aggregate income. This, of course, is not necessarily a problem if the share of college workers increases by the same amount that the share of non-college labor is falling. The  data reveals an heterogeneous behavior between sectors: some show a considerable decline in the total share of labor in spite of the rising participation of college workers, while others show a general rise explained by the surge of the wages of college graduates. 

By and large, the sectors excluded from the analysis because of their questionable assignments of value added show a steady or even rising share of wages. For instance, the labor share in unproductive labor services---which is the sum of Professional, scientific, and technical services, Management of companies and enterprises, Educational services, and Health care and social assistance---show a remarkable rise from about 40 percent in 1950 to about 90 percent since the 1990s. The FIRE sector also shows a slight increase in the labor share since the  1970s, which contrasts with the behavior of all other major sectors in  Figures \ref{fig:lab_shares1} and \ref{fig:lab_shares2}. Even without counting manufacturing, which  shows the steepest decline in the wage share since the late 1970s, we see that sectors like productive services, utilities and information, retail, transportation and warehousing, and wholesale trade all exhibit a considerable decline in the wage share since 1980s. 

These differences in the sectors of the economy are of great importance for understanding the changes in income distribution. Particularly, one of the conclusions that can be drawn from this data is that different forces may be shaping the behavior of income distribution in each sector  depending on the role they play in the production of wealth in the economy.

  \subsection*{Additional Empirical results } 
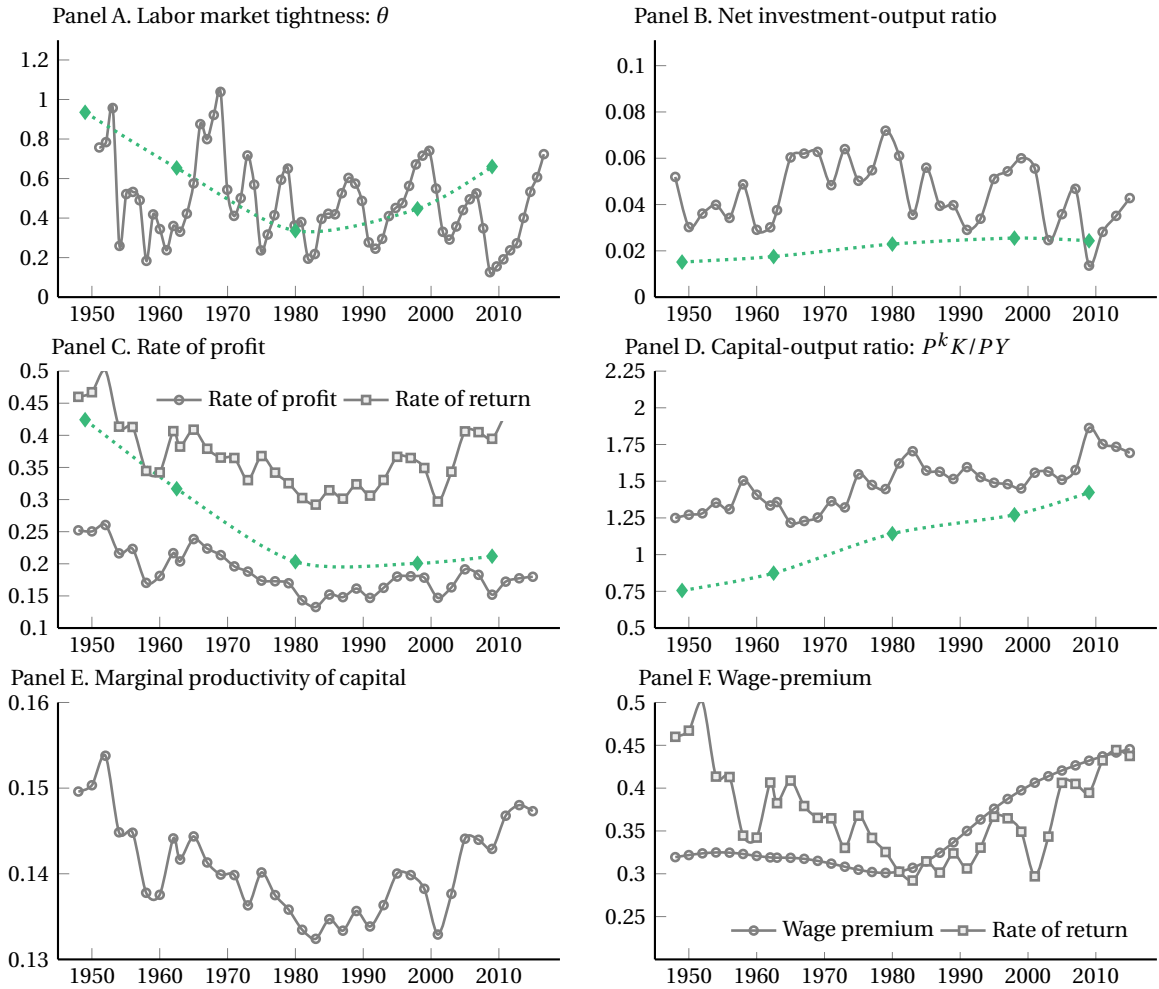
\begin{figure}
\begin{center}
\pgfplotstableread[col sep=comma,]{dt_latex_shares.csv}\datatable
\pgfplotstableread[col sep=comma,]{dt_latex_un.csv}\datatablee
\pgfplotstableread[col sep=comma,]{dt_latex_wages.csv}\datatablew
\pgfplotstableread[col sep=comma,]{dt_results_nash_SEP.csv}\datatableN
\pgfplotstableread[col sep=comma,]{dt_latex_savings.csv}\datatableS
  \begin{tikzpicture}
    \begin{axis}[
  name=plot1,
  width=0.5*\textwidth, height=5 cm,
    y label style={at={(axis description cs:-0.065,.5)},rotate=90,anchor=south},
  xmin=1945,
  xmax=2019,
  ymin=0,
  ymax=1.3,
         axis x line*=bottom,
axis y line*=left,
  tick label style={font=\footnotesize,/pgf/number format/fixed},
  title={\footnotesize Panel A. Labor market tightness: $\theta$ },
      title style={at={(axis description cs:0.325,0.95)}, anchor=south} , 
  xticklabel style={/pgf/number format/set thousands separator={}},
    ytick={0.0,0.2,...,1.2},
  xtick={1950,1960,1970,...,2020}  ]

    \addplot [mark=o, mark size = 1.5pt,mark options={fill=gray!30}, draw=gray,line width=1, smooth] table[x index = {0}, y index = {2},
  each nth point={4}]{\datatablee};

    \addplot [Green!70,smooth,mark=diamond*,mark size=3pt, dotted, line width=1.25,
    mark options={fill=Green!70,draw opacity=0}]  table[x index = {0}, y index = {11}]{\datatableN};

    \end{axis}

      \begin{axis}[
  name=plot2,
  width=0.5*\textwidth, height=5 cm,
    at=(plot1.right of north east), anchor=left of north west,
    y label style={at={(axis description cs:-0.065,.5)},rotate=90,anchor=south},
  xmin=1945,
  xmax=2019,
  ymin=0.0,
  ymax=0.111,
         axis x line*=bottom,
axis y line*=left,
  tick label style={font=\footnotesize,/pgf/number format/fixed},
  title={\footnotesize Panel B. Net investment-output ratio },
      title style={at={(axis description cs:0.32,0.95)}, anchor=south} , 
  xticklabel style={/pgf/number format/set thousands separator={}},
    ytick={0,0.02,...,0.11} ,
  xtick={1950,1960,1970,...,2020}  ]

   \addplot [mark=o, mark size = 1.5pt,mark options={fill=gray!50}, draw=gray,line width=1, smooth] table[x index = {0}, y index = {4},
  each nth point={2}]{\datatablew};

  \addplot  [Green!70,smooth,mark=diamond*,mark size=3pt, dotted,line width=1.25,
    mark options={fill=Green!70,draw opacity=0}]  table[x index = {0}, y index = {8}]{\datatableN};

  \end{axis}

    \begin{axis}[
  name=plot3,
  width=0.5*\textwidth, height=5 cm,
   at=(plot1.below south east), anchor=above north east, 
    y label style={at={(axis description cs:-0.065,.5)},rotate=90,anchor=south},
  xmin=1945,
  xmax=2019,
  ymin=0.1,
  ymax=0.5,
       axis x line*=bottom,
axis y line*=left,
legend columns=2, 
  legend pos=north east,
  legend style={draw=none,font=\footnotesize},
   legend cell align={left}, 
  tick label style={font=\footnotesize,/pgf/number format/fixed},
  title={\footnotesize Panel C. Rate of profit},
      title style={at={(axis description cs:0.2,0.95)}, anchor=south} , 
  xticklabel style={/pgf/number format/set thousands separator={}},
  xtick={1950,1960,1970,...,2020},
  ytick={0.1,0.15,...,0.5}  ]  

  \addplot [mark=o, mark size = 1.5pt,mark options={fill=lightgray!30}, draw=gray,line width= 1, smooth] table[x index = {0}, y index = {3},
  each nth point={2}]{\datatable};
  \addlegendentry{Rate of profit};

            \addplot [mark=square*, mark size = 1.5pt,mark options={fill=lightgray!30}, draw=gray,line width= 1, smooth] table[x index = {0}, y index = {1},
  each nth point={2}]{\datatable};
    \addlegendentry{Rate of return};
   
  \addplot  [Green!70,smooth,mark=diamond*,mark size=3pt, dotted,line width=1.25,
    mark options={fill=Green!70,draw opacity=0}]  table[x index = {0}, y index = {14}]{\datatableN};

  \end{axis}

    \begin{axis}[
  name=plot4,
  width=0.5*\textwidth, height=5 cm,
 at=(plot3.right of north east), anchor=left of north west,
    y label style={at={(axis description cs:-0.065,.5)},rotate=90,anchor=south},
  xmin=1945,
  xmax=2019,
  ymin=0.5,
  ymax=2.25,
         axis x line*=bottom,
axis y line*=left,
  tick label style={font=\footnotesize,/pgf/number format/fixed},
  title={\footnotesize Panel D. Capital-output ratio:  $P^k K/PY$ },
      title style={at={(axis description cs:0.325,0.95)}, anchor=south} , 
  xticklabel style={/pgf/number format/set thousands separator={}},
    ytick={0.5,0.75,...,2.25},
  xtick={1950,1960,1970,...,2020}  ]

  \addplot [mark=o, mark size = 1.5pt,mark options={fill=lightgray!30}, draw=gray,line width= 1, smooth] table[x index = {0}, y index = {4},
  each nth point={2}]{\datatable};

  \addplot  [Green!70,smooth,mark=diamond*,mark size=3pt, dotted,line width=1.25,
    mark options={fill=Green!70,draw opacity=0}]  table[x index = {0}, y index = {13}]{\datatableN};

    \end{axis}

    \begin{axis}[
  name=plot5,
  width=0.5*\textwidth, height=5 cm,
   at=(plot3.below south east), anchor=above north east, 
    y label style={at={(axis description cs:-0.065,.5)},rotate=90,anchor=south},
  xmin=1945,
  xmax=2019,
  ymin=0.13,
  ymax=0.16,
       axis x line*=bottom,
axis y line*=left,
  tick label style={font=\footnotesize,/pgf/number format/fixed},
  title={\footnotesize Panel E. Marginal productivity of capital},
      title style={at={(axis description cs:0.3,0.95)}, anchor=south} , 
  xticklabel style={/pgf/number format/set thousands separator={}},
  xtick={1950,1960,1970,...,2020},
  ytick={0.13,0.14,...,0.16}  ]  

  \addplot [mark=o, mark size = 1.5pt,mark options={fill=lightgray!30}, draw=gray,line width= 1, smooth] table[x index = {0}, y index = {6},
  each nth point={2}]{\datatable};

  \end{axis}

    \begin{axis}[
  name=plot6,
  width=0.5*\textwidth, height=5 cm,
 at=(plot5.right of north east), anchor=left of north west,
    y label style={at={(axis description cs:-0.065,.5)},rotate=90,anchor=south},
  xmin=1945,
  xmax=2019,
  ymin=0.2,
  ymax=0.5,
         axis x line*=bottom,
axis y line*=left,
legend columns=2, 
  legend pos=south east,
  legend style={draw=none,font=\footnotesize},
   legend cell align={left}, 
  tick label style={font=\footnotesize,/pgf/number format/fixed},
  title={\footnotesize Panel F. Wage-premium },
      title style={at={(axis description cs:0.2,0.95)}, anchor=south} , 
  xticklabel style={/pgf/number format/set thousands separator={}},
    ytick={0.25,0.3,...,0.5},
  xtick={1950,1960,1970,...,2020}  ]

  \addplot [mark=o, mark size = 1.5pt,mark options={fill=lightgray!30}, draw=gray,line width= 1, smooth] table[x index = {0}, y index = {8},
  each nth point={2}]{\datatable};
       \addlegendentry{Wage premium}; 
       
       \addplot [mark=square*, mark size = 1.5pt,mark options={fill=lightgray!30}, draw=gray,line width= 1, smooth] table[x index = {0}, y index = {1},
  each nth point={2}]{\datatable};
  \addlegendentry{Rate of return};

    \end{axis}

    \end{tikzpicture}
  \end{center}
  \caption{Additional equilibria. \emph{Notes---}  Panels  B,C,D,E, and F   are  based on  the BLS-BEA integrated data set  \protect  \cite{eldridge2020toward}. Panel A uses the non-farming  vacancy data of  \protect  \citeA{petrosky2021unemployment}. The wage-premium is normalized to the 1981 value of the rate of return.   As in Figure \ref{fig:counterfactual_test}, the green diamonds are the result of adjusting $m$ and $\beta^w$ to the time averages of the capital cost share and the labor share. \label{fig:counter_2}}
  \end{figure}

Figure \ref{fig:counter_2} shows some results which complement the empirical results in the main text. Panel A corroborates the results of Panel E in Figure \ref{fig:counterfactual_test}, showing that the model does capture the main movements of the labor market through time. The results of net investment-output ratio, rate of profit, and the capital-output show a similar pattern as that of the investment-output ratio in Figure \ref{fig:counterfactual_test}. Essentially, these results indicate that the calibration of the depreciation rate was probably too high, given that with a lower $\delta$, the fit of the model in Panel C of Figure \ref{fig:counterfactual_test} would have required a higher capital-output ratio, and this would have implied  higher values of $P^{k} K/ PY$, $(P^{I} X-\delta P^{k} K)/P Y$ (panel B en Figure \ref{fig:counter_2}), on one hand, and a lower value for the rate of profit, on the other
  
  A lower $\delta$, however, would have increased the steady-state values of the savings rate. The reason for this is that---according to Theorem \ref{theorem:equil_harrod}---in the steady-state equilibrium: $s \approx g(1-\Omega^c)/(\delta \mu -(1-\Omega^c)(\tau + \zeta))$. In the paper I made the decision to underpredict the value of $P^{k} K/ PY$ to maintain relatively low rates of savings, even though I am aware that the model should overpredict the value of $s$ given the assumption that all investments are financed from retained earnings.  
  
  Having cleared out why the model underpredicts   the capital(investment)-output ratio, we may turn our attention to Panel C, which displays the rate of profit (measured as profits over the capital stock), the rate of return (measured as profits over costs of production), and the implied equilibrium rate of profit from equation \eqref{subeq:steady_profit_rate}. One of the key results here is that, unlike the rate of return, the rate of profit was kept relatively constant after the 1980s. This is explained by the increase in the rate of automation which, in accordance to Proposition \ref{prop:comp_stat} and  Figure \ref{fig:cap_market_comp_stat}, reduces the rate of profit because it increases the value of capital relative to the final good. This negative effect on $r$ is balanced with the increase in $\mu$ caused by the deterioration of the bargaining power of labor. 
  
The data of the marginal productivity of capital is obtained using \eqref{subeq:steady_marg_prod_cap} and estimating

\begin{equation*}
\hat{Y}_{K_{t}} = \frac{\delta (1+\mu_{t}) \Omega_{I} (1-(1-\pi_{I})e^{-\bar{z}(\upsilon-1)/\upsilon}}{1-(1-\pi_{I})e^{-\bar{z}}}
\end{equation*}
  
From this perspective, the marginal productivity of capital is directly obtained by the rate of return of capital, which---in turn---is determined as a social outcome resulting from the bargaining process of wages between capitalists and workers. Stated differently, the  marginal productivity of capital is here a meaningless concept \emph{if} it is detached from the social elements determining the return of capital.  Note, in addition, that the marginal productivity of capital---unlike the rate of profit---increases considerably after the 1980s, which is precisely what is expected from  Proposition \ref{prop:comp_stat} and  Figure \ref{fig:cap_market_comp_stat}.

Finally, Panel F shows a clear positive correlation between the wage premium and the rate of return of capital. Though the wage premium is not a topic which is directly treated in the paper, the data in Figure \ref{fig:counter_2} provides some evidence to the hypothesis that the  increase of the marginal productivity of skilled workers is responding to external factors like rising sales and  profits, all of which may be the result of political changes favoring top income earners  (see, e.g.,\citeA{piketty2013capital}) . The role that, e.g., managers and CEOs play in the creation of profits is a topic which is well worth studying in future research.

\subsection{Equation \eqref{eq:tvp_ub} }\label{appendix:kalman}

The public policy equation in \eqref{eq:tvp_ub}  is estimated using a Kalman filter and the following priors on the model:

\begin{equation*}
\beta^{b} \sim N(\beta^{b}_{0}, M_{0}), \; \;\; \varphi^{2}_{b} \sim IG(\nu_{0}, s_{0}),\;\; \varphi^{2}_{\beta^{b}_{i}} \sim IG(\nu^{\beta}_{0},s^{\beta}_{0}) \; \text{for} \; i=0,1,...,p+1. 
\end{equation*}

Time-varying models often require information priors to narrow down the uncertainty of the parameters. Here I set $\beta^{b}_{0} =0_{p+1 \times 1}$ to avoid including any bias in the sign of the parameters and set $M_{0}=0.5 \times I_{p+1 \times p+1}$ to convey the message that the initial values are probably close to zero. Working with the assumption that the time-varying parameters are not too far from the fixed parameter solution, I set $s^{\beta}_{0}=(s^{\beta}_{0}-1) \times [0.1^2, 0.01^2, 0.01^2, 0.05^2]$ and $\nu^{\beta}=5$. That is, the expected of the variance of the parameters is equal to two times the vector on the right-hand side of $s^{\beta}_{0}$. Similarly, using the sample variance of UI extensions over labor productivity is about 0.00018, I set $s_{0}=(\nu_{0}/2-1)\times 0.01^2$ with $\nu_{0}=5$. 

Given these priors it is quite simple to estimate the posterior distribution of the parameters using a Kalman filter and a Gibbs sampler. Further computational details can be found in \citeA{prado2010time}. 

  \subsubsection{Analysis of vacancy costs}\label{appendix:cavancy}  It is important to note that the data of rates of return and of  labor market tightness require vacancy costs which are significantly higher than those commonly reported in the literature. Using the calibration in Table \ref{table:calibration_steady_state1} and the results  in Figure \ref{fig:counterfactual_test}, the proportional costs of hiring are on average 4.7 times greater than the average productivity of labor. Though this may be  an  implausibly large value in light of the empirical studies of \citeA{silva2009labor}, who show that recruiting costs are about 14 percent of quarterly pay per hire, and the estimates of \citeA{merz2007labor}, showing that the marginal costs of hiring are about 1.5 times the average productivity of labor, it is necessary if the model is to satisfy the conditions for steady-state growth paths in Theorem \ref{theorem:equil_harrod}.

  \begin{table}\caption{Implied rates of return in the literature}
    \centering 
  \begin{adjustbox}{width=0.75\textwidth}
    \small 
  \begin{tabular}{c c c l}
  \hline \hline
  Vacancy costs ($\kappa/Y_{N}$) & $\mu$ & $\mu^{\min}$ &  Source \\
  \hline
  0.21  &  0.015  & 0.046&  \citeNP{shimer2005cyclical} \\
 0.585 & 0.0124  &  0.0684  & \citeNP{hagedorn2008cyclical}\\
0.433 &  0.0154  & 0.0242  & \citeNP{hall2008limited}\\
  0.356 &  0.017  & 0.0272    & \citeNP{pissarides2009}\\
    0.7 &  0.0926  & 0.1192  & \citeNP{petrosky2018endogenous}\\
0.531 &  0.022  & 0.0311    & \citeNP{petrosky2021unemployment}\\
    \hline 
  \end{tabular}
    \end{adjustbox}
      \caption*{\emph{Notes---} The minimum rate of return is obtained by setting aggregate profits equal to the sum of vacancy costs plus taxes as in equation \eqref{eq:upper_power}.  \label{table:implied_rate_of_return}} 
  \end{table}

  Table \ref{table:implied_rate_of_return} reports some   commonly used calibrations for vacancy costs relative to average labor productivity and shows that under no circumstance can any of the models of the equilibrium unemployment literature satisfy the condition where capitalists are capable of paying  taxes,  vacancy costs and have a remnant for financing their own consumption.  Search and matching models generally avoid this problem by working under the assumption that all households share the ownership of capital, so it makes  no  difference  whether consumption is financed from  wages or profits.  This, however, merely hides the problem since it does not address the key issue  of showing that---given the assumptions of the model---the economy can expand at an increasing scale through time.

\end{document}